%% file: main.tex
\documentclass[12pt]{article}
\pdfoutput=1

\usepackage[utf8]{inputenc}
\usepackage{xcolor}
\usepackage{amsmath,amssymb,mathrsfs,bm,xspace}
\usepackage{mathtools}
\usepackage{amsthm}
\newtheorem{proposition}{Proposition}
\usepackage{graphicx}
\usepackage{float}
\usepackage{tikz}
\usetikzlibrary{arrows.meta,positioning,shapes.geometric,calc}
\newcommand{\falseverdict}[1]{\textcolor{red!80!black}{#1}}
\newcommand{\trueverdict}[1]{\textcolor{green!45!black}{#1}}
\usepackage{comment}
\usepackage{booktabs}
\usepackage{longtable}
\usepackage{multirow}
\usepackage{tabularx}
\usepackage{geometry}
\usepackage{enumitem}
\usepackage[english]{babel}
\usepackage{csquotes}
\MakeOuterQuote{"}
\usepackage[font={small}]{caption}

\definecolor{darkblue}{rgb}{0.1,0.1,0.7}
\usepackage[colorlinks, linkcolor=darkblue, citecolor=darkblue, urlcolor=darkblue, linktocpage, backref=page]{hyperref}
\renewcommand*{\backrefalt}[4]{\ifcase #1 \or (p.~#2)\else (pp.~#2)\fi}
\makeatletter
\patchcmd{\Hy@backout}{{\thepage}{\@currentlabel}{\@currentHref}}{{\thepage}{}{}}{}{\PackageWarning{main}{backref patch failed}}
\makeatother

\definecolor{martenorange}{rgb}{0.91, 0.41, 0.17}

\input{shorthand.tex}

\numberwithin{equation}{section}
\begin{document}
\vspace*{-.6in} \thispagestyle{empty}
\vspace{1cm}
{\Large
	\begin{center}
		{\bf Guaranteed Slater Conditions for the Numerical Conformal Bootstrap}\\[6pt] Fixing Instabilities in SDPB, Navigator Methods, and Skydiving
	\end{center}
}
\vspace{0.6cm}
\begin{center}
	{\bf Marten Reehorst}\\[0.7cm]
	{
{\it Department of Mathematics, King's College London, Strand, London WC2R 2LS, United Kingdom}\\
	}
	\vspace{0.4cm}
\end{center}

\begin{abstract}
The semidefinite programs (SDPs) of the numerical conformal bootstrap, and of S-matrix and quantum-mechanics bootstraps, are solved with primal-dual interior-point methods whose convergence guarantees rest on the primal and dual Slater conditions: the existence of strictly feasible points on both sides of the duality. Many bootstrap SDPs violate them, for instance when positivity forces whole sectors of a multi-correlator functional to vanish, or when the prefactor factored out of the conformal blocks plants a zero at the endpoint of a sector. The consequences range from slow convergence to inconclusive allowed/disallowed verdicts and the breakdown of the Skydive algorithm.

We introduce \enquote{Slater tools}, all acting within the PMP class accepted by SDPB. Diagnostics measure both Slater margins of any bootstrap SDP and, when a margin vanishes, extract the exposing certificate that identifies the degeneracy. Two independent cures follow. A transform enforces each Slater condition by one penalized slack variable and bounds the original optimum from one side; the bound becomes exact once the penalty weight exceeds a finite threshold, and the run itself certifies when it has. Facial reduction instead uses the exposing certificate to remove the degeneracy itself, which allows two-sided bounds.

To illustrate the tools we reproduce, explain and resolve all known types of failures in a mixed-correlator 3d Ising setup. We show that the $X$-Slater condition (SDPB's primal) fails at every disallowed point in feasibility mode, holds in every navigator problem we tested, and can be guaranteed by one additional condition on the navigator function. We show that the folklore `instability remedy' of a twist gap works by removing degeneracies at the endpoints of the positivity conditions. These often stem from a mistake in factoring out the block prefactor but can also arise legitimately in channels that mix equal and unequal operator pairs. Degeneracies away from the endpoints are beyond the twist gap's reach, but not beyond that of our tools. Finally we classify and remove all degeneracies of the published current--scalar bootstrap, among them one at $x=\infty$ caused by unequal polynomial degrees within one matrix block, which no gap could cure.
\end{abstract}

\newpage

{\small
	\parskip=-0.1em
\tableofcontents
}

\newpage

\section{Introduction and summary}
\label{sec:intro}

Since 2008 the numerical conformal bootstrap\footnote{See \cite{Poland:2018epd} for a thorough review of the first decade, up to 2018, \cite{Poland:2022qrs} for the state of the field in 2022, and \cite{Rychkov:2023wsd} for the developments up to 2023.} has used the positivity of crossing-symmetric correlators \cite{Rattazzi:2008pe} to obtain some of the most precise nonperturbative results in quantum field theory, most famously the determination of the 3d Ising critical exponents \cite{ElShowk:2012ht,Kos:2014bka,Kos:2016ysd}. At the computational level, essentially every such computation is a semidefinite program (SDP) with a highly structured sparse form inherited from a polynomial matrix program (PMP). These problems are solved by the arbitrary-precision primal-dual interior-point solver SDPB, which is optimized for this structure \cite{Simmons-Duffin:2015qma,Landry:2019qug}. More recent algorithmic advances build additional machinery on top of the same interior-point core: the navigator function \cite{Reehorst:2021ykw}, which replaces the binary \enquote{allowed/disallowed} oracle by a differentiable objective, and the Skydive algorithm \cite{Liu:2023elz}, which optimizes over external parameters while solving the family of SDPs (hot-starting from point to point).

All of this machinery comes with convergence guarantees, but these guarantees have important qualifications. The polynomial-time convergence of primal-dual interior-point methods, and even the existence of the central path (the curve of strictly feasible points that SDPB and Skydive track towards the optimum), requires the primal and dual \emph{Slater conditions}: on each side of the SDP duality there must exist a feasible point whose semidefinite matrix variable is strictly positive definite. These qualifications matter, because many SDPs that arise in the numerical bootstrap violate the Slater conditions. Several works report that a dual feasible jump never occurred in their setup \cite{Chester:2019ifh,Bartlett-Tisdall:2023ghh,Bartlett-Tisdall:2024mbx,Nancarrow:2022wdr,Ferretti:2026iij}, that jumps were lost in some region of parameter space \cite{Meineri:2025qwz,Nakayama:2025mrm}, that the dual error decreased only gradually \cite{Liu:2023elz}, or that a primal-side verdict by error threshold turned out to be wrong \cite{Go:2019lke}. In each case the remedy was a workaround: a small twist gap, or a fall back to threshold mode, in which the run is stopped once one error falls below a threshold and whose verdicts Section \ref{sec:feascorrupt} shows to be unreliable. The same degeneracies are reported outside the conformal bootstrap. In the S-matrix bootstrap, \cite{Chiang:2026ubv} find leading coefficients that positivity forces to vanish for every functional and \cite{Ferretti:2026iij} lose the dual jump to near-zero eigenvalues; in the quantum-mechanical bootstrap, \cite{Lin:2025srf} trace their instabilities to the lack of strict feasibility, \cite{Lawrence:2024mnj} find that no strictly feasible point exists, and \cite{Nancarrow:2022wdr} find that the dual jump ``will never work'' (Appendix \ref{app:survey}). The tools of this paper diagnose the degeneracy behind such reports and remove it. A characteristic example is a multi-correlator setup in which positivity forces every functional satisfying the constraints to have identically vanishing components in certain directions, for instance all components acting on the crossing equations of one external operator. In such a case, every dual-feasible matrix is confined to the boundary of the positive-semidefinite cone, and the Slater condition of that side fails. Following \cite{Simmons-Duffin:2015qma} we call the two sides $X$ (SDPB's primal, the densities) and $Y$ (SDPB's dual, the functional); Section \ref{sec:XYproblems} gives the dictionary. When a Slater condition fails:
\begin{itemize}
	\item SDPB can slow to a crawl, stall or return an inconclusive verdict;
	\item the Skydive algorithm fails, because its centering step requires driving $XY \to \mu I$, which is impossible when every feasible $Y$ is singular, so that the centering residual diverges and no external step is ever taken;
	\item an uncareful observer -- who relies on threshold-race termination conditions in SDPB -- can mistake a stalled output for a rigorous certificate, leading to misclassification of the point.
\end{itemize}

The purpose of this paper is to make these failures diagnosable and curable in a systematic way. We present a family of \emph{Slater tools}: transformations of the SDP data that either measure the Slater margins (diagnostics) or perturb the problem slightly so that the Slater conditions hold by construction (guarantees); two guarantees bound the original optimum from above and from below. Two design constraints guide the construction:
\begin{enumerate}
	\item The output must again be a valid SDPB input. SDPB runs only on a specific class of SDPs with a particular form: those obtained from a PMP by SDPB's sampling procedure (Section \ref{sec:pmp}). We call a member of this class an \SDP. A transformation is only useful if its output is again an \SDP, which in practice means it acts on the PMP level or on the linear data $(B,b,c)$, leaving the existing structure intact. This rules out the textbook fixes in their textbook form, such as facial reduction acting on the SDP, or the homogeneous self-dual embedding, which adds constraint matrices built from $b$ and $c$. Both are useful for a general-purpose SDP solver but do not produce an \SDP.\footnote{Facial reduction \cite{BorweinWolkowicz:1981,BorweinWolkowicz:1981b,WakiMuramatsu:2013,Pataki:2013,DrusvyatskiyWolkowicz:2017,PermenterParrilo:2018,ZhuPatakiTranDinh:2019} restricts the problem to the minimal face of the cone containing its feasible set. Done at the level of the SDP, it projects the constraint matrices out of the PMP class. Also note that finding the face this way requires solving a sequence of SDPs that could themselves be degenerate. Another alternative, the homogeneous self-dual embedding \cite{YeToddMizuno:1994,deKlerkRoosTerlaky:1997}, used by general-purpose conic solvers, couples the $X$- and $Y$-blocks through constraint matrices built from $b$ and $c$ \cite{PermenterFribergAndersen:2017}.} Using a general-purpose SDP solver is also not an option, since such solvers either cannot deal with the arbitrary-precision requirements of the bootstrap or are far too slow.
	\item The transformed problem must remain informative, in that its optimum should either equal the original one or bound it rigorously, so that one can still rigorously decide whether points are allowed or disallowed.\footnote{Throughout, \enquote{rigorous} refers to inequalities between the optimal values of the \SDPs\ involved. For the purpose of this paper we ignore that these do not result in mathematically proven bounds. See \cite{Reehorst:rigorous} for how to modify bootstrap setups to obtain mathematically rigorous bounds.}
\end{enumerate}

Our main results are the following:
\begin{enumerate}
	\item \textbf{Diagnostics.} We construct SDPB-compatible probes of both Slater conditions. The $Y$-side diagnostic \texttt{ycompute} computes the \emph{$Y$-margin} $\vareps^*$, the largest uniform margin by which the dual matrix can be pushed inside the cone; its sign classifies the original problem as strictly feasible ($\vareps^*>0$), strictly infeasible ($\vareps^*<0$), or on the boundary ($\vareps^*=0$), where weak feasibility and weak infeasibility can be told apart by a second check, or by facial reduction (Sections \ref{sec:attainment} and \ref{sec:feasrigorous}). On the $X$-side we observe that the default bilinear bases of SDPB~3.1 and later \cite{SDPB-repo} make the identity matrix an exact combination of the constraint matrices (the sum over the diagonal constraint rows), so the textbook $X$-slack $X-tI\succeq0$ can be implemented by a shift of variables; the resulting diagnostic \texttt{xcompute} measures the $X$-margin $t^*$ (Section \ref{sec:xside}). When the $Y$-problem is feasible, the $X$-margin is positive exactly when the set of optimal functionals is nonempty and bounded; it is healthy on every navigator problem we tested and vanishes at every disallowed point of the feasibility-mode runs we tested (Section \ref{sec:feasxslater}).
	\item \textbf{Guarantees.} The transformation \texttt{yguarantee} adds one penalized slack variable $\vareps\le 0$ and makes the $Y$-Slater condition hold unconditionally. Its optimal value is a \emph{rigorous upper bound} on the original optimal value (the navigator value) for any penalty $\lambda>0$; absent a duality gap it converges to it as $\lambda\to\infty$ and, when the original $X$-optimum is attained, equals it above a finite threshold $L_*$; a single solve can certify that $\lambda$ is above this threshold for a given problem (Section \ref{sec:yguarantee}). The $X$-side guarantee \texttt{xguarantee} relaxes $X\succeq0$ to $X+\eta I\succeq0$ with a penalty $\nu\eta$. On the functional side this is the trace bound $\Tr Y\le\nu$. The optimal value of \texttt{xguarantee} is a lower bound on the navigator value for any $\nu$, and equals it as soon as $\nu$ exceeds the trace of an optimal $Y$; that the trace bound is then inactive is read off from the solution (Section \ref{sec:xguarantee}). Together they bracket the navigator value from both sides. The two guarantees modify different parts of the data and compose; \texttt{xyguarantee} applies both in one call. For a large enough trace bound $\nu$ this satisfies both Slater conditions whatever the original problem. It returns the navigator value when both optima of the original problem are attained; when one is not, the guarantee bounds it from one side only. Facial reduction can be used to bound the other side (Section \ref{sec:twosided}). An alternative $X$-guarantee that relaxes the linear constraints instead (the \enquote{$B$-slack}) is described in Appendix \ref{app:bslack}. The guarantees are more than a stopgap: a degeneracy that lives in the solution, like the vanishing of a functional on crossing equations that turn out to be inactive, is not visible in the input and can change from one external point or truncation to the next, so it cannot be removed once and for all before a scan. The guarantee makes every point non-degenerate by one transformation of the input, which is what a scan, or a Skydive run that must center at every point it visits, requires.
	\item \textbf{Practical examples.} We diagnose and resolve examples of previously unexplained stalls in SDPB and in the Skydive algorithm (stalls of SDPB are reported in \cite{Go:2020ahx,Chester:2023ehi}, and \cite{DiPietro:2025ozw,Ferretti:2026iij} run a modified SDPB that recovers from them). The $Y$-margin diagnostic returns a margin that is zero at the duality gap of the solve, indicating a $Y$-Slater failure, while the $X$-side margin is healthy. Applying \texttt{yguarantee} restores $Y$-Slater by construction, and with it Skydive's centering (Section \ref{sec:casestudy} and Appendix \ref{sec:skydive-ab}). The construction is not tied to the navigator: it applies to any bounded objective whose non-degenerate version SDPB can solve.
	
	We also examine a common misuse of SDPB where feasibility of a program is decided by a threshold race, i.e.\ by terminating when a primal or dual error reaches a certain threshold first. We show how this can misclassify points in both directions (Section \ref{sec:feascorrupt}) and show that a reliable verdict requires a dual jump to conclude a point is disallowed or a sign-checked primal jump, a primal jump along with a negative $c\cdot x$, to conclude the point allowed (Section \ref{sec:feasjump}). Since only programs where the $X$-Slater condition holds can have a primal jump and only programs where the $Y$-Slater condition holds can have a dual jump, a Slater failure prevents a reliable verdict. It has to be removed, by facial reduction or by posing the question in a form that satisfies the Slater conditions (e.g.\ using a navigator function with the appropriate guaranteeing transform) (Sections \ref{sec:feasxslater} and \ref{sec:feasrigorous}).
	
	\item \textbf{Facial reduction within the PMP class.} Beyond diagnosis and restoration of convergence, we show that the converged \texttt{ycompute} solve identifies which functional components or points are forced to vanish on a degenerate problem, and that these can usually be fixed exactly, so that the reduction is performed on the PMP data itself, as facial reduction that never leaves the class of \SDPs\ (Section \ref{sec:pmpfr}). Carried out on the case study, the reduction preserves the navigator value while reopening the Slater margin from zero to a strictly positive value, and identifies the degeneracy analytically. A second case study demonstrates the same toolchain as a debugging instrument: a deliberately planted wrong-spin sector is detected, localized and explained (Section \ref{sec:faceworked2}).
	\item \textbf{The twist-gap folklore.} Section \ref{sec:twistgap} studies a folklore twist-gap remedy for stalls. This remedy resolves degeneracies at the positivity condition's end point. Such degeneracies occur for two reasons: A common convention for factoring out the positive prefactor of the conformal blocks (the shared prefactor) includes poles the blocks do not have, and so plants exact zero rows at the endpoints of spinning sectors gapped at unitarity. This $Y$-Slater failure is is removed by taking out only the poles each block actually has. A second endpoint degeneracy is genuine, because in channels whose OPE basis mixes equal and unequal pairs of operators the equal-pair rows vanish at the unitarity bound even with the minimal prefactor; no exact reduction within the PMP class removes it, and there a small twist gap, with conserved currents kept as isolated operators, is the right cure. We also show a twist gap can never cure other common degeneracies that occurs away from the endpoint.
	\item \textbf{A published degenerate example.} Section \ref{sec:mixedcurrent} revisits the mixed current--scalar bootstrap of \cite{Reehorst:2019pzi}. Its formulation carries four degeneracies, three of them Slater failures and one a $B$-degeneracy, all of them removable: a redundant set of equations, a zero planted by the normalization of the conserved three-point structures (a convention-planted numerator zero, the mechanism of Section \ref{sec:fakepoles} in another guise), unequal polynomial degrees within one matrix block (a forced contact at $x=\infty$, which a twist gap cannot cure and which the tools of this paper detect while a pointwise check does not), and spurious leading coefficients left by roundoff and block truncation. Each is resolved by one of the facial-reduction moves of Section \ref{sec:pmpfr} or by a different choice of equations. Because the contact at $x=\infty$ is present at every point and every $\Lambda$, no dual jump could ever fire in that formulation, and its allowed regions had to be decided by less reliable threshold races.
\end{enumerate}

Our aim throughout is to understand the degeneracies that actually occur in bootstrap numerics and to resolve them within the $\SDP$ framework on which bootstrap computations depend.

All tools described in this paper are available in the repository \cite{SlaterTools}, a fork of SDPB that adds them as separate programs: \texttt{ycompute}, \texttt{yguarantee}, \texttt{xcompute}, \texttt{xguarantee}, the combined \texttt{xyguarantee}, and the alternative \texttt{xguarantee\_\allowbreak bslack} of Appendix \ref{app:bslack}. Figure \ref{fig:workflow} in Section \ref{sec:toolsummary} arranges them into a single procedure. They are implemented in C++ as composable transformations acting on standard SDPB input directories, and their output is solved by the unmodified SDPB (version 3.1 or later). The computations were performed in the PolyBoot framework \cite{PolyBoot}. The directory \texttt{papers/\allowbreak slater\_\allowbreak conditions} of that repository is organized in run families (F01, F02, \dots) and maps every numerical table and the figure of this paper to the problem definitions, task launchers, job scripts and reference outputs behind it; it marks which launchers and scripts were reconstructed afterwards.

The paper is structured as follows. Section \ref{sec:slater} reviews the \SDPs \, solved by SDPB, the Slater conditions and the central path. Sections \ref{sec:diagnostics} and \ref{sec:guarantees} construct the diagnostics and the guarantees, and Section \ref{sec:pmpfr} develops facial reduction within the PMP class. Sections \ref{sec:casestudy}--\ref{sec:twistgap} apply the tools: to a Skydive stall, to feasibility mode, to a planted configuration error, and to the twist-gap folklore. Section \ref{sec:mixedcurrent} places the degeneracies of a published spinning setup, the mixed current--scalar bootstrap, in this classification and shows how they were removed, and Section \ref{sec:conclusions} concludes. The appendices treat the $Y$-margin at the \SDP\ level (Appendix \ref{app:Y-slack-PMP-to-SDP}), the faithfulness of the PMP-to-\SDP\ translation (Appendix \ref{app:faithful}), and alternative $X$-side constructions together with the lessons from implementations that fail (Appendix \ref{app:xslack-elim}); the run-level detail of the Skydive case study is in Appendix \ref{app:skydive}, and Appendix \ref{app:survey} tabulates the SDPB termination settings used in the published literature.

\section{SDPs, Slater conditions, and the central path}
\label{sec:slater}

\subsection{Bootstrap problems as polynomial matrix programs}
\label{sec:pmp}

In numerical bootstrap problems we generally search for linear functionals subject to positivity conditions on a half-line of operator dimensions:
\beq
	\begin{aligned}
		\text{max } \alpha(V_{\mathrm{obj}}) \text{ over }  \alpha \in \calF_N \text{ such that}\qquad
		&\alpha\!\bigl(V_{\mathrm{Id}}\bigr)=1, \\
		&\alpha\!\bigl(V_{\De,\ell}\bigr)\ge 0
		\qquad \forall \ell,\ \forall \De\ge \De_{\min}(\ell),
	\end{aligned}
	\label{eq:intro-functional}
\eeq
where $\calF_N$ is a finite-dimensional ansatz space of functionals (typically derivatives at the crossing-symmetric point). After the standard approximation of conformal blocks by positive prefactors times polynomials, such problems take the polynomial matrix program (PMP) form \cite{Simmons-Duffin:2015qma}
\beq
	\label{eq:pmp-original}
	\begin{aligned}
		\text{maximize}\quad & b^\top y \\
		\text{subject to}\quad &
		M_0^j(x)+\sum_{n=1}^N y_n M_n^j(x)\succeq 0,
		\qquad \forall x\ge 0,\ \forall j,
	\end{aligned}
\eeq
where the $M_n^j(x)$ are symmetric polynomial-matrix-valued functions and $y\in\bR^N$ are the decision variables.

PMPs are translated into SDPs with a highly structured sparse form; these are the \SDPs\ of the introduction. SDPB provides an efficient solver for this class of problems, but its convergence guarantees rest on the Slater conditions of Section \ref{sec:slaterconds}, and this paper is about what to do when they fail.

\subsection{The \texorpdfstring{$X$}{X}- and \texorpdfstring{$Y$}{Y}-problems}
\label{sec:XYproblems}

Following the conventions of \cite{Simmons-Duffin:2015qma}, an \SDP\ consists of two coupled optimization problems. In the bootstrap literature the problem of finding a positive functional, with solution variables $(y, Y)$, is called \enquote{dual}, while its dual, in terms of $(x,X)$, is called \enquote{primal}. Because the choice of which side of a duality is called primal is purely conventional (and most of the optimization literature uses the opposite convention), we label the two sides $Y$ and $X$ when speaking of the problems and of their Slater conditions. We keep SDPB's words \enquote{primal} and \enquote{dual} for what SDPB itself names that way: its flags, messages, error measures and feasible jumps, which refer to the $X$-side and to the $Y$-side respectively.\footnote{Dictionary: our $Y$-problem is SDPB's \enquote{dual}, which in the optimization literature is often called \enquote{primal}; our $Y$-Slater condition is what the interior-point literature usually calls strict feasibility of the (standard-form) primal problem. We apologize to both communities equally.}

The \emph{$Y$-problem} reads
\beq
	\label{eq:D}
	\begin{aligned}
		\text{maximize}\quad & b^\top y \\
		\text{subject to}\quad
		& B\, y + \bigl(\Tr(A_p\, Y)\bigr)_{p=1}^{P} = c, \\
		& Y \succeq 0,
	\end{aligned}
\eeq
where $y \in \bR^N$ are free variables, $Y \in \bS^K$ is a positive-semidefinite matrix variable, $A_1,\dots,A_P \in \bS^K$ are fixed constraint matrices, $B$ is a $P\times N$ matrix, and $b\in\bR^N$, $c\in\bR^P$ are given vectors. The problem is considered strictly feasible if a solution exists with $Y \succ 0$, weakly feasible if solutions exist but only on the boundary of the cone ($Y \succeq 0$ with $\det Y = 0$), and infeasible if no $Y \succeq 0$ solution exists at all.

The \emph{$X$-problem} is
\beq
	\label{eq:P}
	\begin{aligned}
		\text{minimize}\quad & c^\top x \\
		\text{subject to}\quad
		& X = \sum_{p=1}^P x_p\, A_p, \qquad
		B^\top x = b, \qquad
		X \succeq 0,
	\end{aligned}
\eeq
with variables $x \in \bR^P$ and $X \in \bS^K$. Weak duality $c^\top x \ge b^\top y$ holds for every feasible pair, and under the conditions discussed next the optimal values coincide.

In practice $Y=\bigoplus_j Y^{(j)}$ decomposes into blocks, each arising from one PMP positivity constraint via a bilinear-basis expansion; the $A_p$ inherit this block-diagonal structure and are of low rank within each block (rank one per Gram half on the diagonal rows $r=s$),
\beq
	A_{(j,r,s,k)}
	=
	\sum_{b\in \mathrm{blocks}(j)}
	\mathrm{Block}_b\bigl(
	v_{b,k} v_{b,k}^{\mathsf T} \otimes E^{(r,s)}
	\bigr),
	\label{eq:SDPB-Ap}
\eeq
where the $v_{b,k}$ are bilinear basis vectors evaluated at sample points $x_k$ and $E^{(r,s)}$ is the symmetrized matrix unit. The \emph{bilinear basis} of a block is the set of polynomials $q_m(x)$, in practice orthogonal polynomials, through which SDPB represents positivity on the half-line, and the $v_{b,k}$ collect their values at the sample points; see eqs.\ (2.8), (2.18) and (3.13) of \cite{Simmons-Duffin:2015qma}. Storing only the $v_{b,k}$ and never the $A_p$ makes SDPB fast, but it also rules out naive Slater fixes (Section \ref{sec:xside}).

A \emph{navigator \SDP} \cite{Reehorst:2021ykw} is an \SDP\ of the above form whose optimal value $\vnav$ (the $Y$-side value $v_Y$, a supremum if it is not attained) tracks feasibility of an underlying bootstrap test problem: $\vnav<0$ if and only if the point is allowed. By construction, navigator \SDPs\ are bounded and at least weakly feasible on both sides.

\subsection{Slater conditions and strong duality}
\label{sec:slaterconds}

The two Slater conditions for the pair \eqref{eq:D}--\eqref{eq:P} are:
\begin{itemize}
	\item \emph{$Y$-Slater:} there exist $(y,Y)$ satisfying the constraints of \eqref{eq:D} with $Y \succ 0$;
	\item \emph{$X$-Slater:} there exist $(x,X)$ satisfying the constraints of \eqref{eq:P} with $X \succ 0$.
\end{itemize}
If one of the two holds and that side has a finite optimal value, \emph{strong duality} is guaranteed: the optimal values of \eqref{eq:D} and \eqref{eq:P} coincide, and the optimum on the other side is attained \cite{Slater:1950}; the optimum on the Slater side itself need not be attained and may only be approached as a limit. The same questions, attainment on both sides and absence of a duality gap, arise for the untruncated bootstrap at $\Lambda=\infty$, where Slater's condition becomes an interiority condition in a function space of correlators; \cite{Hercek:2026ext} establish primal attainment and zero duality gap for OPE maximization in 1d CFT by this route, and note that the finite-$\Lambda$ multi-correlator problem, the one SDPB solves, raises the Slater questions treated here. A useful reformulation of the same theorem is that, when the $Y$-problem is feasible (and $B$ has full column rank), $X$-Slater holds if and only if the set of optimal $(y,Y)$ is nonempty and bounded, and symmetrically with the roles exchanged \cite{Todd:2001}.

Throughout, we call an \SDP\ \emph{$Y$-degenerate} if $Y$-Slater fails, \emph{$X$-degenerate} if $X$-Slater fails, and \emph{$B$-degenerate} if $B$ lacks full column rank, i.e.\ if an exact linear relation holds among the functional components; the last is not a face of either cone and is treated separately (Section \ref{sec:feasrigorous}). We call the \SDP\ \emph{degenerate} if any of the three holds. The three conditions are exactly what the central path needs: it exists and is bounded if and only if both Slater conditions hold, and along it the Newton system of SDPB is nonsingular if and only if, in addition, $B$ has full column rank. Another form of degeneracy concerns the limit point rather than the path: the optimum may fail to be unique, or strict complementarity may fail \cite{Alizadeh:1997}. Neither prevents SDPB from finding an optimal solution. The optimal value is unaffected, and the central path still ends at an optimal solution, the analytic center of the optimal face when the optimum is not unique, so a run returns one representative of that face. Nor does it obstruct the Newton steps: at every finite $\mu$ the centering problem is well posed and full $X$- and $Y$-steps remain admissible; what it can do is cap the steps that decrease $\mu$ at a fixed fraction, which slows the final approach from superlinear to linear convergence. This paper therefore does not discuss this form.

When neither condition holds, a \emph{duality gap} may appear: the two optimal values can differ, and the interior-point iterates converge to inconsistent values on the two sides. In this situation the standard optimality certificate breaks down. Normally one verifies optimality of a candidate pair $(x^*,X^*,y^*,Y^*)$ by checking feasibility on both sides together with complementary slackness $\Tr(X^*Y^*)=0$, since weak duality gives $c^{\mathsf T}x - b^{\mathsf T}y = \Tr(XY)\ge 0$ for every feasible pair, so that a vanishing product certifies a vanishing gap. But when a duality gap exists, $\Tr(X^*Y^*)>0$ at the respective optima and complementary slackness never holds, so one cannot certify optimality by exhibiting any matching pair. The optima may, moreover, fail to be attained at all.

Let us stress that Slater is sufficient but not necessary, since strong duality and well-defined solutions can hold without either condition. Slater provides a guarantee checkable \emph{a priori} from the structure of the problem; without it, strong duality may or may not hold.

\subsection{The central path, SDPB, and what breaks without Slater}
\label{sec:centralpath}

The \emph{central path} is a one-parameter family $(x(\mu),y(\mu),X(\mu),Y(\mu))$, $\mu>0$, of strictly feasible points satisfying the feasibility constraints of both problems with $X(\mu)\succ 0$ and $Y(\mu)\succ 0$, together with the modified complementarity condition
\beq
	X(\mu)\, Y(\mu) = \mu\, I.
	\label{eq:complementarity}
\eeq
It arises as the family of stationary points of the Lagrangian augmented by a logarithmic barrier (cf.~\cite{Liu:2023elz}),
\beq
	L_\mu \;=\; c^\top x + b^\top y - x^\top B\, y
	+ \Tr\bigl((X - \textstyle\sum_p x_p A_p)\, Y\bigr)
	- \mu \log\det X,
	\label{eq:lagrangian}
\eeq
and as $\mu\to0$ it converges to the optimal solution. Both Slater conditions are required for the central path to exist. If $X$-Slater fails, $-\log\det X = +\infty$ for every feasible $X$ and the barrier problem has no finite solution; if $Y$-Slater fails, $Y(\mu)\succ0$ cannot be achieved. In either case the central path does not exist.

SDPB is a primal-dual interior-point solver that follows the central path. At each iteration it computes a Newton direction aimed at the central-path point with parameter $\mu_\beta=\beta\mu$, where $\mu=\Tr(XY)/K$ is the current complementarity value (on the central path it equals the path parameter) and $\beta\ge0$ is the centering parameter (eqs.\ (2.36)--(2.38) of \cite{Simmons-Duffin:2015qma}). Two values are used. The predictor uses $\beta_p$, which is $0$ once both sides are feasible and \texttt{infeasibleCenteringParameter} before. The corrector uses $\beta_c$, computed from the predictor step; once both sides are feasible it is clamped to $[\texttt{feasibleCenteringParameter},1]$, before that it is only bounded below by \texttt{infeasibleCenteringParameter}. Where the distinction does not matter we write $\beta$ for either. SDPB monitors three error measures (the primal error, i.e.\ the $X$-side infeasibility, the dual error, i.e.\ the $Y$-side infeasibility, and the normalized duality gap) and terminates with \texttt{PrimalDualOptimal} when all three fall below their thresholds (by default $10^{-30}$). When both Slater conditions hold, the central path exists and path-following methods enjoy the classical polynomial-time guarantee: for short-step methods the complementarity parameter decreases by a constant factor ${\sim}(1-c/\sqrt K)$ per iteration and $O(\sqrt K \log (1/\vartheta))$ iterations suffice \cite{NesterovTodd:1997,NesterovTodd:1998}. SDPB is an infeasible-start predictor--corrector method \cite{Mehrotra:1992} with long steps, for which this bound is not proven. On healthy problems, however, all three error measures decrease together at a geometric rate, as expected of a method tracking the central path; without Slater this behavior is lost.

When the central path does not exist this guarantee is lost, and the three errors no longer decrease in lockstep. The characteristic signature of a $Y$-Slater failure is that the dual error decreases only gradually (instead of the sudden "dual feasible jump" seen in healthy problems) as the iterates force $Y$ ever closer to singularity \cite{Liu:2023elz}, while the $X$-side appears to converge normally, giving a misleading impression of progress. At the same time the Schur complement used in the Newton system (the matrix SDPB factorizes by Cholesky at every iteration, before the $B$ constraints are included) becomes ill-conditioned, degrading the computed directions.\footnote{The convergence proofs assume exact arithmetic. Finite-precision error (principally in the Cholesky factorization of the Schur complement) is a controllable nuisance, since SDPB works at arbitrary precision and the working precision can be increased when necessary. A Slater failure is different in kind, because the central path does not exist and higher working precision cannot restore it. The two effects compound. When Slater barely holds, the Schur complement is already ill-conditioned and much higher precision is needed, which is why the magnitude of a positive Slater margin matters as well as its sign.}

The Skydive algorithm \cite{Liu:2023elz} suffers a sharper failure. Skydive simultaneously optimizes the SDP variables $\xi=(x,y,X,Y)$ and external parameters $p$ (e.g.\ operator dimensions), navigating along the \emph{central section} where $\nabla_\xi L_\mu = 0$ at fixed $(p,\mu)$. A key step is \emph{centering}: iterating the SDP variables until $\lVert XY-\mu I\rVert$ is small. If $Y$-Slater fails, no feasible $Y$ is invertible, $XY=\mu I$ cannot be even approximately satisfied at finite $\mu$, and the centering residual cannot converge; the algorithm stalls before taking a single step in $p$. We will see this textbook failure mode realized, in a rather extreme form, in Section \ref{sec:casestudy}.

Table \ref{tab:guarantees} summarizes what each combination of Slater conditions buys.

\begin{table}[htpb]
\centering
\small
\begin{tabular}{lccc}
	\toprule
	 & Strong duality & Central path & Polynomial convergence \\
	\midrule
	Neither Slater & not guaranteed & does not exist & not guaranteed \\
	One Slater & \checkmark & does not exist & not guaranteed \\
	Both Slater & \checkmark & exists & $O(\sqrt{K}\log(1/\vartheta))$ \\
	\bottomrule
\end{tabular}
\caption{What the Slater conditions guarantee. In the "one Slater" case the \SDP\ has no duality gap and, when the optimal value is finite, the optimum on the opposite side is attained, but interior-point convergence rates are no longer guaranteed. In the "neither" case a duality gap is possible, the optimum on either side may not be attained, and the complementary-slackness certificate breaks down. The iteration bound is the classical one for short-step path-following methods, with $\vartheta$ the target accuracy. These are statements about guarantees. Without Slater things can still work, but one cannot know in advance.}
\label{tab:guarantees}
\end{table}

\subsection{How Slater fails in the bootstrap}
\label{sec:howfails}

The feasible set of an SDP lies in a unique minimal face of the positive-semidefinite cone.\footnote{A \emph{face} of a convex cone $K$ is a subcone $F$ with the property that whenever a sum of two elements of $K$ lies in $F$, both summands do. For the positive-semidefinite cone the faces are exactly the sets $\{Y\succeq0:\mathrm{range}\,Y\subseteq V\}$, one for each subspace $V$. The \emph{minimal face} of a subset is the smallest face containing it, and a face is \emph{proper} if it is not the whole cone.} When the $Y$-Slater condition fails, this face is proper: there exists a nonzero vector $v$ with
\beq
	Y v = 0
	\qquad\text{for every feasible } Y\succeq 0,
	\label{eq:commonnull}
\eeq
i.e.\ all feasible $Y$ share a common null direction.\footnote{Let $Y_*$ be a feasible point of maximal rank. For any feasible $Y$, the midpoint $(Y_*+Y)/2$ is feasible and its range contains $\mathrm{range}\,Y_*+\mathrm{range}\,Y$, so maximality forces $\mathrm{range}\,Y\subseteq\mathrm{range}\,Y_*$. Hence every null vector of $Y_*$ is a common null vector of the feasible set, and one exists because $Y_*\not\succ0$.} Similarly, $X$-Slater failure means a common null direction for all feasible $X$.

In bootstrap applications, $Y$-Slater failure arises when the positivity constraints force certain components of the functional to vanish identically. The multi-correlator situation described in the introduction is the canonical example. Suppose the constraint vectors that involve the additional crossing equations (in the mixed Ising system, the $\e\e$-components of the $\bZ_2$-even vectors and the $\bZ_2$-odd vectors) admit an \emph{exposing certificate} (Section \ref{sec:exposing}). In bootstrap language this is a positive homogeneous solution of the truncated crossing equations: a nonnegative combination of these constraint vectors, with no identity component, that vanishes. Then every feasible functional must vanish on its support, and when the supported vectors span the components feeding those equations, these components are forced to zero, and the problem is the smaller subsystem written in a larger form. Once those components are forced to zero, the point is disallowed if and only if the subsystem excludes it. The situation has been noticed in published scans under the name \enquote{trivial mix}: the functional components on the mixed-correlator equations are numerically suppressed to zero, the bound reverts to the single-correlator one, and no dual jump occurs in that region \cite{Nakayama:2025mrm}. The converse fails, because a point disallowed by the subsystem does not by itself make the full problem degenerate. If the other sectors admit a strictly positive functional, adding a small multiple of it to a strictly excluding subsystem functional gives a strictly feasible point. In our examples exposing certificates occur both at allowed and at disallowed points (Section \ref{sec:casestudy} and Section \ref{sec:exp-twist}).

This is the one mechanism that is invisible in the input. The other $Y$-Slater failures we encounter are forced contacts of the block polynomials, at a point of the half-line or at $x=\infty$, which Section \ref{sec:pmpfr} classifies by location: a zero at the endpoint of a sector planted by a convention (the shared block prefactor of Section \ref{sec:fakepoles}, or the normalization of conserved three-point structures in Section \ref{sec:mixedcurrent}), a genuine endpoint zero in channels that mix equal and unequal operator pairs (Section \ref{sec:mixedentry}), and a contact at $x=\infty$ when the entries of a matrix block cannot reach its declared degree (Section \ref{sec:exp-twist}, Appendix \ref{app:faithful}). Each produces a common null vector \eqref{eq:commonnull} in the same way, but its origin is visible in the PMP data and a static check of the input catches it (Sections \ref{sec:fakepolefr} and \ref{sec:conclusions}). Rank deficiency of $B$, the $B$-degeneracy of Section \ref{sec:slaterconds}, is a different matter: it is not a face of either cone, and Section \ref{sec:mixedcurrent} has an example.

On the $X$-side an exposing certificate is a direction along which the $X$-variable can grow indefinitely without violating the constraints or changing the objective; we call such directions \emph{zero-cost rays}. When $Y$-Slater fails, the interior-point iterates drift along them, because keeping $XY\approx\mu I$ while every $Y$ is nearly singular in the common null directions forces $X$ to grow in those directions (Section \ref{sec:centralpath}). $X$-Slater itself fails in feasibility mode at every disallowed point, where the $X$-problem has no feasible point at all, and more generally whenever a functional exists that is positive on every unitary block (Section \ref{sec:feasxslater}). In navigator problems it held in every case we tested, for a reason given in Section \ref{sec:xside}.

There is a structural reason why failures of the first kind cannot be predicted from the setup. Exact linear relations among the truncated constraint vectors exist automatically, at any truncation order, because each sector's vectors live in a space whose dimension is set by the number of functional components that sector feeds, and the sampled family contains far more vectors than that (the counting is made explicit in Section \ref{sec:faceworked2}). A degeneracy of this type occurs exactly when one of these relations has all-nonnegative coefficients, and exactly one of two things always holds: either some nonzero nonnegative combination of the constraint vectors vanishes, or some functional is strictly positive on all of them (this dichotomy, \emph{Gordan's alternative}, recurs in Sections \ref{sec:feascorrupt} and \ref{sec:faceworked2}). Which of the two is realized depends on the numerical values of the constraint vectors rather than on the structure of the problem, so it cannot be read from the PMP input (Section \ref{sec:exposing}) and changes with the external point, the truncation order and the formulation. Whether the crossing equations added by a multi-correlator setup actively constrain the functional, or instead force parts of it to vanish, is therefore a question about the solution of the problem. It is answerable by one diagnostic solve per formulation (Section \ref{sec:diagnostics}) but not in advance.

\section{Diagnosing Slater failure}
\label{sec:diagnostics}

\subsection{Measuring the \texorpdfstring{$Y$}{Y}-margin (\texttt{ycompute})}
\label{sec:ycompute}\label{sec:weightedslack}

Throughout this section it is useful to keep the block structure of the \SDP\ in mind. A bootstrap \SDP\ is not one large positivity condition but a long list of independent ones: one \emph{block} per (sector, spin) pair, plus isolated blocks for external or normalization structures. Each block $j$ is the constraint that the polynomial matrix $M^j(x;y)=M^j_0(x)+\sum_n y_n M^j_n(x)$ be positive semidefinite on the half-line. The blocks share nothing except the functional components $y$; all the geometry of the feasible set arises from the same $y$ having to keep every block positive at once. Global quantities (the navigator value, the Slater margin defined next) are single numbers, but their failure modes are local, in the sense that they are caused by specific blocks.

For a given PMP \eqref{eq:pmp-original} it is easy to write a related PMP that measures the $Y$-side margin: introduce one extra variable $\vareps$, subtract a positive multiple of the identity from every block, and maximize,
\beq
	\label{eq:pmp-Y-slack}
	\text{maximize } \vareps
	\qquad\text{subject to}\qquad
	M_0^j(x) + \sum_{n=1}^N y_n M_n^j(x)
	- \vareps\, \rho_j(x)\, I \succeq 0
	\qquad\forall x\ge0,\ \forall j,
\eeq
where $\rho_j$ is a fixed positive polynomial weight, one per block, whose degree equals the degree $d_j$ of the block; the canonical choice is
\beq
	\rho_j(x) = (1+x)^{d_j}.
	\label{eq:slack-weight}
\eeq
We call the optimal value $\vareps^*$ the \emph{$Y$-margin}, and \texttt{ycompute} the tool that computes it; the constant-weight $Y$-margin $\vareps^*_1$, obtained with $\rho_j=1$ instead (option \texttt{--rhoDegree 0}), is used below only for comparison.\footnote{A shift of this form with $\vareps$ held at a small constant was used as a cure by \cite{Nancarrow:2022wdr}, who add $\epsilon x^{d}$ to the diagonal of a block, an ``infinitesimal $\epsilon$-shift'' with no value stated, ``to push the solution away from being positive semidefinite to being slightly positive definite'' after a dual jump was lost to mismatched leading powers (the mechanism of Section \ref{sec:mixedcurrent}). Here $\epsilon$ has to be guessed: large enough for the jump to fire, small enough that the relaxation of the bound, of order $\epsilon$ times an unknown constant, stays below the accuracy wanted. Maximizing $\vareps$ instead turns the cure into a measurement: $\vareps^*$ is the margin and its sign classifies the problem; the guarantees of Section \ref{sec:guarantees} certify from the run itself whether the value returned is the original one; and the reduction of Section \ref{sec:pmpfr} removes the contact exactly, with no relaxation at all.} The sign of $\vareps^*$ classifies the original problem:\footnote{Throughout the paper, measured quantities are compared against two distinct scales. A converged solve determines its optimum only to the terminal duality gap, so any reported value that vanishes at the exact optimum (in particular the margin $\vareps^*$ at a Slater failure) is indistinguishable from zero below that scale; we then say it is zero at the duality gap of the solve. Quantities the algorithm satisfies exactly, such as affine residuals after a full Newton step, are instead zero at the working precision, i.e.\ at its roundoff scale (e.g.\ a primal error of $10^{-173}$ in our runs), typically dozens of orders below the duality gap. We say which of the two applies wherever the distinction matters; a third, empirically distinct scale (face-block residuals that settle far below even the duality gap) is discussed where it appears, in Section \ref{sec:threeuses}.}
\begin{equation*}
	\begin{aligned}
	\vareps^* > 0&:\ \text{strictly $Y$-feasible ($Y$-Slater holds, with margin $\vareps^*$)},\\
	\vareps^* = 0&:\ \text{weakly $Y$-feasible ($Y$-Slater fails)},\qquad
	\vareps^* < 0:\ \text{$Y$-infeasible}.
	\end{aligned}
\end{equation*}
The middle verdict presumes that the supremum is attained (Section \ref{sec:attainment}). In general semidefinite programming $\vareps^*=0$ without attainment is possible, and then the problem is weakly infeasible: no feasible functional exists at all. Attainment is guaranteed when the diagnostic's own $X$-side Slater condition holds, which is testable with the $X$-side diagnostic of Section \ref{sec:xdiag} applied to the \texttt{ycompute} \SDP; we find that condition to hold in all navigator problems and at allowed feasibility points, and to fail at disallowed feasibility points (Section \ref{sec:feasrigorous}).

\paragraph{The need for a weight.} Taking $\rho_j=1$ (subtracting $\vareps I$) does not work, because the $Y$-Slater condition concerns the Gram matrices of the \SDP\ rather than the values of the polynomial matrices at points. For each block, SDPB's matrix variable is a pair of positive-semidefinite matrices $(Y_1,Y_2)$ representing the block polynomial through the bilinear pairing $M(x)=\Tr_q\bigl(Y_1(Q_{\delta_1}(x)\otimes I)\bigr)+x\,\Tr_q\bigl(Y_2(Q_{\delta_2}(x)\otimes I)\bigr)$ ($\Tr_q$ the partial trace over the bilinear-basis index), with $Q_\delta(x)$ built from the bilinear basis (eq.~(2.8) of \cite{Simmons-Duffin:2015qma}); we call $(Y_1,Y_2)$ the block's \emph{Gram matrices}, in analogy with the Gram-matrix representation of sums of squares. Strict feasibility means a positive-definite Gram pair reproducing the constraint, and pointwise strict positivity of a polynomial matrix does not, by itself, imply the existence of one. The gap between the two notions sits at $x=\infty$.

A toy example makes this concrete. Consider a single $1\times1$ constraint of degree $2$ whose polynomial happens to be constant, $p(x)=1$. SDPB represents a degree-2 scalar constraint on the half-line as
\beq
	p(x) \;=\; \begin{pmatrix}1 & x\end{pmatrix} G_1 \begin{pmatrix}1 \\ x\end{pmatrix} \;+\; x\, g_2,
	\qquad G_1\succeq0,\quad g_2\ge0,
	\label{eq:toy-gram}
\eeq
with $Y$-block $(G_1,g_2)$. Matching coefficients of $p=1$ gives $(G_1)_{11}=1$, $2(G_1)_{12}+g_2=0$, and $(G_1)_{22}=0$. Positive semidefiniteness of $G_1$ with a vanishing corner forces $(G_1)_{12}=0$, hence $g_2=0$, so the Gram representation is unique and singular. Thus the constraint ``$1\succeq0$'', strictly satisfied with margin $1$ at every point of the half-line, is only weakly feasible at the Gram level, and $Y$-Slater fails. A constant slack cannot see this, since replacing $1$ by $1+\abs\vareps$ only changes $(G_1)_{11}$ and the same computation forces the Gram to stay singular for every $\vareps$. An unweighted diagnostic would therefore incorrectly report a margin of $1$ for a problem whose $Y$-Slater condition fails.

The underlying criterion is that a strictly feasible Gram of a degree-$d$ block exists precisely when the polynomial matrix is uniformly positive at finite $x$ and its leading coefficient matrix is positive definite, i.e.\ strictly positive on the half-line including its point at infinity.\footnote{This is part (2) of the Lemma in Appendix \ref{app:faithful}, which proves it and develops it into a classification of what the PMP-to-\SDP\ translation can and cannot break.} A constant slack has degree $0$ and therefore buys margin at every finite $x$ and none at all at $x=\infty$, so it is blind to degeneracies of the leading behavior (in bootstrap language, of the $\De\to\infty$ tail of a constraint). The full-degree weight \eqref{eq:slack-weight} repairs this. In the toy example, $1-\vareps(1+x)^2 = (1+\abs\vareps) + 2\abs\vareps\, x + \abs\vareps\, x^2$ is represented by $G_1 = \mathrm{diag}(1+\abs\vareps,\ \abs\vareps)\succ0$, $g_2 = 2\abs\vareps>0$, a strictly feasible Gram. In general, for any $y$ and $\vareps$ sufficiently negative, $M(x;y)-\vareps\rho(x)I$ is uniformly positive at finite $x$ and has positive-definite leading coefficient, so a strictly feasible Gram always exists and the sign classification above is exact.

More precisely, the requirement on the weight is that $\rho_j(x) I$ lie in the interior of the fixed-degree cone of matrix polynomials nonnegative on the half-line. Equivalently, it must admit some positive-definite Gram representation at degree $d_j$, which holds iff $\rho_j$ is strictly positive on $[0,\infty)$ including nonvanishing leading behavior; $(1+x)^{d_j}$ is the canonical such choice.

\paragraph{Concrete implementation.} At the PMP level the modification is one extra variable, $y'=(y_1,\dots,y_N,\vareps)$, one extra matrix $M^j_{N+1}(x)=-\rho_j(x)\,I$ for every block, and the objective $b'=(0,\dots,0,1)$. At the \SDP\ level it touches only one column. Let block $j$ carry sample points $x_{j,0},\dots,x_{j,d_j}$ and sample scalings (the per-row rescalings of eqs.\ (3.12)--(3.13) of \cite{Simmons-Duffin:2015qma}, which produce an isomorphic but better-conditioned \SDP), with constraint rows labeled $p=(j,r,s,k)$ as in \eqref{eq:SDPB-Ap}. Current versions of SDPB distinguish two sets of scalings: $s_{j,k}=\chi_j(x_{j,k})$, from the full prefactor, multiplies the rows of $B$ and $c$, while $\tilde s_{j,k}=\tilde\chi_j(x_{j,k})$, from a \emph{reduced} prefactor $\tilde\chi_j$ with fewer poles, is the scaling with which the bilinear basis is sampled and hence enters the $A_p$.\footnote{This \enquote{pole reduction} lowers the number of sample points from $d+1$ to $d+1-(\#\text{poles}(\chi_j)-\#\text{poles}(\tilde\chi_j))$, with $d$ the degree before reduction. The constraint actually imposed is Gram-representability of the degree-reduced polynomial interpolating $P(x)\,\chi_j(x)/\tilde\chi_j(x)$ at the sample points. All statements of this paper about \enquote{the PMP} refer to this reduced PMP, and $d_j$ denotes its degree. Without pole reduction $\tilde s=s$.} The column must be built with the scalings of the $A_p$:
\beq
	B' = \bigl[\,B \;\big|\; w\,\bigr],
	\qquad
	w_{(j,r,s,k)} = \rho_j(x_{j,k})\;\delta_{rs}\; \tilde s_{j,k} \,,
	\label{eq:weighted-column}
\eeq
so that it represents $\rho_j(x)I$ through a Gram matrix, which we denote $\widehat Y$, and $c$ is unchanged.\footnote{The reduced scalings matter, since the full scalings give a column that is not an interior direction.} Everything else (the sample points, the bilinear bases, every $A_p$, every other column of $B$) is unchanged.

Two remarks on the weight. First, no exponentials appear anywhere in this construction, despite the exponentially damped positivity prefactors $\chi_j$ of the underlying bootstrap problem, because the weight is a polynomial and only its values at the existing sample points enter the data. (The prefactors are the positive factors $\chi_\ell(\De)=r_*^{\De}/\prod_i(\De-\De_{*i})$, with $r_*$ the radial coordinate of \cite{Hogervorst:2013sma}, stripped from the polynomial approximation of conformal blocks, eqs.\ (3.7) and (A.2) of \cite{Simmons-Duffin:2015qma}.) The choice $\rho_j=\chi_j$ is not available, since the prefactor is not a polynomial and $-\vareps\,\chi_j I$ would leave the PMP class. It is also not needed, because the prefactor multiplies the positivity constraint as a whole and cancels from it, and the Gram-level degeneracy lives entirely in the polynomial part. Second, degree bookkeeping is the main structural constraint. At $\deg\rho_j=d_j$ the augmented constraint lives in the same polynomial space as the original block, so the sample points, bilinear bases and block sizes are unchanged; a higher-degree weight would enlarge the \SDP. Since block $j$ carries exactly $d_j+1$ sample points, any assignment of numbers to the diagonal slack column (the $\vareps$-column of the \SDP) entries is the sampling of a unique polynomial of degree at most $d_j$, so every such column is automatically PMP-representable. The weight is thus a choice of units for the slack, since the value of $\vareps^*$ depends on $\rho_j$ and the sample scalings and only its sign is invariant under admissible changes of weight; \eqref{eq:slack-weight} is the choice that makes the units nondegenerate on every block including at $x=\infty$.

In Appendix \ref{app:Y-slack-PMP-to-SDP} we show that after a change of variables the $Y$-margin is the largest \emph{conic margin} of the feasible set in the fixed direction $\widehat Y$, i.e.\ a generalized minimal eigenvalue, which makes the classification above manifest; the weight \eqref{eq:slack-weight} is what makes $\widehat Y$ positive definite, so that a positive margin certifies an interior point. Because $c$ is untouched, the $Y$-feasible set of the diagnostic retains the structure of the original, plus the new $\vareps$ direction which can always buy strict positivity. In practice \texttt{ycompute} converges quickly on every navigator \SDP\ we have tested, including those on which the original problem stalls.

One caveat is that the $Y$-margin PMP can be unbounded. If there is a direction $\Delta y$ and $\omega>0$ with $\sum_n (\Delta y)_n M_n^j(x)\succeq \omega \rho_j(x) I$ for all $x,j$ (a uniformly positive-definite homogeneous direction), then $\vareps$ can be pushed to $+\infty$ along it. The equivalent \SDP-level statement, in terms of \emph{recession directions} (directions along which the functional can be moved indefinitely while staying feasible), is given in Appendix \ref{app:Y-slack-PMP-to-SDP}. For the complete navigator \SDPs\ of this paper we have not encountered unboundedness (the deliberately restricted sub-problems of Section \ref{sec:faceworked2}, which omit sectors, can be unbounded). In feasibility mode it is common at disallowed points, where a strictly positive functional with vanishing identity component is such a direction (Section \ref{sec:feasxslater}). \texttt{ycompute} then reports a diverging margin, the $\vareps\le0$ cutoff of \texttt{yguarantee} gives a finite number, and in feasibility mode ($b=0$) the divergence itself signals that the $X$-Slater condition fails.

\subsection{Attainment and the four conic outcomes}
\label{sec:attainment}

The $Y$-margin classifies the functional set of any \SDP\ into four conic outcomes. If $\vareps^*>0$ the set is strictly feasible, and if $\vareps^*<0$ it is empty, with no functional satisfying positivity even approximately (strictly infeasible). The value $\vareps^*=0$ covers two cases. If the supremum in \eqref{eq:pmp-Y-slack} is attained, the set is nonempty but every functional in it sits on the boundary of the cone, i.e.\ on a face in the language of Section \ref{sec:pmpfr} (weakly feasible). If it is not attained, no functional satisfies positivity, yet there are functionals violating it by less than any given tolerance (weakly infeasible). The distinction is a conic phenomenon that has no analogue in linear programming. The standard example is the constraint set $\bigl\{\left(\begin{smallmatrix}0&1\\1&a\end{smallmatrix}\right)\succeq0,\ a\in\mathbb R\bigr\}$, which is infeasible (the top-left corner vanishes while the off-diagonal does not) but comes within any $\omega>0$ of feasibility as $a\to\infty$. No finite-precision solver can distinguish the two middle rows by inspecting near-feasible iterates, and the computed value $\vareps^*\approx0$ cannot distinguish them either, since what separates them is \emph{attainment} of the supremum in \eqref{eq:pmp-Y-slack}, which is not a number any solve returns. 

Attainment is certified by an auxiliary strict-feasibility check, which we call the \emph{attainment check} and which is in effect a second margin computed on a different object. The relevant question is whether the constraint data admits a nonzero \emph{recession direction}, a direction along which the functional can be moved indefinitely while staying feasible: a $\Delta y$ with $\sum_n(\Delta y)_n M^j_n(x)\succeq0$ for all $x,j$ (the zero-slack-gain counterpart of the unboundedness directions of Appendix \ref{app:Y-slack-PMP-to-SDP}). If no such direction exists, the level sets $\{y:M(x;y)-\vareps\rho(x) I\succeq0\ \forall x\}$ are compact, so the supremum is attained; moreover a weakly infeasible set can only be approached along such a direction ($a\to\infty$ in the toy example above), so the absence of recession directions rules out the weakly infeasible case. Equivalently, on the $X$-side, the condition is strict feasibility of the diagnostic's own $X$-problem \eqref{eq:ycompute-X}. The useful direction of that equivalence is elementary and turns the check into a concrete procedure. In \SDP\ form a recession direction is a pair $(\Delta y,\Delta Y)$ with $\Delta Y\succeq0$ and $B\,\Delta y+\bigl(\Tr(A_p\,\Delta Y)\bigr)_p=0$, the Gram representation of $\sum_n(\Delta y)_n M_n\succeq0$. Pairing these constraints with any strictly feasible point $x$ of \eqref{eq:ycompute-X} gives
\begin{equation*}
	0 \;=\; \sum_p x_p\Bigl[(B\,\Delta y)_p+\Tr(A_p\,\Delta Y)\Bigr]
	\;=\; (B^\top x)\cdot\Delta y+\Tr(X\,\Delta Y)
	\;=\; \Tr(X\,\Delta Y),
\end{equation*}
and $X\succ0$ forces $\Delta Y=0$, hence $B\,\Delta y=0$. Two ingredients thus certify attainment.

\paragraph{(i) $B$ has full column rank,} so that $B\,\Delta y=0$ forces $\Delta y=0$. Full column rank is an exact linear-algebra fact about the data, checkable by a rank computation at high precision, and it can fail only for a degenerate PMP in which some direction of functionals acts as zero on every sampled constraint.

\paragraph{(ii) A strict-feasibility witness for \eqref{eq:ycompute-X},} which the diagnostic solve supplies itself. Every interior-point iterate is strictly inside its cone, so any iterate whose affine residuals ($B^\top x=0$, $w^\top x=1$) are zero at the working precision (for instance any iterate after a primal jump) is such a witness, with its margin read off as $\lambda_{\min}(X)$ against the data norms. No auxiliary SDP is needed when the diagnostic solve supplies such an iterate; otherwise \texttt{xcompute} on the \texttt{ycompute} \SDP\ measures the same margin (Section \ref{sec:ycompute}).

The two cases are therefore told apart by one more margin computation, on the homogeneous constraint set, which asks a different geometric question and need not be marginal when the original one is. When the attainment check passes, weak infeasibility is excluded a priori and $\vareps^*=0$ can only mean weak feasibility, so that the sign of $\vareps^*$ alone decides.\footnote{The attainment check is itself a strict-inequality statement, so if its margin is also zero at the duality gap, attainment is uncertified and the ambiguity of the middle rows reappears, the same problem that limits textbook facial reduction. Likewise, the computed sign of a small $\vareps^*$ is trustworthy only when its magnitude stands clear of the duality gap of the diagnostic solve, as for any other measured margin.} In feasibility mode (Section \ref{sec:feasibility}) the check passes at allowed points and cannot pass at disallowed ones, where a vanishing margin is instead resolved by facial reduction (Section \ref{sec:feasrigorous}).

\subsection{The \texorpdfstring{$X$}{X}-side: representability of the identity}
\label{sec:xside}

One would like to run the mirror-image diagnostic on the $X$-problem: replace $X\succeq0$ by $X - t I\succeq 0$ and maximize $t$,
\beq
	\label{eq:DS}
	\begin{aligned}
		\text{maximize}\quad & t \\
		\text{subject to}\quad & X = \sum_{p=1}^P x_p A_p, \qquad
		B^\top x = b, \qquad
		X - t I \succeq 0,
	\end{aligned}
\eeq
so that for fixed $(x,X)$ the best $t$ is $\lambda_{\min}(X)$ and the optimum measures how deep inside the cone the $X$-feasible set reaches. SDPB never stores the $A_p$; they are generated from the sampled bilinear basis vectors through \eqref{eq:SDPB-Ap}, one row $p$ per (block, matrix entry, sample point). A new dense constraint matrix $A_{P+1}=-I$ is therefore not something one can append. For each block the representable matrices span
\beq
	\calR_b
	\;:=\;
	\mathrm{span}\Bigl\{
	v_{b,k} v_{b,k}^{\mathsf T} \otimes E^{(r,s)}
	\Bigr\}_{k,r,s}
	\;\subset\; \bS^{K_b},
	\label{eq:Rb}
\eeq
a proper subspace which, by dimension counting, would not be expected to contain the identity. Because of how the bilinear bases are chosen, however, it does. Since SDPB~3.1 \cite{SDPB-repo}, \texttt{pmp2sdp} constructs, for each block, the two bilinear bases $q_m$ and $q'_m$ as orthonormal polynomials with respect to the sampled measures $\sum_k\tilde s_k\,\delta(x-x_k)$ and $\sum_k\tilde s_k\,x_k\,\delta(x-x_k)$, and samples them with the same weights, $v_{k}=\sqrt{\tilde s_k}\,q(x_k)$ and $v'_k=\sqrt{\tilde s_kx_k}\,q'(x_k)$.\footnote{The exact identity requires \SDPs\ written by \texttt{pmp2sdp} of SDPB~3.1 or later, which introduced the sampled-measure orthonormalization. Earlier versions orthonormalized a single basis with respect to the continuous measure $\int_0^\infty\chi(x)\,x^m\,dx$ and used it for both halves; then $\sum_kv_kv_k^{\mathsf T}\neq1$ (we measure $|\sum_kv_kv_k^{\mathsf T}-1|\sim10^3$ on \SDPs\ produced by such a pipeline). The same happens if a PMP supplies its own bilinear bases. Nothing below depends on the sum being exactly the identity. The diagonal-row sum $X_0:=\sum_{p\in\mathrm{diag}}A_p=\bigoplus_j\bigl(\sum_kv_kv_k^{\mathsf T}\bigr)\otimes I$ is positive definite whenever the sampled basis vectors span each block (a requirement of any well-posed discretization), and the shift \eqref{eq:identity-shift} produces $X-tX_0$ instead of $X-tI$. All tools of this paper then run unchanged, with the margins measured in units of $X_0$ and the trace in \eqref{eq:trace-linear} replaced by $\Tr(X_0Y)$. Other choices of $X_0$ are discussed in Appendix \ref{sec:X0idea}.} Orthonormality is then the statement $\sum_k v_kv_k^{\mathsf T}=1$ and $\sum_kv'_kv'^{\mathsf T}_k=1$, so that the sum of the $A_p$ over the diagonal rows $r=s$ is the identity:
\beq
	\sum_{p\,\in\,\mathrm{diag}} A_p \;=\; I .
	\label{eq:identity-rep}
\eeq
We verified \eqref{eq:identity-rep} on all \SDPs\ of this paper to the working precision (a residual of $10^{-189}$ in our runs); the tools below recompute and report the residual. The identity can therefore be used through a shift of variables rather than a new row: with $\mathbf 1_{\mathrm{diag}}$ the indicator vector of the diagonal rows and $h:=B^\top\mathbf 1_{\mathrm{diag}}$,
\beq
	z := x - t\,\mathbf 1_{\mathrm{diag}}
	\qquad\Longrightarrow\qquad
	X - tI = \sum_p z_pA_p ,\qquad B^\top x = B^\top z + t\,d .
	\label{eq:identity-shift}
\eeq
The variable $t$ now appears only in the linear constraints, i.e.\ as a new $1\times1$ block, and \eqref{eq:DS} is an ordinary SDPB input. Note also the $Y$-side meaning of $h$: for any $(y,Y)$ satisfying the constraints of \eqref{eq:D},
\beq
	\Tr Y = \sum_{p\in\mathrm{diag}}\Tr(A_pY) = c_{\mathrm{diag}} - d\cdot y ,
	\qquad c_{\mathrm{diag}}:=\sum_{p\in\mathrm{diag}}c_p ,
	\label{eq:trace-linear}
\eeq
so the trace of the Gram matrix is a linear function of the functional. Both facts are used repeatedly below.

\subsection{The \texorpdfstring{$X$}{X}-Slater diagnostic (\texttt{xcompute})}
\label{sec:xdiag}

With \eqref{eq:identity-shift}, the $X$-slack problem \eqref{eq:DS} becomes: keep $B$ and all $A_p$, set $c\to0$ on the original rows, and add one $1\times1$ block for the margin variable, with $B$-row $h$ and objective coefficient $-1$. We give the margin variable a floor, $t\ge-t_0$ with a fixed $t_0>0$, implemented by a block that holds $t'=t+t_0\ge0$, which shifts $b$ by $t_0\,h$ and the objective by the constant $t_0$, so that $t^*$ is still minus the optimal $X$-objective. We call $t^*$ the \emph{$X$-margin}. Its sign classifies the $X$-side just as $\vareps^*$ classifies the $Y$-side:
\begin{equation*}
	\begin{aligned}
	t^*>0&:\ \text{$X$-Slater holds},\qquad t^*=0:\ \text{$X$-feasible but not strictly},\\
	t^*<0&:\ \text{$X$-infeasible}.
	\end{aligned}
\end{equation*}
The floor makes the diagnostic trustworthy in the case that matters, $t^*=0$. With $t\ge0$ instead, a strictly feasible point of the diagnostic would need $X\succ tI$ with $t>0$, so the diagnostic would satisfy its own $X$-Slater condition exactly when the original problem does, and the verdict $t^*=0$ would be computed without a central path. With the floor, any feasible $X$ together with $t=-t_0/2$ is strictly feasible, so the diagnostic satisfies $X$-Slater whenever the original $X$-problem is feasible at all, and the floor does not change the optimum ($t^*\ge0>-t_0$). Conversely, a run that ends at the floor, or whose $Y$-objective diverges, signals that the original $X$-side is infeasible or that $t_0$ is too small to reach its (negative) margin.

$t^*=+\infty$ is possible, and means that the affine set $\{\sum_px_pA_p: B^\top x=b\}$ contains a positive definite direction, the strongest form of $X$-Slater. To obtain a finite number, and a diagnostic whose own $Y$-Slater condition holds, we add the $\vareps$-layer of Section \ref{sec:yguarantee} to the diagnostic (option \texttt{--lambda}); the layer is the $Y$-guarantee applied to the diagnostic \SDP\ itself, whose reference matrix has a unit entry on the block of $t$. On the $X$-side it imposes $\langle X-tI,\widehat Y\rangle+t\le\lambda$, which caps $t$ directly, since $X-tI\succeq0$. The layer's $X$-side variable, the cutoff multiplier $x_c=\lambda-\langle X-tI,\widehat Y\rangle-t\ge0$ (Section \ref{sec:yguarantee}), reports whether the cap is active ($x_c=0$) or not. The cap must be chosen large enough that the $X$-side stays feasible (a too-small cap shows up as a diverging $Y$-objective), and the sign of $t^*$ does not depend on it. The cap also sets the scale of the floor, since the interior point $t=-t_0/2$ must keep the cap's slack positive, which requires $t_0\,(\Tr\widehat Y-1)<2(\lambda-\langle X,\widehat Y\rangle)$ for some feasible $X$; our default $t_0=\lambda/(2\Tr\widehat Y)$ leaves room for $\langle X,\widehat Y\rangle$ up to at least $3\lambda/4$. With both, the diagnostic satisfies both Slater conditions, and $t^*>-t_0$ (floor inactive) confirms the setup from the same run.

Only the sign of $t^*$ (negative, zero or positive) is used as a verdict. Whenever the affine set contains a positive definite direction the uncapped margin is $+\infty$, the cap binds ($x_c=0$), and the capped diagnostic returns a value that grows linearly with $\lambda$; at fixed $\lambda$ this value is a normalized depth per unit of weighted trace (used in Section \ref{sec:feascorrupt}) rather than a margin of the original problem. This is the situation at the allowed feasibility points of Table \ref{tab:feas-xslater}, since in feasibility mode $b=0$ makes the $X$-feasible set a cone, so that $t^*$ is either $0$ or $+\infty$ and the capped value is exactly proportional to $\lambda$. A zero margin, by contrast, stays at the duality gap for every $\lambda$. We checked both behaviors on the feasibility points of Table \ref{tab:floor-law} at $\lambda=10^3,10^{10},10^{20},10^{30}$ (duality gap ${\sim}5\times10^{-46}$): at the disallowed point $\abs{t^*}\le3\times10^{-46}$ throughout, with a large cutoff multiplier, while at the allowed point closest to the boundary $t^*$ grows from $1.4\times10^{-24}$ to $1.4\times10^{3}$, proportional to $\lambda$ to all printed digits. A larger cap therefore only sharpens the separation of zero from positive values, and when a value lies close to the gap a second solve at another $\lambda$ settles it.

The $X$-margin has a transparent meaning on the functional side, through the reformulation of Slater's theorem in Section \ref{sec:slaterconds}: when the $Y$-problem is feasible, $X$-Slater holds exactly when the set of optimal functionals is nonempty and bounded. For navigator \SDPs\ this is the typical situation (Sections \ref{sec:casestudy}--\ref{sec:twistgap}). In feasibility mode, where every excluding functional is optimal, the set of optimal functionals is unbounded exactly when a nonzero functional that is positive semidefinite on all blocks and has vanishing identity component exists (a recession direction), and the $X$-Slater condition then fails (Section \ref{sec:feasxslater}). As a test of the diagnostic itself we planted an $X$-Slater failure in the healthy three-parameter navigator of Section \ref{sec:casestudy}:\footnote{The planted component breaks the construction that guarantees a navigator is bounded from above, so the result is not a true navigator function.} a new functional component with zero objective that enters a single block through the polynomial represented by the identity Gram matrix, which forces that block of every feasible $X$ to vanish. On the original problem the diagnostic returns $t^*=+\infty$ without the $\vareps$-layer, and $3.4\times10^{-14}$ with it ($\lambda=10^3$; the value then sits at the cap, in the units of the layer, and only whether it lies above the duality gap matters). On the planted problem it returns $t^*=-2.9\times10^{-32}$, zero at the duality gap of the solve (${\sim}10^{-31}$), in $176$ iterations, as many as the healthy problem needs. Without the floor the verdict at the disallowed feasibility points of Section \ref{sec:feasxslater} comes out as $t^*\approx10^{-75}$, in $5$--$45\%$ more iterations than with it. The diagnostic then has no central path, and the tiny value is an artifact of the boundary rather than a measured margin.

The $+\infty$ of the healthy navigator has a simple origin, which also shows how $X$-Slater can be built into a navigator. If the normalization vector is the crossing image of a positive-definite density, $\vec M=-\sum_p x^\circ_pV_p$ with $\sum_px^\circ_pA_p\succ0$, then raising the navigator multiplier by $s$ and the density by $s\,x^\circ$ preserves the $X$-side equations identically, so $x^\circ$ is a positive-definite recession direction of the $X$-feasible set. $X$-Slater then holds, with $t^*=+\infty$, at every point where that set is nonempty, i.e.\ wherever the navigator is finite, whatever the polynomial matrices do at particular points, since planted zeros, shared prefactors and degree mismatches constrain only the $Y$-side. The $\Sigma$-navigator of \cite{Reehorst:2021ykw} is of this form when its sum covers every block with enough points to make the sampled Gram matrix positive definite. The implementation used here sums over the continuum blocks only, so the isolated external-exchange block of the three-parameter formulation is not covered, and its positive margin is observed rather than built in. The planted failure above is the complementary situation, a functional component on which the normalization vanishes, which a normalization covering every block excludes. Where a navigator fails $X$-Slater, extending its normalization in this way is a cheaper remedy than the transform of Section \ref{sec:xguarantee}.

Together, \texttt{ycompute} and \texttt{xcompute} check both Slater conditions, at the cost of one SDPB solve each. In the next section we use the same two representations (the weighted slack column on the $Y$-side and the identity \eqref{eq:identity-rep} on the $X$-side) to guarantee the Slater conditions.

\section{Guaranteeing Slater conditions}
\label{sec:guarantees}

The diagnostic tools introduced above identify when a problem is degenerate; this section presents the corresponding remedies. We seek a modified \SDP\ that (i) satisfies one or both Slater conditions \emph{by construction}, (ii) is again an \SDP, and (iii) has an optimal value that rigorously bounds, and ideally equals, the original navigator value. We present constructions for both sides; which side fails in which kind of run is discussed in Section \ref{sec:feasxslater}.

\subsection{\texttt{yguarantee}: the bounded \texorpdfstring{$Y$}{Y}-margin navigator}
\label{sec:yguarantee}
We assume in this section that the problem is a navigator \SDP\ (Section \ref{sec:XYproblems}), or more generally that both sides are at least weakly feasible with finite optimal values. In feasibility-mode runs with a trivial objective only weak $X$-feasibility is guaranteed, and the guarantees do not combine with the feasibility flags (Section \ref{sec:feasrigorous}); there the classifier is the $Y$-margin of Section \ref{sec:ycompute}, or the question is posed as a navigator problem.

We keep the slack variable $\vareps$ of \texttt{ycompute} but cut it off at zero and penalize it, so that the optimizer is driven back to the original problem:
\beq
	\label{eq:pmp-nav-cap}
	\begin{aligned}
		\text{maximize}\quad & b^\top y + \lambda\,\vareps \\
		\text{subject to}\quad
		& M_0^j(x)+\sum_{n=1}^N y_n M_n^j(x)-\vareps\,\rho_j(x)\, I \succeq 0
		\qquad \forall x\ge 0,\ \forall j, \\
		& \vareps \le 0,
	\end{aligned}
\eeq
with penalty weight $\lambda>0$. The cutoff $\vareps\le0$ is implemented in PMP form by one additional $1\times1$ scalar block, so \eqref{eq:pmp-nav-cap} remains a standard SDPB input. In \SDP\ data terms, \texttt{yguarantee} appends the slack column to $B$, appends the cutoff block, extends $b\to(b,\lambda)$, and leaves $c$ untouched for all original blocks.

The $Y$-guarantee has three properties.

\paragraph{(1) $Y$-Slater holds.} For any $y$, taking $\vareps$ sufficiently negative makes every block uniformly positive definite at finite $x$ and positive definite in its leading coefficient, which by the criterion of Section \ref{sec:ycompute} guarantees a strictly feasible Gram; the cutoff block contributes $-\vareps>0$. Hence a strictly feasible $Y'\succ0$ always exists, whatever the original problem; the weight $\rho_j$ was designed to provide this property.

\paragraph{(2) The optimal value is a rigorous upper bound, for any $\lambda>0$.} Any $Y$-feasible point $(y^*,Y^*)$ of the original \SDP\ embeds into the \texttt{yguarantee} feasible set with $\vareps=0$ and unchanged objective $b^\top y^*$. Maximizing over a superset can only increase the value:
\beq
	\boxed{\;v_{\mathrm{yg}} \;\ge\; \vnav \qquad\text{for every }\lambda>0.\;}
	\label{eq:yg-upper}
\eeq
Thus, a negative \texttt{yguarantee} value shows that $\vnav<0$, i.e.\ an allowed point. A positive value is inconclusive for exclusion until $\lambda$ has been verified to be large enough (see below). The inequality can be strict, because the $\vareps$-slack enlarges the feasible set, and when $\lambda$ is too small the gain in $b^\top y$ from the enlargement outweighs the penalty.

\paragraph{(3) The penalty threshold.} Define the gain function
\beq
	g(\vareps) := \sup\bigl\{b^\top y : \text{$Y$-feasible with slack } \vareps\bigr\},
	\qquad \vareps\le0,\qquad g(0)=v_Y=\vnav ,
	\label{eq:gainfunction}
\eeq
so that $v_{\mathrm{yg}}(\lambda)=\sup_{\vareps\le0}\bigl[g(\vareps)+\lambda\vareps\bigr]$. Since relaxing the constraints enlarges the feasible set, $g$ is nonincreasing in $\vareps$, and it is concave, being the optimal-value function of a convex program in $(y,\vareps)$. The statements below are collected in a proposition; the rest of this subsection proves and interprets them. Write $v_X$ for the value of the original $X$-problem (a weak-duality upper bound on $v_Y$, equal to it absent a duality gap).
\begin{proposition}\label{prop:yg}
Assume both sides of the original \SDP\ are at least weakly feasible with finite values.
\begin{enumerate}[label=(\roman*)]
	\item $v_{\mathrm{yg}}(\lambda)\ge\vnav$ for every $\lambda>0$, $v_{\mathrm{yg}}$ is nonincreasing in $\lambda$, and $v_{\mathrm{yg}}(\lambda)\to v_X$ as $\lambda\to\infty$.
	\item $v_{\mathrm{yg}}(\lambda)=\vnav$ if and only if $\lambda\ge L_*$, where
	\beq
		L_* := \sup_{\vareps<0} \frac{g(\vareps)-g(0)}{-\vareps}
		\label{eq:Lstar}
	\eeq
	is the maximal slope of the gain function. Absent a duality gap, $L_*$ equals the minimum of $\langle X^*,\widehat Y\rangle$ over the optima $X^*$ of the original $X$-problem, and is finite if and only if that optimum is attained.
	\item Let $\lambda_{\mathrm{feas}}:=\min\{\langle X,\widehat Y\rangle:\ X\ \text{feasible for the original $X$-problem}\}$. For $\lambda<\lambda_{\mathrm{feas}}$ the guaranteed problem is unbounded; for $\lambda\ge\lambda_{\mathrm{feas}}$ its value is finite and its $X$-side attains its optimum. If the returned $X$ has cutoff multiplier $x_c=\lambda-\langle X,\widehat Y\rangle>0$ (defined in \eqref{eq:xc} below), then $X$ is an optimum of the original $X$-problem and $v_{\mathrm{yg}}(\lambda)=v_X$; absent a duality gap, therefore, $v_{\mathrm{yg}}(\lambda)=\vnav$ and $L_*\le\lambda-x_c$.
\end{enumerate}
\end{proposition}
\noindent Statement (ii) follows directly from the definitions: $v_{\mathrm{yg}}=\vnav$ exactly when $g(\vareps)-g(0)+\lambda\vareps\le0$ for every $\vareps<0$.\footnote{This is an instance of the classical \emph{exact penalty} phenomenon of constrained optimization: replacing a constraint by a linear (nonsmooth) penalty on its violation reproduces the constrained optimum exactly for every finite penalty weight above a threshold set by the problem's optimal dual data. Finite exactness of such penalties goes back to Zangwill \cite{Zangwill:1967}; Bertsekas gave necessary and sufficient conditions with the sharp threshold for convex programs \cite{Bertsekas:1975exact}, and Han and Mangasarian a general treatment \cite{HanMangasarian:1979}. Here the penalized constraint is $\vareps=0$, the gain function \eqref{eq:gainfunction} is its (concave) optimal-value function, and the threshold $L_*$ is the maximal slope of that function, i.e.\ its multiplier.} The threshold has a direct $X$-side meaning. Write the slack column as $w_p=\Tr(A_p\widehat Y)$, with $\widehat Y\succ0$ as in Section \ref{sec:ycompute}. The $X$-side of \eqref{eq:pmp-nav-cap} is the original $X$-problem with one extra variable, the \emph{cutoff multiplier} $x_c\ge0$ (the $X$-side variable of the cutoff block), tied to the original ones by
\beq
	x_c \;=\; \lambda-\langle X,\widehat Y\rangle ,
	\label{eq:xc}
\eeq
and an original $X$-optimum $X^*$ is feasible, hence optimal, for the guaranteed $X$-problem exactly when $\langle X^*,\widehat Y\rangle\le\lambda$, i.e.\ $x_c\ge0$. Hence (absent a duality gap)
\beq
	L_* \;=\; \min\bigl\{\langle X^*,\widehat Y\rangle:\ X^*\ \text{optimal for the original $X$-problem}\bigr\},
	\label{eq:Lstar-X}
\eeq
which is finite if and only if the original $X$-optimum is attained. Attainment is not guaranteed on the problems of interest, since $X$-Slater guarantees attainment on the $Y$-side only, and when $Y$-Slater fails the set of $X$-optima may be empty (and is unbounded when it is not).\footnote{Attainment is subtle already before truncation: at $\Lambda=\infty$ \cite{Hercek:2026ext} prove that the optimum on the solution side is attained for 1d OPE maximization, while the existence of extremal functionals remains open; at finite $\Lambda$ with one correlator the functional side is attained \cite{ElShowk:2012hu} but the solution side need not be.}

What does hold whenever the original $X$-problem is feasible is convergence, statement (i). Define $G(\vareps):=\inf_x\bigl[c^\top x+|\vareps|\,\langle X,\widehat Y\rangle\bigr]$ over the fixed $X$-feasible set, for all $\vareps\le0$, so that $G(0)=v_X$. For $\vareps<0$ the relaxed $Y$-problem is strictly feasible (add $|\vareps|\widehat Y$ to any feasible Gram matrix), so strong duality gives $g(\vareps)=G(\vareps)$ there. $G$ is an infimum of affine functions of $\vareps$, hence upper semicontinuous at $0$: $\limsup_{\vareps\to0^-}g(\vareps)\le v_X$. On the other hand $G(\vareps)\ge G(0)$ for $\vareps<0$, since the added term is nonnegative. Hence $\lim_{\vareps\to0^-}g(\vareps)=v_X$ (with a duality gap, $g$ jumps at $\vareps=0$ from $v_X$ down to $v_Y$, and $L_*=\infty$). Since the chord slopes of a concave function are monotone, $v_{\mathrm{yg}}=\sup_{\vareps\le0}\bigl[g(\vareps)-\lambda|\vareps|\bigr]\to v_X$ as $\lambda\to\infty$, and $v_X=\vnav$ absent a duality gap, in particular under $X$-Slater. Exactness at finite $\lambda$ additionally requires $L_*<\infty$, and a single run can establish it (below).

When the $Y$-side optimum of the guaranteed problem is attained, which is the case when the guaranteed problem satisfies $X$-Slater (Table \ref{tab:guarantee-logic}), write $\bar\vareps$ for its optimal slack, to distinguish it from the $Y$-margin $\vareps^*$ of \texttt{ycompute}. Below $L_*$, $\bar\vareps<0$ lies where $-\lambda\in\partial g(\bar\vareps)$ and moves toward zero as $\lambda$ grows; above it, $\bar\vareps=0$ exactly and $v_{\mathrm{yg}}=\vnav$ (in a finite-precision solve the computed $\bar\vareps$ is then zero at the duality gap of the solve, $\lambda|\bar\vareps|\sim$ gap, so that it appears to scale as $1/\lambda$). The size of the deviation follows from concavity of $g$: at the optimum the slope of $g$ has magnitude $\lambda$, so the chord slope $(g(\bar\vareps)-g(0))/|\bar\vareps|$ lies between $\lambda$ and $L_*$, and
\beq
	0\;\le\; v_{\mathrm{yg}}-\vnav \;=\; \bigl[g(\bar\vareps)-g(0)\bigr]-\lambda|\bar\vareps| \;\le\; (L_*-\lambda)\,|\bar\vareps| .
	\label{eq:yg-bias}
\eeq
In particular $\lambda|\bar\vareps|$ is \emph{not} a bound on the bias, which exceeds it whenever the chord slope exceeds $2\lambda$.\footnote{For example, $g=10|\vareps|$ for $|\vareps|\le1$ with slope $1$ beyond and $\lambda=2$ give $\bar\vareps=-1$, a bias of $8$, and $\lambda|\bar\vareps|=2$.}

What a single run can establish is that its $\lambda$ lies above the threshold $L_*$, where the bias vanishes, and \eqref{eq:xc} provides the certificate, statement (iii). The guaranteed problem satisfies $Y$-Slater, so whenever its value is finite ($\lambda\ge\lambda_{\mathrm{feas}}$) its $X$-side attains its optimum, and that $X$-side is the original $X$-problem restricted by $\langle X,\widehat Y\rangle\le\lambda$. If the returned minimizer has $x_c=\lambda-\langle X,\widehat Y\rangle$ positive well above the solver tolerance, the restriction is inactive, so by convexity $X$ is a minimizer of the unrestricted original $X$-problem. The $X$-optimum is then attained and $v_{\mathrm{yg}}=v_X$; absent a duality gap, $v_{\mathrm{yg}}=\vnav$ and $L_*\le\lambda-x_c$ is finite. (When the $Y$-side is attained too, the cutoff pair is complementary, $x_c\,\bar\vareps=0$, so $x_c>0$ forces $\bar\vareps=0$ directly.) On a $Y$-degenerate problem the attained $X$-optimal set is unbounded, since $X$ can grow along the exposing certificate (Section \ref{sec:exposing}), so the returned $\langle X,\widehat Y\rangle$ is set by the cap (close to $\lambda$ in our runs, Table \ref{tab:lambda-sweep}) and is not $L_*$, which is the minimum over that set. Conversely, $x_c\approx0$ with $\bar\vareps$ clearly negative signals a run below the threshold; increase $\lambda$. Both numbers are printed by every solve ($x_c$ is the last block of $x$, $\bar\vareps$ the last entry of $y$), so the check costs nothing.\footnote{The certificate is a numerical statement at the solver tolerance, so a run at the threshold, $\lambda=L_*$, returns $x_c\approx0$ and $\bar\vareps\approx0$ and is inconclusive, while the upper bound \eqref{eq:yg-upper} holds regardless. Since the location of $L_*$ depends on the units of the slack column (Section \ref{sec:exp-survey}), we recommend the a posteriori check rather than a fixed default.}

The unmodified navigator is recovered as $\lambda\to\infty$ rather than as $\lambda\to0$. At $\lambda=0$ the slack column and cutoff block are still part of the constraints and $\vareps$ is unpenalized, so the deformation is maximal; generically the problem is then unbounded. More precisely, the guaranteed problem is bounded exactly when its $X$-side is feasible, i.e.\ when $\lambda\ge\lambda_{\mathrm{feas}}$ (the minimum is attained, since $\widehat Y\succ0$), the asymptotic slope of $g$; at $\lambda=1$ both case-study points of Section \ref{sec:exp-survey} are below this bound. Large $\lambda$ removes the effect of the guarantee on the value, and $Y$-Slater holds for every $\lambda$, but for $\lambda\ge L_*$ the optimum sits at $\bar\vareps=0$, on the original degenerate face, so the late iterations again approach the degenerate geometry. This trade-off between bias and conditioning matters most close to the navigator zero-crossing, and we return to it in Section \ref{sec:exp-survey} and Appendix \ref{sec:skydive-ab}.

\subsection{A no-go: no \texorpdfstring{$Y$}{Y}-Slater-restoring lower bound}
\label{sec:nogo}

The bound \eqref{eq:yg-upper} controls $\vnav$ from above only, and no modification that is sound in the sense below can control it from below while restoring $Y$-Slater. The lower bounds of the next two subsections are restrictions of the $Y$-side, or hold only after an a posteriori check, and do not contradict this (see the end of this subsection). Suppose we look for the mirror image of \texttt{yguarantee}: a modified \SDP\ that (i) satisfies $Y$-Slater by construction and (ii) has optimal value $\le\vnav$ by construction. Requirement (ii) is soundness: we call a modification \emph{sound} when every feasible point of the modified problem maps to a feasible point of the original problem with no smaller objective, so that the modified optimum is a valid bound on the original one. The natural candidate flips the sign of the slack, demanding a margin instead of permitting a deficit,
\beq
	M_0^j(x)+\sum_n y_n M_n^j(x) - \vareps\, W^j(x) \succeq 0,
	\qquad \vareps\ge0,
	\label{eq:flipped-slack}
\eeq
with $W^j\succeq0$ on the half-line. This modification is sound, since any feasible $(y,\vareps)$ has $M(x;y)\succeq \vareps W(x)\succeq 0$, so that $y$ is feasible for the original problem with the same objective, and the optimal value can only drop. Now suppose the modified problem had a strictly feasible point, i.e.\ a positive-definite Gram for $M-\vareps W$. Then
\begin{equation*}
	M \;=\; \underbrace{(M-\vareps W)}_{\text{positive-definite Gram}} \;+\; \underbrace{\vareps W}_{\text{Gram-representable}},
\end{equation*}
and since Gram representations add, $M$ itself acquires a positive-definite Gram, so the original problem would satisfy $Y$-Slater. Contrapositively, when the original $Y$-Slater condition fails, the flipped-slack problem has no strictly feasible point either, for any choice of weight $W$. The argument is general, because any modification whose constraints imply the original block constraints, as soundness demands, inherits their degeneracy.\footnote{Detours through the $X$-side fare no better. Relaxing the $X$-cone, $X\succeq-\eta W$, lower-bounds $v_X$ ($=\vnav$ when $X$-Slater holds), but its dual is the same $Y$-feasible set with objective $b^\top y - \eta\Tr(WY)$, so the geometry SDPB must navigate is untouched. Relaxing the cone by a penalized $\eta I$ (the $X$-guarantee of Section \ref{sec:xguarantee}) or the linear constraint $B^\top x=b$ (the $B$-slack of Appendix \ref{app:bslack}) restricts the $Y$-side by a half-space, and a restriction cannot create interior. Every perturbative route either leaves the degenerate $Y$-geometry in place or shrinks it further.}

The asymmetry has a simple origin. An upper bound comes from a relaxation, which can create interior where there was none (as \texttt{yguarantee} does), whereas a lower bound comes from a restriction, and a subset of a set with empty interior has empty interior in the same ambient space. The Slater condition, however, is a statement relative to an ambient space: a proper face of the positive-semidefinite cone has empty interior in $\bS^K$ but a nonempty \emph{relative interior} (the interior taken within the smallest linear subspace containing the set) in its own span. Restricting the problem to the face, by changing variables rather than adding penalties, therefore escapes the no-go argument; this is facial reduction \cite{BorweinWolkowicz:1981,BorweinWolkowicz:1981b}. As Section \ref{sec:pmpfr} shows, our diagnostics supply the data needed to perform it without leaving the PMP class. The lower bounds of Sections \ref{sec:xguarantee}--\ref{sec:twosided} respect the no-go: \texttt{xguarantee} restricts the $Y$-side and inherits its degeneracy, and \texttt{xyguarantee} does restore $Y$-Slater, but its value is a lower bound only after the check $x_c>0$ passes rather than by construction, and it reaches $\vnav$ only when the $X$-optimum is attained (Section \ref{sec:twosided}).

\subsection{The \texorpdfstring{$X$}{X}-guarantee (\texttt{xguarantee})}
\label{sec:xguarantee}

The identity representation \eqref{eq:identity-rep} makes the textbook $X$-side regularization available: relax the cone to $X+\eta I\succeq0$ with a penalized $\eta\ge0$,
\beq
	\label{eq:xguarantee}
	\begin{aligned}
		\text{minimize}\quad & c^\top x + \nu\,\eta \\
		\text{subject to}\quad & X = \sum_p x_pA_p,\qquad B^\top x = b,\qquad X+\eta I\succeq0,\qquad \eta\ge0 .
	\end{aligned}
\eeq
With $z:=x+\eta\,\mathbf 1_{\mathrm{diag}}$ one has $X+\eta I=\sum_pz_pA_p$, $B^\top x=B^\top z-\eta\,h$ and $c^\top x=c^\top z-\eta\,c_{\mathrm{diag}}$, so in SDPB form \eqref{eq:xguarantee} leaves all original data untouched and adds a single $1\times1$ block for $\eta$, with $B$-row $-h$ and $c_\eta=\nu-c_{\mathrm{diag}}$.
\begin{itemize}
	\item \emph{$X$-Slater holds by construction.} Take any $x$ with $B^\top x=b$ (such $x$ exist because $B$ has full column rank, which SDPB requires anyway) and any $\eta>\max(0,-\lambda_{\min}(X))$.
	\item \emph{On the $Y$-side the new block is a trace bound.} It reads $\nu-c_{\mathrm{diag}}+d\cdot y\ge0$, i.e.\ by \eqref{eq:trace-linear}
	\beq
		\Tr Y\;\le\;\nu .
		\label{eq:trace-bound}
	\eeq
	\item \emph{Value.} The $X$-feasible set is enlarged and the $Y$-feasible set restricted, so $v_{\mathrm{xg}}\le\vnav$. The inequality is an equality as soon as some optimal $Y$ of the original problem has $\Tr Y<\nu$, and this is checked a posteriori: if the solution has $Y_\eta=\nu-\Tr Y^*>0$, the constraint \eqref{eq:trace-bound} is inactive at the optimum and can be dropped without changing it (the problem is convex), so $v_{\mathrm{xg}}=\vnav$. If $Y_\eta$ is zero, raise $\nu$. For bootstrap logic this mirrors \texttt{yguarantee}, in that a positive $v_{\mathrm{xg}}$ implies $\vnav>0$ (a disallowed point) for every $\nu$, while a negative value is inconclusive until the check $Y_\eta>0$ has passed.
\end{itemize}
The $X$-guarantee composes with the $Y$-guarantee, and \texttt{xyguarantee} applies both in one call. Applied to the output of \texttt{yguarantee}, whose cutoff block is a $1\times1$ diagonal block and hence part of the identity, it adds the trace bound to the guaranteed problem. The $X$-Slater condition again holds by construction, and the $Y$-Slater condition holds as soon as $\nu$ exceeds the trace of some strictly feasible $Y$ of the $Y$-guaranteed problem, which is implied by the exactness condition above. The value of the doubly guaranteed problem lies between $v_{\mathrm{xg}}$ and $v_{\mathrm{yg}}$, and the two checks $x_c>0$, $Y_\eta>0$ of the same run tell which end it is, or whether it is $\vnav$; Section \ref{sec:twosided} turns this into a procedure.

The trace interpretation also explains why the alternative $B$-slack guarantee of Appendix \ref{app:bslack} needs penalties of order $10^{16}$, since its half-space is, up to off-diagonal terms, a bound on $\Tr Y$ in SDPB's scaled units.

\paragraph{Tests.} On the healthy three-parameter navigator of Section \ref{sec:casestudy} ($\vnav=5.8788\times10^{-4}$) the $X$-guarantee reproduces the navigator value at $\nu=10^{20}$, where $\Tr Y^*=1.2\times10^{12}$ leaves the bound inactive by eight orders of magnitude; at $\nu=10^3$ and $10^{10}$ it returns values below $\vnav$, each flagged by $Y_\eta=0$. On the planted $X$-Slater failure of Section \ref{sec:xdiag} it reproduces that problem's $\vnav$ at $\nu=10^{20}$, with $\Tr Y=0.93\,\nu$, so $Y_\eta>0$ and the value is exact; the bound keeps finite the unbounded optimal set that constitutes the $X$-Slater failure. Combined with the $Y$-guarantee, the doubly guaranteed problem returns $\vnav$ on both. On the $Y$-degenerate two-parameter problem at $(0.51,\,2.5)$, the doubly guaranteed problem returns the $Y$-guarantee's upper bound, coinciding with the original optimum when $\lambda$ is above that point's threshold $L_*$ (Section \ref{sec:exp-survey}).

\subsection{Bounds, exactness and attainment}
\label{sec:twosided}

Table \ref{tab:guarantee-logic} collects what each guarantee does to the two Slater conditions and what its value means. Two thresholds appear: $L_*$ of Section \ref{sec:yguarantee}, finite exactly when the $X$-optimum of the original problem is attained, and its mirror $\nu_*$, the smallest trace of an optimal $Y$, finite exactly when the $Y$-optimum is attained. Throughout this subsection we assume that the original problem has no duality gap, which either Slater condition guarantees and which therefore holds for every navigator of this paper. (With a gap, $L_*=\infty$ even when the $X$-optimum is attained, and the check $x_c>0$ certifies $v_{\mathrm{yg}}=v_X$ only.)
\begin{table}[htpb]
\centering
\small
\renewcommand{\arraystretch}{1.3}
\setlength{\tabcolsep}{4pt}
\begin{tabularx}{\textwidth}{@{}>{\raggedright\arraybackslash}p{2.3cm}>{\raggedright\arraybackslash}X>{\raggedright\arraybackslash}X>{\raggedright\arraybackslash}p{2.2cm}>{\raggedright\arraybackslash}X@{}}
	\toprule
	 & own $Y$-Slater & own $X$-Slater & value & exact when (check) \\
	\midrule
	\texttt{yguarantee} & always & iff original $X$-Slater and $\lambda>\langle X^\circ,\widehat Y\rangle$ for some strictly feasible $X^\circ$ & $v_{\mathrm{yg}}\ge\vnav$ & $\lambda\ge L_*$ ($x_c>0$); possible iff the $X$-optimum is attained \\
	\texttt{xguarantee} & iff original $Y$-Slater and $\nu>\Tr Y^\circ$ for some strictly feasible $Y^\circ$ & always & $v_{\mathrm{xg}}\le\vnav$ & $\nu\ge\nu_*$ ($Y_\eta>0$); possible iff the $Y$-optimum is attained \\
	\texttt{xyguarantee} & for $\nu$ large enough, whatever the original problem & always & between $v_{\mathrm{xg}}$ and $v_{\mathrm{yg}}$ & $=v_{\mathrm{yg}}$ if $Y_\eta>0$, $=v_{\mathrm{xg}}$ if $x_c>0$, $=\vnav$ if both (both optima attained) \\
	\bottomrule
\end{tabularx}
\caption{What the guarantees deliver. \enquote{Own} Slater conditions are those of the transformed problem, which decide whether SDPB converges on it. The checks are read off the same solve; a penalty too small for the transformed problem to be feasible shows up as a diverging objective.}
\label{tab:guarantee-logic}
\end{table}

The bracket
\beq
	v_{\mathrm{xg}}(\nu)\;\le\;\vnav\;\le\;v_{\mathrm{yg}}(\lambda)\qquad\text{for all }\nu,\lambda>0
	\label{eq:xy-bracket}
\eeq
holds for every penalty. The practical procedure follows from the table. For $\nu$ large enough \texttt{xyguarantee} satisfies both Slater conditions whatever the original problem, so SDPB converges on it, and the same run tells which end its value is: the upper end $v_{\mathrm{yg}}(\lambda)$ if the trace bound is inactive ($Y_\eta>0$), the lower end $v_{\mathrm{xg}}(\nu)$ if the cap is inactive ($x_c>0$), and $\vnav$ itself if both are. If only one check passes, the other penalty is raised. Which Slater condition of the original problem fails does not decide the outcome; attainment does, and each check speaks about the opposite side. By complementary slackness, $x_c>0$ at $\nu>\nu_*$ exhibits an attained $X$-optimum, and $Y_\eta>0$ at $\lambda>L_*$ an attained $Y$-optimum, so a single run returns $\vnav$ exactly when both optima are attained. For a $Y$-degenerate problem that satisfies $X$-Slater, such as every navigator of this paper (Section \ref{sec:feasxslater}), the $Y$-optimum is attained automatically, and the problem is solved exactly whenever its $X$-optimum is. If the $X$-optimum is not attained, which can happen only when $Y$-Slater fails, $x_c$ vanishes for every $\lambda$ once $\nu>\nu_*$, so the runs return upper bounds $v_{\mathrm{yg}}(\lambda)$ that converge from above, and no lower end. The lower bound is then supplied by facial reduction of the $Y$-side (Section \ref{sec:sandwich}). In the mirror case, where the $Y$-optimum is not attained, only lower bounds are read off, and a reduction of the $X$-side supplies the upper bound. Both reductions give bounds that are valid whether or not the face is identified correctly.

As an example we take the $Y$-degenerate two-parameter navigator of Section \ref{sec:casestudy} at the disallowed point $(0.51,\,2.5)$ ($\Lambda=5$, $\nu=10^{20}$). At $\lambda=10^6$, below $L_*$, \texttt{xyguarantee} converges to $1.7783612\times10^{-3}$ with $x_c=6\times10^{-20}$ and $Y_\eta>0$, so the value is the upper bound only. At $\lambda=10^8$ and $10^{10}$ it converges to $1.778356673\times10^{-3}$ with $x_c=4.2\times10^5$ and $4.5\times10^7$ and $Y_\eta>0$ (the optimal Gram matrix has trace $0.40$), so both checks pass and the value is $\vnav$, although the original problem violates $Y$-Slater. With the trace bound inactive these coincide with the \texttt{yguarantee} values of Table \ref{tab:lambda-sweep}, as they must. Lowering $\nu$ below the trace of every feasible Gram matrix of the $Y$-guaranteed problem ($\nu=0.1$) makes the transformed problem infeasible, and the solve diverges.

These statements concern the values SDPB computes. For bounds that hold in exact arithmetic, the numerical point realizing each end must in addition be rounded into a strictly feasible neighbourhood and verified exactly (e.g.\ polynomial positivity in the Bernstein basis, as done for the 3d Ising bounds of \cite{Reehorst:rigorous}), which requires the Slater condition of the side that supplies the point: the $X$-side for the upper end, the $Y$-side for the lower end. In exact arithmetic the width of the verified interval bounds every source of error at once (a penalty below its threshold, a misidentified face, numerical noise of the solve), with no component requiring separate proof; for the computed values it bounds the first two, up to the duality gaps of the solves.

\subsection{Choosing the penalty weight: a $\lambda$ sweep}
\label{sec:exp-survey}

\textbf{Setup.} We take the two $\Lambda=5$ case-study points of Section \ref{sec:casestudy}, the allowed point $(0.5181,\,1.412625)$ and the disallowed point $(0.51,\,2.5)$; the two-parameter navigator's $Y$-Slater condition fails at both (Table \ref{tab:diagnosis-case-study}). We apply \texttt{yguarantee} with the column \eqref{eq:weighted-column} at $\lambda\in\{10^0,\dots,10^6,10^8,10^{10},10^{12}\}$, and solve each transformed \SDP\ to duality gap $10^{-30}$, recording the guaranteed navigator $v_{\mathrm{yg}}(\lambda)$ and the terminal slack $\bar\vareps(\lambda)$. The results are collected in Table \ref{tab:lambda-sweep}.

\begin{table}[htpb]
\centering
\renewcommand{\arraystretch}{1.15}
\small
\setlength{\tabcolsep}{3pt}
\begin{tabular}{r|lll|lll}
\toprule
 & \multicolumn{3}{c|}{allowed point} & \multicolumn{3}{c}{disallowed point} \\
$\lambda$ & $v_{\mathrm{yg}}$ & $\bar\vareps$ & $x_c$ & $v_{\mathrm{yg}}$ & $\bar\vareps$ & $x_c$ \\
\midrule
$10^0$ & \multicolumn{3}{c|}{unbounded} & \multicolumn{3}{c}{unbounded} \\
$10^1$ & $-4.2654\times10^{-6}$ & $-2.0\times10^{-7}$ & $1\times10^{-26}$ & $3.3215\times10^{-3}$ & $-4.0\times10^{-5}$ & $5\times10^{-29}$ \\
$10^2$ & $-7.3618\times10^{-6}$ & $-8.0\times10^{-10}$ & $9\times10^{-25}$ & $3.0345\times10^{-3}$ & $-1.8\times10^{-7}$ & $7\times10^{-27}$ \\
$10^3$ & $\vnav$ & $-3.3\times10^{-34}$ & $\mathbf{5.0}$ & $2.9520\times10^{-3}$ & $-7.5\times10^{-8}$ & $2\times10^{-26}$ \\
$10^4$ & $\vnav$ & $-1.8\times10^{-35}$ & $\mathbf{73}$ & $2.3808\times10^{-3}$ & $-6.1\times10^{-8}$ & $3\times10^{-26}$ \\
$10^5$ & $\vnav$ & $-1.3\times10^{-36}$ & $\mathbf{8.5\times10^{2}}$ & $1.7804\times10^{-3}$ & $-4.7\times10^{-11}$ & $3\times10^{-23}$ \\
$10^6$ & $\vnav$ & $-1.8\times10^{-37}$ & $\mathbf{8.7\times10^{3}}$ & $1.7783612\times10^{-3}$ & $-2.0\times10^{-14}$ & $6\times10^{-20}$ \\
$10^8$ & $\vnav$ & $-2.1\times10^{-39}$ & $\mathbf{8.7\times10^{5}}$ & $\vnav$ & $-5.3\times10^{-39}$ & $\mathbf{4.2\times10^{5}}$ \\
$10^{10}$ & $\vnav$ & $-8.8\times10^{-42}$ & $\mathbf{8.6\times10^{7}}$ & $\vnav$ & $-2.4\times10^{-41}$ & $\mathbf{4.5\times10^{7}}$ \\
$10^{12}$ & $\vnav$ & $-1.8\times10^{-43}$ & $\mathbf{8.6\times10^{9}}$ & $\vnav$ & $-1.8\times10^{-43}$ & $\mathbf{4.6\times10^{9}}$ \\
\bottomrule
\end{tabular}
\caption{$\lambda$-sweep of \texttt{yguarantee} (column \eqref{eq:weighted-column}) at the two $\Lambda=5$ case-study points: guaranteed navigator $v_{\mathrm{yg}}$, terminal slack $\bar\vareps$, and the cutoff multiplier $x_c$ of \eqref{eq:xc}, each solve to duality gap $10^{-30}$. $x_c$ in bold: clearly positive, certifying $\bar\vareps=0$ and $v_{\mathrm{yg}}=\vnav$ from that run alone. Entries $\vnav$ agree with the navigator value to all computed digits. }
\label{tab:lambda-sweep}
\end{table}

\textbf{Findings.} \emph{(i) The penalty threshold $L_*$ and its location.} At the allowed point $L_*$ lies between $\lambda=10^2$ and $10^3$: at $\lambda=10$ and $10^2$ the value is still perturbed, but already negative, which certifies the point allowed by the one-sided bound \eqref{eq:yg-upper}; from $\lambda=10^3$ on $v_{\mathrm{yg}}$ reproduces $\vnav$ to all computed digits. The $\lambda=10^3$ run alone locates the threshold more sharply, since its $x_c=5.0$ gives $\langle X,\widehat Y\rangle=995$, so $L_*\le995$ by \eqref{eq:Lstar-X}. At the disallowed point $L_*$ lies between $10^6$ and $10^8$: the value approaches $\vnav$ from above through $\lambda=10^3$--$10^6$, where the slack is still in use, and is exact from $10^{8}$ on. The cutoff multiplier tracks the threshold: at both points $x_c$ is negligible ($\le6\times10^{-20}$, against $\lambda\ge10$) in every run below $L_*$ and of order one or larger in every run above it, so a single run tells which side it is on. $L_*$ therefore depends on the problem and the point, by several orders of magnitude; no fixed default covers all cases.

\emph{(ii) The duality-gap signature.} Above $L_*$ the product $\lambda|\bar\vareps|$ is pinned at the duality-gap scale, ${\sim}10^{-31}$, and $|\bar\vareps|\propto1/\lambda$; below $L_*$ the product is far above the gap. The product also locates a run relative to $L_*$, but less sharply than $x_c$.

\emph{(iii) Too small a penalty.} At $\lambda=1$ the guaranteed problem is unbounded at both points, because the gain from relaxing outweighs the penalty all the way to $\vareps\to-\infty$ ($\lambda<\lambda_{\mathrm{feas}}$ in the notation of Proposition \ref{prop:yg}), and SDPB reports this as a complementarity blow-up.

\subsection{Summary of the toolbox}
\label{sec:toolsummary}

Table \ref{tab:tools} summarizes the tools, their data modifications, their guarantees, and the a posteriori check that goes with each parameter.

\begin{table}[htpb]
\centering
\small
\renewcommand{\arraystretch}{1.25}
\setlength{\tabcolsep}{4pt}
\begin{tabularx}{\textwidth}{@{}>{\raggedright\arraybackslash}p{2.2cm}>{\raggedright\arraybackslash}X>{\raggedright\arraybackslash}p{1.95cm}>{\raggedright\arraybackslash}p{1.95cm}>{\raggedright\arraybackslash}X>{\raggedright\arraybackslash}X@{}}
	\toprule
	Tool & Data modification & $Y$-Slater & $X$-Slater & Value & Parameter check \\
	\midrule
	\texttt{ycompute} & slack column \eqref{eq:weighted-column}; $b\to \hat e_{N+1}$ & guaranteed & attainment check$^{\ddagger}$ & $\vareps^*$ (diagnostic) & --- \\
	\texttt{xcompute} & $c\to0$; $1\times1$ block for $t+t_0\ge0$ ($B$-row $h$, $c=-1$); $b\to b+t_0 h$; optional $\vareps$-layer & guaranteed with $\vareps$-layer & guaranteed$^{\S}$ & $t^*$ (diagnostic) & $t^*>-t_0$; cap $\lambda$ large enough that the $X$-side stays feasible \\
	\texttt{yguarantee} & slack column, cutoff block; $b\to(b,\lambda)$ & guaranteed & as original$^{*}$ & $\ge\vnav$; $\to\vnav$ as $\lambda\to\infty$; $=\vnav$ for $\lambda\ge L_*$ & $x_c>0$ \eqref{eq:xc} \\
	\texttt{xguarantee} & $1\times1$ block ($B$-row $-h$, $c=\nu-c_{\mathrm{diag}}$) & as original$^{*}$ & guaranteed & $\le\vnav$; $=\vnav$ if $\Tr Y^*<\nu$ & $Y_\eta=\nu-\Tr Y^*>0$ \\
	\texttt{xyguarantee} & both of the above, $Y$-side first & guaranteed$^{\dagger}$ & guaranteed & $=v_{\mathrm{yg}}$ if $Y_\eta>0$, $=v_{\mathrm{xg}}$ if $x_c>0$, $=\vnav$ if both & both checks \\
	\bottomrule
\end{tabularx}
\caption{The Slater tools. \enquote{Diagnostic} tools measure a margin; \enquote{guarantee} tools solve the original problem with Slater conditions enforced. Each parameter check is read off the solution of the same run. The limit $\lambda\to\infty$ assumes no duality gap, and equality with $\vnav$ at finite $\lambda$ requires $L_*<\infty$ (Section \ref{sec:yguarantee}). $^\dagger$For $\nu$ above the trace of some strictly feasible $Y$, in particular whenever the trace bound is inactive. $^{*}$For a large enough penalty: \texttt{yguarantee} needs $\lambda>\langle X^\circ,\widehat Y\rangle$ for some strictly feasible $X^\circ$ \eqref{eq:xc}, \texttt{xguarantee} needs $\nu>\Tr Y^\circ$ for some strictly feasible $Y^\circ$. $^{\ddagger}$The $X$-side of \texttt{ycompute} is the homogeneous problem normalized by $w^\top x=1$ \eqref{eq:ycompute-X}; its Slater condition is the attainment check of Section \ref{sec:attainment} rather than that of the original problem. $^{\S}$Through the floor $t\ge-t_0$: whenever the original $X$-problem is feasible, with the cap large enough when the $\vareps$-layer is used (Section \ref{sec:xdiag}); the cutoff multiplier vanishes whenever the uncapped margin is $+\infty$ (e.g.\ at allowed feasibility points), since the cap then binds by design. Alternative $X$-side constructions (the $B$-slack guarantee and reference directions $X_0\neq I$) are collected in Appendix \ref{app:xslack-elim}.}
\label{tab:tools}
\end{table}

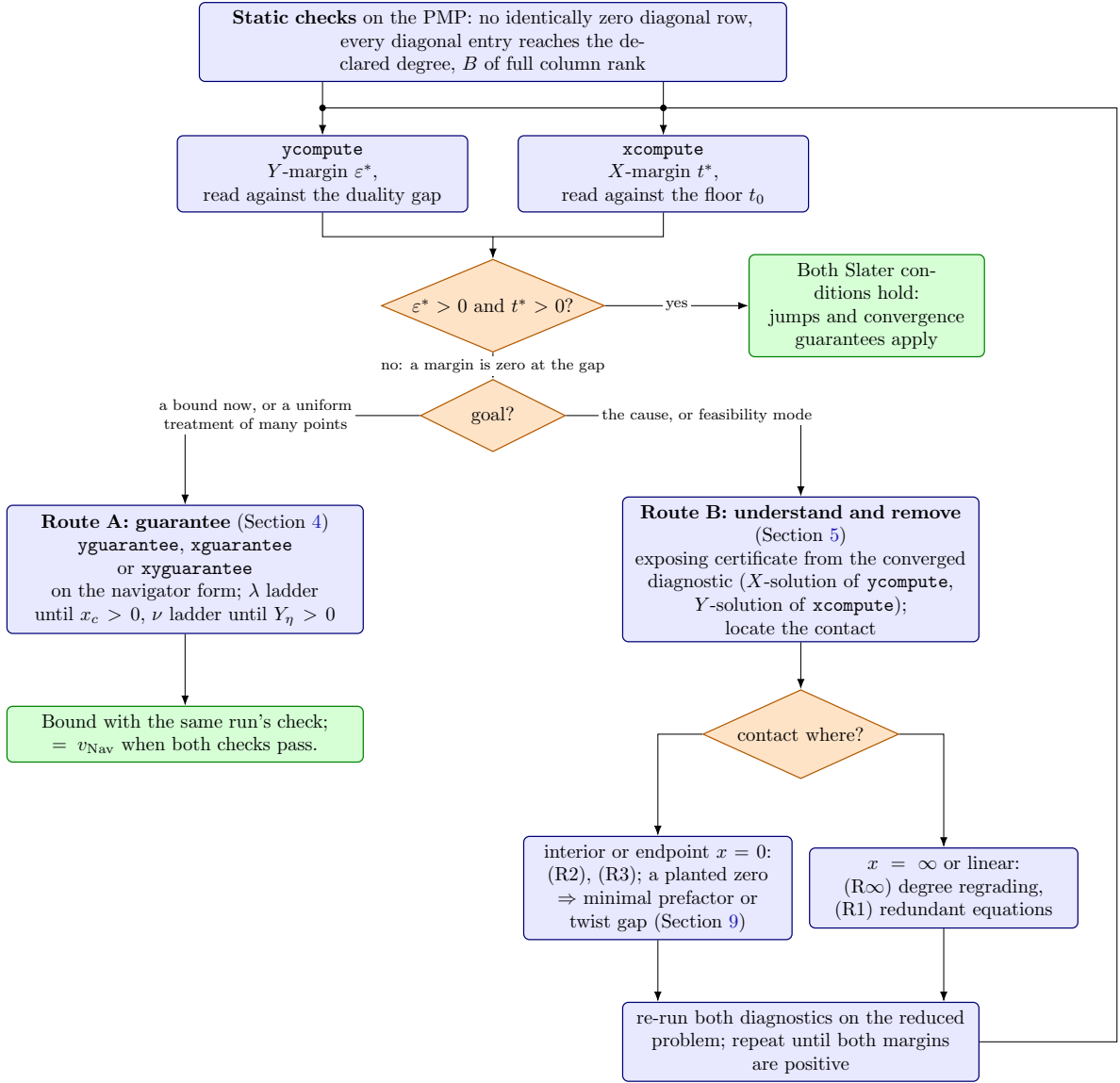
\begin{figure}[htpb]
\centering
\resizebox{\textwidth}{!}{%
\begin{tikzpicture}[
  font=\footnotesize, >=Latex, line width=0.5pt,
  box/.style={draw, rounded corners=3pt, align=center, inner sep=4pt, minimum height=9mm, line width=0.6pt},
  dec/.style={draw, diamond, aspect=2.4, align=center, inner sep=1pt, fill=orange!22, draw=orange!70!black, line width=0.6pt},
  plain/.style={box, fill=blue!9, draw=blue!45!black},
  good/.style={box, fill=green!16, draw=green!50!black},
  lbl/.style={font=\scriptsize, fill=white, inner sep=1.5pt}]
\node[plain, text width=104mm] (static) at (0,0) {\textbf{Static checks} on the PMP: no identically zero diagonal row,\\ every diagonal entry reaches the declared degree, $B$ of full column rank};
\node[plain, text width=50mm] (yc) at (-31mm,-24mm) {\texttt{ycompute}\\ $Y$-margin $\vareps^*$,\\ read against the duality gap};
\node[plain, text width=50mm] (xc) at (31mm,-24mm) {\texttt{xcompute}\\ $X$-margin $t^*$,\\ read against the floor $t_0$};
\node[dec, text width=32mm] (sign) at (0,-48mm) {$\vareps^*>0$ and $t^*>0$?};
\node[good, text width=40mm] (ok) at (68mm,-48mm) {Both Slater conditions hold:\\ jumps and convergence\\ guarantees apply};
\node[dec, text width=18mm, aspect=2] (route) at (0,-68mm) {goal?};
\node[plain, text width=62mm] (guar) at (-56mm,-96mm) {\textbf{Route A: guarantee} (Section~\ref{sec:guarantees})\\ \texttt{yguarantee}, \texttt{xguarantee} or \texttt{xyguarantee}\\ on the navigator form; $\lambda$ ladder\\ until $x_c>0$, $\nu$ ladder until $Y_\eta>0$};
\node[good, text width=62mm] (gres) at (-56mm,-126mm) {Bound with the same run's check;\\ $=\vnav$ when both checks pass.};
\node[plain, text width=62mm] (cert) at (56mm,-96mm) {\textbf{Route B: understand and remove}\\ (Section~\ref{sec:pmpfr})\\ exposing certificate from the converged\\ diagnostic ($X$-solution of \texttt{ycompute},\\ $Y$-solution of \texttt{xcompute}); locate the contact};
\node[dec, text width=28mm, aspect=2.2] (kind) at (56mm,-126mm) {contact where?};
\node[plain, text width=46mm] (fin) at (30mm,-154mm) {interior or endpoint $x=0$:\\ (R2), (R3); a planted zero\\ $\Rightarrow$ minimal prefactor or\\ twist gap (Section~\ref{sec:twistgap})};
\node[plain, text width=46mm] (inf) at (82mm,-154mm) {$x=\infty$ or linear:\\ (R$\infty$) degree regrading,\\ (R1) redundant equations};
\node[plain, text width=62mm] (rerun) at (56mm,-182mm) {re-run both diagnostics on the reduced\\ problem; repeat until both margins\\ are positive};
\draw[->] (static.south -| yc.north) -- (yc.north);
\draw[->] (static.south -| xc.north) -- (xc.north);
\draw[->] (yc.south) |- ($(sign.north)+(0,4mm)$) -- (sign.north);
\draw (xc.south) |- ($(sign.north)+(0,4mm)$);
\draw[->] (sign) -- node[lbl]{yes} (ok);
\draw[->] (sign) -- node[lbl]{no: a margin is zero at the gap} (route);
\draw[->] (route.west) -| node[lbl, pos=0.35, align=center]{a bound now, or a uniform\\ treatment of many points} (guar.north);
\draw[->] (route.east) -| node[lbl, pos=0.3]{the cause, or feasibility mode} (cert.north);
\draw[->] (guar) -- (gres);
\draw[->] (cert) -- (kind);
\draw[->] (kind.west) -| (fin.north);
\draw[->] (kind.east) -| (inf.north);
\draw[->] (fin.south) -- (fin.south |- rerun.north);
\draw[->] (inf.south) -- (inf.south |- rerun.north);
\draw (rerun.east) -| ($(inf.east)+(7mm,0)$) |- (-31mm,-12mm);
\fill (-31mm,-12mm) circle (0.6mm); \fill (31mm,-12mm) circle (0.6mm);
\end{tikzpicture}}
\caption{The tools as a procedure. Route A (left) buys a bound with its own check from one transformed solve; Route B (right) removes the degeneracy itself with the moves of Section \ref{sec:pmpfr} and is the only route for feasibility-mode questions. The static checks at the top catch the planted zeros of Section \ref{sec:fakepoles} and the degree deficits of Section \ref{sec:mixedcurrent} before any solve. After a reduction the loop returns to both diagnostics.}
\label{fig:workflow}
\end{figure}

Figure \ref{fig:workflow} arranges the tools into a procedure for a problem that does not converge, or that never produces a jump. The entry point is a static check of the PMP: no identically vanishing diagonal row, every diagonal entry reaching its block's declared degree, and $B$ of full column rank (Sections \ref{sec:fakepoles}, \ref{sec:mixedcurrent} and \ref{sec:attainment}). The diagnostics then measure both margins against the duality gap of their solves. If one of them is zero there are two routes. Route A applies a guarantee to the navigator form and costs one transformed solve per penalty value; it returns a bound whose status is read from the same run, reaches the navigator value once the parameter checks pass, and does not apply to the feasibility flags (Section \ref{sec:feasrigorous}). Route B reads the exposing certificate off the converged diagnostic, locates the contact and removes it with the moves of Section \ref{sec:pmpfr}; the inspection of the certificate is manual, but the route removes the cause, is exact within the PMP class, and is the only one available for a feasibility-mode question. A re-run of the diagnostic on the reduced problem closes the loop. Route A is the right choice whenever the degeneracy is a property of the solution rather than of the input, as in Case Study I, where inactive crossing equations force part of the functional to vanish. Such a face can be identified only after the fact, point by point, and need not persist from one external point or truncation to the next, so a scan, or a Skydive run that must center at every point it visits, cannot rely on its removal. The guarantee gives a uniform treatment: one transformation of the input makes every point of the scan non-degenerate by construction, with the check of its parameter read from each run. Route B is the choice when the cause is visible in the input, a planted zero or a degree deficit, and when the question is one the flags must answer.

\section{Facial reduction within the PMP class}
\label{sec:pmpfr}

Section \ref{sec:nogo} established that when $Y$-Slater fails, no modification whose value is a lower bound by construction can restore it, and Section \ref{sec:twosided} that the combined guarantee reaches the lower end only when the $X$-optimum is attained. Facial reduction, which restricts the problem to the face on which its feasible set lies, removes the degeneracy itself and gives a lower bound that needs neither a penalty nor attainment. This is Route B of Figure \ref{fig:workflow}. Two obstacles must be overcome to perform facial reduction for \SDPs. Textbook reductions operate at the Gram level, where the constraint matrices projected onto the face fall outside the polynomial structure on which SDPB's input format and solver depend.\footnote{Similar concerns have been addressed before for sum-of-squares programs, where the reduction can be kept within the polynomial structure by deleting monomials from the Gram basis \cite{Lofberg:2009,WakiMuramatsu:2010,PermenterParrilo:2014}. We found the simpler approach presented in this section sufficient and did not pursue this direction.} Moreover, the facial-reduction program can be as ill-conditioned as the original \SDP. This section shows how to resolve both problems. Reductions within the PMP class have been carried out by hand in individual setups, which we learned of during the final preparation of this paper through the census of Appendix \ref{app:survey}: \cite{Nancarrow:2022wdr} note that one can ``scan for the mismatching powers'' and eliminate the forced-zero functional component before running SDPB, and \cite{Chiang:2026ubv} solve the zero constraints forced by the leading coefficients exactly and reduce the functional dimension. Outside the PMP class, \cite{Lawrence:2024mnj} recover the implied affine constraints from a joint null space of the constraint matrices for an interior-point solver of their own. What this section adds is the general procedure: the face is found by a solve rather than by inspection, and every move is proven sound within the \SDP\ class.

The degeneracies of a bootstrap \SDP\ are polynomial-level phenomena: forced \emph{contacts}, i.e.\ points on the half-line, or directions at $x=\infty$, where the block polynomials of every feasible functional are forced to touch zero. That the degeneracies take this form is not obvious; we prove it in Appendix \ref{app:faithful} by classifying what the PMP-to-\SDP\ rewrite can and cannot break. The face can therefore be removed by operations on the PMP data itself, so that the result is again an \SDP. The face-finding problem is the \texttt{ycompute} diagnostic we have already solved, whose slack restores its own $Y$-Slater condition by construction. We first give a pedagogical overview of the whole procedure (Section \ref{sec:threeuses}), then make it precise: what the exposing certificate is and why inspecting the PMP input does not reveal it (Section \ref{sec:exposing}), why \texttt{ycompute}'s converged output is the certificate (Section \ref{sec:certfromycompute}), and the reduction moves and their soundness (Sections \ref{sec:pmpmoves}--\ref{sec:sandwich}).

\subsection{Beyond diagnosis: three further uses of the same solves}
\label{sec:threeuses}

The converged \texttt{ycompute} solve at zero margin contains, in its output, everything needed to (1) remove the degeneracy by facial reduction without leaving the PMP class, (2) understand it, i.e.\ name the sectors, equations and constraint vectors responsible, and (3) obtain a lower bound that does not depend on attainment of the $X$-optimum (Sections \ref{sec:nogo} and \ref{sec:twosided}). This subsection states the ideas in pedagogical form; the remainder of the section develops the theory. The applications in Sections \ref{sec:casestudy} (an intrinsic face of a mixed-correlator formulation), \ref{sec:feasworked} (the same face in feasibility mode), \ref{sec:faceworked2} (a planted configuration error) and \ref{sec:fakepolefr} (a block-prefactor convention) carry the procedure out and show what it reveals in each case (Section \ref{sec:faceworked2} stops after the diagnosis, which already identifies the mistake).

We focus on the $Y$-side problem, since its failure breaks the navigator problems of this paper and it is the side facial reduction has to repair. The relevant data are returned by a converged \texttt{ycompute} solve at $\vareps^*\approx0$. Like every interior-point solve it produces a \emph{pair} of solutions, one for the problem as posed (the ``$Y$ side'', the functional $y^*$ and its Gram matrices) and one for its conic dual (the ``$X$ side'', a weight $\hat x_p$ for every constraint row, combining the constraint matrices into a positive-semidefinite $\sum_p\hat x_pA_p$). At $\vareps^*=0$ these two outputs have the following meaning.
\begin{itemize}
	\item \emph{The $Y$-side output is a most-generic feasible functional.} An SDP's optimal solutions form a face of the feasible set, usually a single point, but here, at $\vareps^*=0$, the optimal set of \eqref{eq:pmp-Y-slack} is the whole original feasible set. An interior-point solver does not land on an arbitrary member of it. Throughout the run the iterates maximize a barrier term $\mu\sum_j\log\det Y^j$ alongside the objective, which rewards keeping every eigenvalue of every block as large as possible. As $\mu\to0$ the iterates therefore approach an optimal point at which no eigenvalue vanishes unless it vanishes on the whole optimal set, namely a point of maximal rank among all optimal solutions.\footnote{In optimization terms, the limit lies in the \emph{relative interior} of the optimal set, and, when a strictly complementary solution exists (an optimal pair $(X,Y)$ with $X+Y\succ0$, i.e.\ whose ranks add up to the full dimension), it is the \emph{analytic center} of the optimal set: the unique maximizer of $\sum_j\log\det Y^j$ over the optimal face, i.e.\ the optimal point whose eigenvalue product is largest. These are theorems about the central path and idealized path-following methods; that SDPB's infeasible-start predictor--corrector iterates at a finite duality gap share the property is an assumption, which we check empirically by the complementarity cross-check below. For linear programming, convergence of the central path to the analytic center of the optimal face goes back to McLinden \cite{McLinden:1980} and Megiddo \cite{Megiddo:1989}, and G\"uler and Ye proved that the iterates of a broad class of interior-point \emph{algorithms} (not just the idealized path) converge to the relative interior of the optimal face \cite{GulerYe:1993}. For semidefinite programming the central path again converges; its limit is the analytic center of the optimal face whenever a strictly complementary solution exists \cite{GoldfarbScheinberg:1998,LuoSturmZhang:1998}, and without strict complementarity it can differ from the analytic center but remains a maximal-rank point of the relative interior of the optimal set \cite{HalickaDeKlerkRoos:2002}, which is the only property the argument here uses.} Applied to the diagnostic at $\vareps^*=0$, this means that the returned $y^*$ is a maximal-rank feasible functional, as strictly positive as a feasible functional can be. Therefore every zero of $y^*$'s block polynomials is universal: if $M^j(x;y^*)$ has a null direction at some $\bar x$, or vanishes identically across a whole block, then the same is true for every feasible functional. One solve turns ``there exists a degeneracy somewhere'' into an explicit list of forced vanishings.
	\item \emph{The $X$-side output is an exposing certificate of the face.} In bootstrap language it is a positive homogeneous solution of the truncated crossing equations: after extracting its discrete support (Section \ref{sec:certfromycompute}), a nonnegative density on finitely many constraint vectors, with no identity component, whose weighted sum vanishes. The weights of the individual sampled rows need not be positive, since positivity is a property of the combination (Section \ref{sec:exposing}). Its existence is the reason no strictly positive functional can exist, since any candidate would have to be strictly positive on vectors that sum to zero, and its support tells us which vectors combine to zero. Section \ref{sec:exposing} gives the general definition, and Section \ref{sec:certfromycompute} makes the identification precise.
\end{itemize}

\paragraph{(1) Facial reduction within the PMP class.} Standard facial reduction \cite{BorweinWolkowicz:1981} restricts the problem to the face by a Gram-level change of variables, which destroys the polynomial structure and with it SDPB compatibility (Section \ref{sec:intro}). The route that stays inside the PMP class works at the level of the functional components instead, in five steps:
\begin{enumerate}
	\item \emph{Detect.} Run \texttt{ycompute}; $\vareps^*\approx0$ means the feasible set lies on a face.
	\item \emph{Localize.} The $Y$-margin $\vareps^*$ is a single global number, but the face it detects is supported on specific blocks. To locate them, compute for each block $j$ of the converged solve its \emph{block residual}
	\beq
		r_j \;:=\; \frac{\max_{r,s,k}\,\abs{(c-By)_{(j,r,s,k)}}}{\lVert(c_j,B_j)\rVert}\,,
		\label{eq:saturation}
	\eeq
	the largest of block $j$'s constraint residuals $c-By$ (the sampled values of the block's polynomial matrix at the returned functional, i.e.\ how far that block's positivity constraint is from being saturated), measured in units of the block's own data norm. A healthy block has $r_j$ of the order set by the extremal-functional contact structure ($10^{-2}$--$10^{-9}$ in our examples). A block that vanishes identically on the face is pinned far below the duality gap of the solve, across all its sample points ($10^{-65}$ against a gap of $10^{-30}$ in the second diagnostic round of Section \ref{sec:faceworked}), because the interior-point iterates approach the face faster than the gap closes. We make no quantitative use of this pinning scale; only the large separation between the two classes of blocks enters, as a classifier.

	Two limitations should be kept in mind. First, $r_j$ detects blocks that vanish identically; a face that only touches a block at isolated points or along a direction (a rank drop rather than a vanishing block) is found from the certificate rather than from $r_j$ (the first round of Section \ref{sec:faceworked} is of this kind). Second, $r_j$ has an intermediate regime. It is measured against the block's data norm, not against the size of the components that feed the block, so a block fed only by components that are themselves tending to zero shows a residual far above the pinning scale ($10^{-10}$--$10^{-17}$ in Section \ref{sec:faceworked}) although it lies on the face. When the face has \emph{singularity degree} (the number of facial-reduction steps needed to reach the minimal face) larger than one, such second-level contacts are approached only as a fractional power of the duality gap (Hölder exponent $2^{-d_{\mathrm{sing}}}$ at singularity degree $d_{\mathrm{sing}}$ \cite{Sturm:2000}, a square root at degree one and a fourth root at degree two) and are not pinned; they are exposed by the next diagnostic round (Section \ref{sec:faceworked}).
	\item \emph{Identify the forced linear conditions.} On the saturated rows (all rows of the blocks flagged in step 2, or the rows named by the certificate when the face is a rank drop), the statement ``every feasible functional saturates these rows'' is a system of affine equations on $y$, $(By)_p=c_p$ (linear when, as in our examples, $c$ vanishes on them). Compute numerically the rank $q$ of the span of these saturated rows. In practice the relations turn out to be rational (e.g.\ whole components or fixed integer combinations of components forced to vanish), so they can then be recognized exactly. This yields $q$ exact affine conditions $\ell_a(y)=0$.
	\item \emph{Reduce.} Change variables to a basis of the solution space of $\ell_a(y)=0$: the new PMP has $N-q$ components, every block's matrices transform by the same substitution, and blocks whose data now vanishes identically (typically the saturated ones) drop as tautologies. We describe this in more detail in Section \ref{sec:pmpmoves}; it is a PMP-to-PMP operation on the data files, and no conformal blocks are recomputed.
	\item \emph{Audit.} Solve the reduced problem and re-run \texttt{ycompute} on it. Success is unambiguous: the navigator value is unchanged within the duality gap (a numerical check, backed by the one-sided bound of Section \ref{sec:sandwich}, which is a theorem), and the reduced $Y$-margin is now positive, so the degeneracy has been removed. If the margin is still zero, iterate, since faces can be nested (the singularity degree defined above).
\end{enumerate}
The reduction is sound for any choice of imposed conditions (Section \ref{sec:sandwich}). Because it restricts the functional side, $v_{\mathrm{red}}\le\vnav$ holds regardless, and a mistake in step 3 can only make the bound weaker, never invalid.

\paragraph{(2) Reading off the physics (or the bug).} The same data explains the origin of the degeneracy. The saturated blocks of step 2 name sectors and spins; the exposing certificate names the linear combination of their crossing vectors that vanishes. Comparing that combination against the definitions of the vectors in the crossing-equation file then typically reveals the mechanism: components of equations that no sector can populate, sign structures that make two families of constraint vectors linearly dependent, external-block contractions that decouple a row. The procedure is the same solution extraction as in the extremal-functional method \cite{ElShowk:2012hu}, applied to the \texttt{ycompute} solution.

\paragraph{(3) Two-sided bounds without attainment.} Every reduced-feasible functional is an original-feasible one, so the reduced value is a lower bound with no attainment assumption, and the reduced problem has interior once the face is removed (Section \ref{sec:sandwich}); together with $v_{\mathrm{yg}}\ge\vnav$ this is the two-sided bound \eqref{eq:sandwich}.

\subsection{Exposing certificates and their invisibility in the PMP data}
\label{sec:exposing}

Facial reduction is driven by \emph{exposing certificates}: vectors $\hat x\in\bR^P$ with
\beq
	S := \sum_p \hat x_p A_p \;\succeq\; 0,
	\qquad B^\top \hat x = 0,
	\qquad c^\top \hat x = 0,
	\qquad S \neq 0 .
	\label{eq:exposing-def}
\eeq
The consequence is immediate: for every $Y$-feasible $(y,Y)$,
\beq
	\Tr(SY) = \sum_p \hat x_p\,\Tr(A_p Y) = \sum_p \hat x_p\bigl(c_p - (By)_p\bigr) = c^\top\hat x - (B^\top\hat x)^\top y = 0,
	\label{eq:exposing-consequence}
\eeq
and a vanishing trace of a product of two positive-semidefinite matrices forces $SY=0$. The individual weights $\hat x_p$ carry no sign, since positivity is a property of the combination $S$, and the condition that makes the certificate informative is $S\neq0$ rather than merely $\hat x\neq0$. Hence the range of $S$ lies in the null space of every feasible $Y$. The certificate gives the common null directions of \eqref{eq:commonnull} explicitly, and restricting to the orthogonal complement of $\mathrm{range}(S)$ is one facial reduction step.

One might hope to find such certificates by inspecting the PMP data for redundancies, such as identical constraint rows or linear dependences among them. In our experience this search comes up empty even on problems whose $Y$-Slater condition fails badly, and there is a structural reason why it generically does. Linear dependence among constraint rows is a rank statement about the data $(A_p, B)$, namely an $X$-side linear-algebra degeneracy, detectable by Gaussian elimination and absent from generic SDPB data. The exception is a constraint row that vanishes identically, the endpoint degeneracy of Section \ref{sec:fakepoles}, which a static check catches. An exposing certificate is a different object, a positivity-weighted combination of constraints, $S\succeq0$, whose weights $\hat x$ are the solution of an optimization problem and are generically irrational numbers with no closed form. A good analogy is the extremal functional spectrum, which is also not visible in the crossing equations and appears only in the solved dual. The face must therefore be extracted from the solver output rather than read from the PMP input.

\subsection{The certificate from \texttt{ycompute}}
\label{sec:certfromycompute}

The required solver output is one we already have. Consider the $X$-problem of the \texttt{ycompute} \SDP: appending the slack column $w$ of \eqref{eq:weighted-column} (entries $\rho_j(x_k)\delta_{rs}$ times the reduced sample scalings $\tilde s_{j,k}$) and the objective $b'=(0,\dots,0,1)$ while leaving $c$ untouched yields, on the $X$ side,
\beq
	\label{eq:ycompute-X}
	\begin{aligned}
		\text{minimize}\quad & c^\top x \\
		\text{subject to}\quad & \sum_p x_p A_p \succeq 0, \qquad
		B^\top x = 0, \qquad
		w^\top x = 1 .
	\end{aligned}
\eeq
Its optimal value equals $\vareps^*$, and because the $Y$-side of the diagnostic satisfies Slater unconditionally (Section \ref{sec:ycompute}), the $X$-optimum is attained (Section \ref{sec:slaterconds}). At $\vareps^*=0$, attainment means that the optimizer $\hat x$ satisfies $\sum_p\hat x_p A_p\succeq0$, $B^\top\hat x=0$, $c^\top\hat x=\vareps^*=0$, and $w^\top\hat x=\langle S,\widehat Y\rangle=1$ with $\widehat Y\succ0$ as in Section \ref{sec:ycompute} (so $S\neq0$). These are the conditions \eqref{eq:exposing-def}, so the $X$-side solution of \texttt{ycompute} at vanishing $Y$-margin is an exposing certificate for the face.

The certificate has a PMP-level reading. Substituting the rank-one structure \eqref{eq:SDPB-Ap} into \eqref{eq:exposing-consequence}, the statement $\Tr(SY)=0$ for all feasible $Y$ becomes the statement that a fixed positive linear functional annihilates the polynomial matrix $M(x;y)$ for every feasible $y$. Extracting the discrete support of that functional (the finitely many points where its mass sits, found by the same operation as extremal-functional spectrum extraction \cite{ElShowk:2012hu}) resolves it into finitely many \emph{contact pairs} $(\bar x_i, \bar v_i)$ with
\beq
	M^{j_i}(\bar x_i;\, y)\;\bar v_i = 0
	\qquad\text{for every feasible } y,
	\label{eq:contact-pairs}
\eeq
possibly together with conditions at $x=\infty$ (degenerate leading coefficients, Appendix \ref{app:faithful}). These pairs describe the face in PMP language, as points on the half-line and directions in block space where positivity is saturated by force. We call this collection (the pairs $(\bar x_i,\bar v_i)$, together with any conditions at $x=\infty$) the \emph{contact data} of the face.

\paragraph{Exact statement vs.\ numerical output.} An exact optimizer $\hat x$ of \eqref{eq:ycompute-X} is an exact exposing certificate, while the solver returns an approximation to it, positive semidefinite and orthogonal to the feasible set only up to the solver tolerance. The approximation is enough because nothing downstream consumes the certificate itself. The moves of Section \ref{sec:pmpmoves} consume only the contact data, which stabilizes long before the certificate entries converge and can usually be snapped to values that are exact by construction (gap endpoints, unitarity bounds, $x=\infty$, components forced to vanish by symmetry). Even where the certificate itself is hard to obtain exactly, as in the multi-correlator mechanism of Section \ref{sec:howfails}, where its weights are the solution of an optimization problem with no closed form (Section \ref{sec:exposing}), the contact data it points to is exact, namely a list of functional components forced to vanish, handled exactly by (R2). In the endpoint mechanisms of Section \ref{sec:twistgap} the contact sits at a sector endpoint, whose location is exact as well. Exact contact data is therefore the common case; the reduction is then exact, and $v_{\mathrm{red}}=\vnav$ once the audit or an exactly verified certificate confirms the face. The moves are sound for any choice of contact data, and the audit (step 5 of Section \ref{sec:threeuses}, justified in Section \ref{sec:sandwich}) confirms the identification a posteriori; no final claim rests on the numerical certificate.

Facial reduction is considered numerically delicate \cite{DrusvyatskiyWolkowicz:2017}, because the auxiliary problem that finds the certificate must be solved accurately and deciding the rank of a numerically computed certificate is ill-conditioned. Auxiliary problems that are themselves strictly feasible have been constructed for this step \cite{CheungSchurrWolkowicz:2013}; \texttt{ycompute} is such a problem within the PMP class. Its $\vareps$-direction restores its own $Y$-Slater condition by construction, so identifying the face costs one SDPB solve of the same size as the original. Its own $X$-Slater condition is not automatic, since it is the attainment check of Section \ref{sec:attainment}, which can fail (Section \ref{sec:feasxslater}). The rank decision is sidestepped as well, because the contact data can usually be fixed exactly (previous paragraph).

The $Y$-side output of the same run is a maximal-rank point of the face (Section \ref{sec:threeuses}), whose null space gives the forced directions; a single converged diagnostic thus yields both a maximal-rank point of the face and its exposing certificate, and the two can be cross-checked via complementarity.\footnote{If strict complementarity fails, $\mathrm{range}(S)$ can be a proper subspace of the common null space, and one reduction step removes only part of the degeneracy. This is the singularity-degree phenomenon \cite{Sturm:2000} (Section \ref{sec:threeuses}), and it is handled by iterating: reduce, re-run the \texttt{ycompute} on the reduced problem, and repeat until $\vareps^*>0$. Each round removes at least one direction of the Gram cone (a functional component, or a degree deficit repaired by (R$\infty$)), so the iteration terminates; the number of rounds is at least the singularity degree, with equality when every certificate has maximal rank. That a single step need not suffice is shown by the case study of Section \ref{sec:faceworked}, which needed two steps with our certificates, and by its feasibility-mode version, which needed three (Section \ref{sec:feasworked}).}

\subsection{Reduction moves that stay in the PMP class}
\label{sec:pmpmoves}

We choose the contact data as exact rational or algebraic numbers, informed by (but not copied blindly from) the numerics. Given such data, the following moves perform the reduction without leaving the PMP class; (R2) is a special case of (R1).

\paragraph{(R1) Affine restriction of the functional.} Imposing linear conditions $\ell_a(y)=0$ means reparameterizing $y = y_0 + Fu$ with the columns of $F$ a basis of the solution space: the new PMP has matrices $\widetilde M_0 = M_0 + \sum_n (y_0)_n M_n$ and $\widetilde M_m = \sum_n F_{nm} M_n$, and objective $\widetilde b = F^\top b$ (plus the constant $b^\top y_0$). A PMP maps to a PMP; the move is sound for any choice of conditions, because it restricts the feasible set.

\paragraph{(R2) A global null direction.} If the contact direction $\bar v$ is $x$-independent (the multi-correlator mechanism of Section \ref{sec:howfails}, where whole components of the functional are forced to vanish identically), the condition $M(x;y)\bar v\equiv0$ is a finite list of linear equations on $y$ (one per polynomial coefficient per row), computable exactly from the PMP data, and therefore an instance of (R1). On the restricted subspace the $\bar v$-row and column of the block vanish identically, so with $V$ a constant basis of the complement,
\begin{equation*}
	M(x;y)\succeq0 \ \ \forall x
	\quad\Longleftrightarrow\quad
	V^\top M(x;y)\, V \succeq 0 \ \ \forall x ,
\end{equation*}
and the block dimension drops by one. The congruence step is an equivalence rather than a further restriction, since conjugation by a constant matrix preserves polynomiality and no feasible point is lost.

\paragraph{(R3) A contact at finite $\bar x$.} For an isolated contact at $\bar x>0$ with direction $\bar v$, first impose (R1) the linear conditions $M(\bar x;y)\bar v=0$ and the tangency $\partial_x\bigl(\bar v^\top M \bar v\bigr)(\bar x)=0$.\footnote{For feasible $y$ these hold automatically, because a nonnegative scalar polynomial with an interior zero vanishes to even order there, and positive semidefiniteness propagates the zero to the whole column via $\abs{M_{1j}}^2\le M_{11}M_{jj}$. Imposing them as constraints selects the sub-family whose contact sits at the \emph{chosen} $\bar x$.} After a constant congruence sending $\bar v\to e_1$, the first column of $M$ vanishes at $\bar x$ and the $(1,1)$ entry doubly so. Now conjugate by $U(x)=\mathrm{diag}\bigl((x-\bar x)/(1+x),1,\dots,1\bigr)$:
\beq
	\widehat M(x;y) := U(x)^{-1}\, M(x;y)\, U(x)^{-1},
	\label{eq:quotient-congruence}
\eeq
whose entries are $(1+x)^2M_{11}/(x-\bar x)^2$, $(1+x)M_{1j}/(x-\bar x)$, and $M_{jk}$, all polynomial by the imposed vanishings, of the same degrees as before, and all still linear in $y$. The factors $1+x$ are needed because the bare congruence by $\mathrm{diag}(x-\bar x,1,\dots,1)$ lowers the degree of the first row and column, and by Consequence (b) of Appendix \ref{app:faithful} a block whose entries cannot reach its declared degree is Gram-degenerate at $x=\infty$, so it would trade the finite contact for one at infinity. For $x\ge0$, $x\neq\bar x$ the congruence is invertible, so positivity is equivalent; continuity extends the equivalence to $\bar x$. The result is again a PMP of the same degree, in which the forced zero at $\bar x$ has been removed.\footnote{At the endpoint $\bar x=0$ contacts of odd order are possible. The scalar case is easy: $p(x)=x\,q(x)$ with $q$ polynomial and linear in $y$, and $p\succeq0$ on $[0,\infty)$ iff $q\succeq0$ on $[0,\infty)$ ($q$ declared at degree $d-1$, or multiplied by $1+x$; see the degree bookkeeping below). Odd-order \emph{matrix} endpoint contacts are different, because the symmetric congruence \eqref{eq:quotient-congruence} divides the corner by $x^2$, and when only some rows of a block vanish to first order no rational congruence removes the contact; the balanced form needs $\sqrt x$. Section \ref{sec:mixedentry} gives the physically relevant instance, where no exact reduction within the PMP class exists (only a restriction beyond the face, which gives a valid but possibly weaker bound, Section \ref{sec:sandwich}) and a twist gap is the practical cure. The same holds for odd-order matrix contacts at infinity in (R$\infty$) below.}

\paragraph{(R$\infty$) A contact at infinity.} Write $M(x;y)=\sum_{i=0}^{d}M_{(i)}(y)\,x^{i}$. A direction $\bar v$ in which the leading coefficient matrix is forced to degenerate is the mirror of (R3) under $x\to1/x$: the substitution $u=1/x$, $\widetilde M(u):=u^{d}M(1/u;y)$, turns the block into a polynomial of the same degree with reversed coefficients and the contact at $x=\infty$ into an endpoint contact at $\bar u=0$, with $\widetilde M(0)=M_{(d)}$. The forced conditions are again linear in $y$ and computable exactly from the PMP coefficients ($M_{(d)}(y)\,\bar v=0$, an instance of (R1)). Since this is an endpoint, the contact may be of odd order, so the subleading condition $\bar v^\top M_{(d-1)}(y)\,\bar v=0$ is not automatic (a nonnegative polynomial on the half-line may vanish to first order at the endpoint) and must be read off the certificate rather than assumed. The removal move is then the endpoint move of (R3) applied at $\bar u=0$, which in the original variable is again a degree-preserving congruence. After a constant congruence sending $\bar v\to e_1$, an even-order contact at infinity means $\deg M_{11}\le d-2$ and $\deg M_{1j}\le d-1$ on the restricted family; conjugating by $U(x)=\mathrm{diag}\bigl(1/(1+x),1,\dots,1\bigr)$ gives the entries $(1+x)^2M_{11}$, $(1+x)M_{1j}$, $M_{jk}$. These are polynomial, of degree at most $d$, linear in $y$, and positive semidefinite on the half-line exactly when $M$ is, since $1+x>0$ there. The first row and column now reach the declared degree, and the contact at infinity is gone. (In the compactified variable $\zeta=x/(1+x)$ of Appendix \ref{app:faithful} this is the (R3) congruence at $\bar\zeta=1$.) In the scalar case the move is multiplication of the constraint by $(1+x)^{k}$, with $k$ the degree deficit, and this covers odd orders as well. A planted instance, with its certificate and its repair, is in Section \ref{sec:exp-twist}.

The degree bookkeeping is essential because, by the criterion of Appendix \ref{app:faithful}, a block kept at declared degree $d$ whose entries can no longer reach degree $d$ is irreparably Gram-degenerate (no reweighting cures it), so the reduction must restore the full degree of every entry (or, where all entries fall short together, lower the declared degree of the block); imposing the vanishing conditions alone is not enough.

\paragraph{Locating the contact from the diagnostic.} The full-degree weight \eqref{eq:slack-weight} gives the slack term leading coefficient $-\vareps I$, so the diagnostic treats the point at infinity on the same footing as the finite sample points, and its converged output localizes the contact. A comparison with the constant-weight $Y$-margin $\vareps^*_1$ can separate the two cases in one direction (Appendix \ref{app:faithful}, Consequence (c)), but the direct readout is the solve itself. Finite-$x$ contacts appear as zeros of $M^j(x;y^*)$ at finite $\bar x$, visible in the sampled residuals of the saturated blocks and extracted as in \eqref{eq:contact-pairs}. A contact at infinity appears as a singular leading coefficient matrix $M_{(d)}(y^*)$; since the returned functional has maximal rank, the null space of $M_{(d)}(y^*)$ consists, up to the duality gap of the solve, of the forced directions $\bar v$. A block whose sampled residuals are healthy (not pinned) but whose leading coefficient degenerates is an infinity-only failure; the converse pattern is a finite-$x$ face.

\subsection{Soundness and the two-sided bound}
\label{sec:sandwich}

All moves act on the functional side of the problem: (R1) and (R2) restrict $y$ to a subspace, (R3) and (R$\infty$) are equivalences on the restricted family, and blocks are deleted only when their data vanishes identically on it, an exactly checkable condition. Consequently every reduced-feasible functional is an original-feasible functional with the same objective, whatever contact data we chose, and since the navigator is the optimum over feasible functionals,
\beq
	v_{\mathrm{red}} \;\le\; \vnav \qquad \text{for every choice of imposed conditions.}
	\label{eq:red-sound}
\eeq
This one-sided soundness is a theorem about the moves rather than about the quality of the identification, so a mistake can never produce an invalid lower bound.\footnote{On the $X$-side representation of the same problem the imposed $y$-conditions appear as a projection of the equality constraints (a relaxation of the minimization), so both sides of the duality move down together, consistently with \eqref{eq:red-sound}.}

If the imposed relations are not actually forced by feasibility, i.e.\ if we misinterpreted the data and chose the wrong face, then feasible functionals have been cut away, and we can have $v_{\mathrm{red}}<\vnav$.

If the identification was correct, the imposed relations are forced by original feasibility, no feasible point is lost, every move is an equivalence, and $v_{\mathrm{red}}=\vnav$ exactly.

Combined with the $Y$-guarantee this gives the two-sided bound
\beq
	\boxed{\;v_{\mathrm{red}} \;\le\; \vnav \;\le\; v_{\mathrm{yg}}.\;}
	\label{eq:sandwich}
\eeq
Once the reduction is complete, the reduced problem has its interior back, so SDPB computes $v_{\mathrm{red}}$ reliably; its own \texttt{ycompute} confirms this.

\paragraph{The mirror: reduction of the $X$-side.} An $X$-degeneracy, in which every feasible density lies on a proper face of its cone, is reduced in the same way from the other side. Its exposing certificate is a nonzero functional direction $H(x)=\sum_n(\Delta y)_nM_n(x)\succeq0$ on $x\ge0$ with no objective and no normalization component, and every feasible density is orthogonal to it. The patterns mirror (R1)--(R$\infty$). Where $H$ is strictly positive on a whole block (including its leading coefficient), that block carries no density and is deleted. A constant null space $W$ of $H$ restricts the block's densities to $W$, the constant congruence $M\mapsto W^{\mathsf T}MW$. Isolated contacts of $H$ at points $\bar x_i$, including $x=\infty$, restrict the continuum to isolated operators at $\bar x_i$ with matrices along the null vectors, i.e.\ to blocks of degree zero (at infinity, the leading coefficient). Each move deletes densities, and on the functional side removes or weakens constraints, so every reduced-feasible density is an original-feasible density with the same objective, and
\beq
	v^{X}_{\mathrm{red}} \;\ge\; \vnav \qquad \text{for every choice of imposed conditions,}
	\label{eq:red-sound-X}
\eeq
the mirror of \eqref{eq:red-sound}. A face that combines these patterns within one block, with one row supported only at a point while another keeps its continuum, has cross terms that the product structure of SDPB's blocks cannot represent (the same obstruction as in Section \ref{sec:mixedentry}); dropping them restricts the densities further, which keeps \eqref{eq:red-sound-X} valid and costs only tightness. The same containment principle holds on the functional side, where a reduced problem only needs to lie inside the face rather than equal it for its bound to be valid, and equality matters only for tightness. With the $X$-guarantee the mirror gives $v_{\mathrm{xg}}\le\vnav\le v^{X}_{\mathrm{red}}$. We did not meet a natural $X$-degeneracy in an optimization problem (Section \ref{sec:feasxslater}; the planted one of Section \ref{sec:xdiag} is the block-deletion pattern above), so we give no worked example.

\section{Case Study I: curing a Skydive centering stall}
\label{sec:casestudy}

The constructions above would be of limited interest if Slater failures were exotic; they are not, although whether they occur can depend on the truncation. Every failure mode discussed in this paper except one can be reproduced in the $\s$--$\e$ system below, and we draw our experiments from it wherever possible. (The exception, the mixed-channel endpoint degeneracy of Section \ref{sec:mixedentry}, needs two scalars in the same representation and is tested in an $\e$--$\e'$ system.)

The same modes, and the stalls they cause, are on record elsewhere. The multi-correlator degeneracy of Section \ref{sec:howfails} is the known unreliability of feasibility-mode runs at points disallowed by a subsystem of the correlators; \cite{Reehorst:2021ykw} observes that such runs ``may not converge due to precision issues'' and offers the navigator as the more robust alternative. An endpoint degeneracy of Section \ref{sec:twistgap} (the shared-prefactor one or, in multi-operator systems, the genuine mixed-channel one) is the likely reason a small twist gap is imposed in \cite{Chester:2019ifh,Erramilli:2022kgp,Liu:2023elz,Kousvos:2024dlz,DiPietro:2025ozw,Chang:2024whx,Chester:2020iyt}; where a motive is stated \cite{Chester:2019ifh,Erramilli:2022kgp,DiPietro:2025ozw} it is numerical stability, \cite{Chester:2019ifh} reports that the gap is what restored the dual feasible jumps, and \cite{Liu:2020tpf} imposes a gap of $10^{-4}$ above the unitarity bound because without it the solver showed ``a steady decay of \texttt{dualError} without vanishing.'' The diagnostics and cures presented here apply to those setups unchanged; we have encountered and resolved the same failures in our own larger production setups.

For our case study we consider a standard $\s$--$\e$ mixed-correlator setup (the \emph{$\s$--$\e$ system} below) used to study the 3d Ising model. At $\Lambda=5$ classical SDPB optimization of the navigator converges at every point we tested (the navigator solves of this section run to duality gap $10^{-20}$, the diagnostic solves to ${\sim}10^{-30}$, and Skydive's inner solver to its $10^{-3}$ target). Skydive, on the other hand, fails in the centering stage at every point we tested. A failure of the Slater conditions removes the convergence guarantee without forcing classical optimization to fail, and here it does not; but it prevents Skydive from taking any step. We use $\Lambda=5$ because it is the cheapest setting in which the failure occurs and can be taken apart completely.

All numbers in this section come from PolyBoot runs; the reproduction directory of \cite{PolyBoot} lists their launchers under the run families F01--F04.\footnote{$\Lambda$ is the standard derivative-order truncation of the functional ansatz: the functionals are spanned by the derivatives $\partial_z^m\partial_{\bar z}^n$ at the crossing-symmetric point with $m\ge n$ and $m+n\le\Lambda$ (appendix A of \cite{Simmons-Duffin:2015qma}), so the number of functional components grows quadratically with $\Lambda$.}

Let us spell out the two bootstrap problems that will be compared.  We use the standard mixed-correlator 3d Ising system built from
\[
	\langle\s\s\s\s\rangle,\qquad
	\langle\s\s\e\e\rangle,\qquad
	\langle\e\e\e\e\rangle,
\]
with $\s$ odd and $\e$ even under $\bZ_2$.  The crossing equation has a $\bZ_2$-even matrix sector and a $\bZ_2$-odd scalar sector,
\beq
	\vec V_{0,0}
	+\sum_{\mathcal{O}^+}\Tr\!\left[P_{\mathcal{O}}\vec V_{+,\Delta,\ell}\right]
	+\sum_{\mathcal{O}^-}p_{\mathcal{O}}\vec V_{-,\Delta,\ell}=0,\qquad
	P_{\mathcal{O}}=
	\begin{pmatrix}
	\lambda_{\s\s\mathcal{O}}\\
	\lambda_{\e\e\mathcal{O}}
	\end{pmatrix}
	\begin{pmatrix}
	\lambda_{\s\s\mathcal{O}} & \lambda_{\e\e\mathcal{O}}
	\end{pmatrix}.
	\label{eq:ising-case-crossing}
\eeq
Apart from the external scalars $\s$ and $\e$, we impose the simple gap assumptions used in the navigator tests: all other even and odd scalars have $\Delta\ge3$, while spinning operators obey the 3d unitarity bound.  The navigator is the $\Sigma$-navigator of \cite{Reehorst:2021ykw} (Section 2.2 there), in which one adds the deformation $\lambda_{\mathrm{nav}}\vec M_\Sigma$ to the crossing equation, with $\vec M_\Sigma$ a fixed negative combination of constraint vectors above the gaps (eq.~(2.25) there), and minimizes $\lambda_{\mathrm{nav}}$ (eqs.~(2.21)--(2.22)). Equivalently, on the functional ($Y$) side, one maximizes $\vec\alpha\cdot\vec V_{0,0}$ over functionals normalized by $\vec\alpha\cdot\vec M_\Sigma=-1$ and positive on all continuum sectors and on the external-exchange block of Table \ref{tab:case-study-setups} (the dual form, eq.~(2.11) there). See eqs.~(2.14)--(2.20) of \cite{Reehorst:2021ykw} for the crossing equations.

\begin{table}[htpb]
\centering
\small
\renewcommand{\arraystretch}{1.25}
\begin{tabularx}{\textwidth}{@{}>{\raggedright\arraybackslash}p{4.3cm}X@{}}
	\toprule
	setup and scanned parameters & distinguished external-exchange block \\
	\midrule
	two-parameter truncation; $(\De_\s,\De_\e)$
	& $\vec\alpha\!\cdot\!\left(\vec V_{+,\De_\e,0}
	   + \begin{psmallmatrix}1&0\\0&0\end{psmallmatrix}
	   \vec V_{-,\De_\s,0}\right)\succeq0$ \\[4pt]
	three-parameter OPE-ratio setup; $(\De_\s,\De_\e,x_{\mathrm{ope}})$, $x_{\mathrm{ope}}=\lambda_{\e\e\e}/\lambda_{\s\s\e}$
	& $(1,x_{\mathrm{ope}})\,
	   \vec\alpha\!\cdot\!\left(\vec V_{+,\De_\e,0}
	   + \begin{psmallmatrix}1&0\\0&0\end{psmallmatrix}
	   \vec V_{-,\De_\s,0}\right)\,
	   \binom{1}{x_{\mathrm{ope}}}\ge0$ \\
	\bottomrule
\end{tabularx}
\caption{The two \SDPs\ compared in the case study.  The two-parameter version keeps only the matrix positivity condition after grouping the leading $\e$ and $\s$ exchanges.  The three-parameter version fixes the same leading exchange to a rank-one OPE vector and treats its ratio $x_{\mathrm{ope}}$ as an external coordinate, $x_{\mathrm{ope}}=\tan\theta$ in terms of the angle parametrization used in the navigator and skydiving papers.}
\label{tab:case-study-setups}
\end{table}

We consider both the two-parameter problem over the $(\De_\s,\De_\e)$ plane and the three-parameter problem where the external operator OPEs are scanned over in terms of the ratio $x_{\mathrm{ope}}$, i.e.\ $(\De_\s,\De_\e,x_{\mathrm{ope}})$.  We will see below that at $\Lambda=5$ scanning over $x_{\mathrm{ope}}$ restores a positive $Y$-margin.

\subsection{Symptom}
\label{sec:symptom}

We start with the allowed point $(\De_\s,\De_\e)=(0.5181,\,1.412625)$, inside the Ising island. In the two-parameter navigator, the Skydive inner solver \texttt{dynamical\_sdp} runs for its full budget of 500 iterations and exits with \texttt{maxIterations exceeded}. The duality gap converges, reaching the $10^{-3}$ target by iteration 240 and staying there. What never converges is the centering residual $\|XY-\mu I\|$, which Skydive reports as $R$-err and requires below $10^{-10}$ before it will take a step in the external parameters. Table \ref{tab:control-trajectory} in Appendix \ref{app:skydive} shows the trajectory: $R$-err descends with everything else until $\mu$ stops decreasing at ${\sim}2\times10^{-6}$, then reverses and diverges, climbing many orders of magnitude while the duality gap stays on target.

The same qualitative failure is not isolated to this point. Repeating the raw two-parameter Skydive at nearby allowed or disallowed points always gave the same signature in our reproduction set: the duality gap reached the requested scale, the centering residual stayed enormous, and no external step was returned. Thus at $\Lambda=5$ the pathology is generic for this two-parameter formulation.

\subsection{Diagnosis}
\label{sec:diagnosis}
\noindent The pattern of Section \ref{sec:centralpath} (healthy gap, divergent centering) points at a Slater failure, and the diagnostics of Section \ref{sec:diagnostics} settle the question. Table \ref{tab:diagnosis-case-study} summarizes the runs that matter for the case study.
\begin{table}[htpb]
\centering
\small
\renewcommand{\arraystretch}{1.25}
\setlength{\tabcolsep}{4pt}
\begin{tabularx}{\textwidth}{@{}>{\raggedright\arraybackslash}Xcccc@{}}
	\toprule
	setup and point & $Y$-margin $\vareps^*$ & duality gap & $X$-margin $t^*$ & verdict \\
	\midrule
	two-parameter, allowed $(0.5181,1.412625)$ & $3.1\times10^{-31}$ & $6\times10^{-31}$ & $3.4\times10^{-14}$ & $Y$-Slater fails \\
	two-parameter, disallowed $(0.51,2.5)$ & $3.8\times10^{-31}$ & ${\sim}10^{-30}$ & --- & $Y$-Slater fails \\
	three-parameter, $(0.5181,1.412625,0.5)$ & $2.4\times10^{-8}$ & ${\sim}10^{-30}$ & $3.4\times10^{-14}$ & $Y$-Slater holds \\
	\bottomrule
\end{tabularx}
\caption{Slater diagnostics for the case study. The two-parameter navigator has a $Y$-margin that is zero at the duality gap at both points, while its $X$-margin at the allowed point is positive. Restoring the OPE-ratio coordinate in the external block gives a positive margin in the corresponding three-parameter setup. $t^*$ is capped by the $\vareps$-layer of \texttt{xcompute} ($\lambda=10^3$); the value $3.4\times10^{-14}$ is this cap for both formulations (cf.\ Section \ref{sec:xdiag}), so only its sign is a verdict. Dashes: not computed.}
\label{tab:diagnosis-case-study}
\end{table}

Every $Y$-feasible point of this two-parameter \SDP\ is, at the duality gap of the solve, on the boundary of the PSD cone. The complementarity product $XY$ then cannot be driven to $\mu I$ for any positive $\mu$, and the centering residual has no reason to converge; empirically it does not. The point is allowed ($\vnav<0$), so although the degeneracy is of the multi-correlator type of Section \ref{sec:howfails}, it is not tied to exclusion by a subsystem of the correlators; at $\Lambda=5$ the two-parameter family itself has a collapsed $Y$-margin. Since the verdict gives no hint of the degeneracy, the diagnostics should be run routinely.

\subsection{The cure}
\label{sec:cure-isolated}
Since only the $Y$-side fails in the two-parameter example (Table \ref{tab:diagnosis-case-study}), the $Y$-guarantee (one extra functional component and one extra $1\times1$ block) is the minimal fix. Wiring the guarantee into the full Skydive loop requires transforming not just the base \SDP\ but the whole family of parameter-derivative \SDPs\ that \texttt{dynamical\_\allowbreak sdp} consumes. Fortunately, the augmentation in \texttt{yguarantee} (the slack column, built from the sample points and reduced sample scalings, and the cutoff block) is independent of the external parameters, so its parameter derivative vanishes identically, and
\beq
	\frac{d}{dp}\bigl(\text{guaranteed \SDP}\bigr)
	\;=\;
	\Bigl[\ \frac{d}{dp}\bigl(\text{original \SDP}\bigr)\ \Big|\ 0\ \Bigr] :
	\label{eq:constellation}
\eeq
the guaranteed derivative is the original derivative, zero-padded to the augmented shape. No new derivative computations are needed, the transform adds no derivative error, and its wall-time cost (a few one-second file transformations per outer iteration) is negligible against the solves.

Table \ref{tab:endtoend} (Appendix \ref{app:skydive}) compares the first Skydive iteration (centering plus one external step, i.e.\ the stage at which the raw run fails) on the original \SDP\ and on the guaranteed one. At the allowed point the raw run diverges as before, while the guaranteed run turns the divergence into a monotone descent, reaching $R$-err $1.4\times10^{-7}$ at iteration $138$ and $8.7\times10^{-13}$ after the final centering step, $139$ iterations in all, and then takes an external step. From the disallowed starting point $(0.51,\,2.5)$ the outcome is the same at $\lambda=10^3$ and at $\lambda=10^{10}$: the raw run exhausts its $500$ iterations without a step, and both guaranteed runs center to threshold and step. The dependence of the guaranteed value on $\lambda$ is studied in Section \ref{sec:exp-survey}.

\subsection{The face, extracted and removed}
\label{sec:faceworked}

We now carry out the program of Section \ref{sec:pmpfr} on the \SDP\ of the case study, end to end: read the face off a single converged diagnostic solve, perform the reduction exactly, iterate once (two reduction steps are needed), and verify that the navigator value is preserved while the $Y$-margin reopens from zero at the duality gap to $6.1\times10^{-8}$. The reduction itself uses only solver output and exact rational surgery on the \SDP\ data files; no conformal blocks are regenerated.

\subsubsection{The \texorpdfstring{\SDP}{SDP} under study}

The \SDP\ is the two-parameter navigator of Table \ref{tab:case-study-setups} at the allowed point $(\De_\s,\De_\e)=(0.5181,\,1.412625)$, at $\Lambda=5$, with the external contribution entering as an isolated $2\times2$ block in the basis $(\lambda_{\s\s\e},\lambda_{\e\e\e})$. It has $N=29$ functional components (six derivative components for each of the five crossing equations, labeled by their external operators $\s\s\s\s$, $\e\e\e\e$, $\s\e\s\e$, $\e\s\s\e$ and $\s\s\e\e$, minus one component eliminated by the normalization) and $24$ blocks: the external $2\times2$ at a single sample point, eight $2\times2$ even-channel spin blocks, and fifteen $1\times1$ blocks (the $\bZ_2$-odd channel, spins $0$ to $14$). The diagnostic verdict of Section \ref{sec:diagnosis} stands, with $\vareps^*=3.1\times10^{-31}$, and the constant-weight $Y$-margin vanishes as well ($\vareps^*_1=2.2\times10^{-31}$), so the degeneracy is pointwise (Appendix \ref{app:faithful}, Consequence (c)) rather than a translation artifact at $x=\infty$; the readout below shows that it is a contact along the whole half-line.

\subsubsection{Reading the converged diagnostic}

\paragraph{$Y$-side.} Reconstructing the block polynomial matrices $M(x;y^*)$ at the converged \texttt{ycompute} functional (by the maximal-rank property of Section \ref{sec:threeuses}, its zeros are the forced zeros) shows the following pattern. The external block is singular, with eigenvalues $(2.5\times10^{-23},\,0.033)$ and null direction $(1.4\times10^{-11},\,1)$, the pure $\lambda_{\e\e\e}$ direction. Every even-channel $2\times2$ block is rank one along the whole half-line: $\det M(x;y^*)/\lVert M\rVert^2\lesssim6\times10^{-17}$ at every $x$ scanned from $0$ to $35$, with $\abs{M_{\e\e}/M_{\s\s}}\lesssim4\times10^{-16}$ and $\abs{M_{\s\e}/M_{\s\s}}\lesssim10^{-8}$, where the rows of each $2\times2$ block are labeled $\s$ and $\e$ after its OPE basis $(\lambda_{\s\s\mathcal O},\lambda_{\e\e\mathcal O})$, so that $M_{\s\e}$ is the off-diagonal entry. The $\e$-row of the even sector is thus forced to vanish, with the off-diagonal approaching zero along the parabola $\abs{M_{\s\e}}\simeq\sqrt{M_{\s\s}M_{\e\e}}$ characteristic of an interior-point approach to a face. As a control, at random $y$ the same determinants are $O(1)$, so the rank-one structure is forced by feasibility rather than built into the data. The fifteen $1\times1$ blocks are not pinned at the duality gap, but they are not healthy either, since they are fed only by the $\s\e\s\e$, $\e\s\s\e$ and $\s\s\e\e$ components of $y^*$, which sit at $10^{-10}$--$10^{-17}$ against $5\times10^{-2}$--$8\times10^{-6}$ for the $\s\s\s\s$ components. As the second reduction step below shows, these blocks lie on a second-level face, which interior-point iterates approach only like a fractional power of the duality gap \cite{Sturm:2000}.

\paragraph{$X$-side.} The exposing certificate of Section \ref{sec:certfromycompute} confirms this and upgrades it from ``property of one feasible point'' to ``property of the feasible set''. The $X$-side solution at the external block, in the row order $(\s\s,\s\e,\e\e)$, is
\begin{equation*}
	\hat x_{\mathrm{ext}} = \bigl(9.5\times10^{-22},\ 2.8\times10^{-11},\ 1\bigr)
	\quad\Longrightarrow\quad
	S_{\mathrm{ext}} \simeq e_\e e_\e^\top ,
\end{equation*}
the exposing certificate of the $\lambda_{\e\e\e}$ contact; here $e_\e$ is the unit vector along the $\e$-row (the $\lambda_{\e\e\e}$ direction). On the even-channel blocks the certificate matrices $S_j$ are positive semidefinite of rank equal to the full bilinear-basis dimension, with every range eigenvector lying in the $\e$-row to one part in $10^{9}$: $S_j\simeq(\text{positive definite})\otimes e_\e e_\e^\top$. On the even blocks the ranks of $S_j$ and of the $Y$-solution add up to the full dimension, and the exposed subspace is the constant direction $e_\e$ in every even block, which is the clean case (R2), with no $x$-dependent congruence needed. Strict complementarity does not hold globally, however, because on the fifteen $1\times1$ blocks both the certificate and the $Y$-solution are small ($Y$ at $10^{-10}$--$10^{-17}$) and neither is bounded away from zero. This failure of strict complementarity is the signature of a face of singularity degree larger than one, and it is why a second step will be needed. The present certificate is supported entirely in the $\e$-row, consisting of an OPE-squared weight $\lambda_{\e\e\e}^2$ at the external $\De_\e$ plus positive even-channel densities, pairing to zero with every feasible functional. Its weight grows by roughly an order of magnitude from one even-spin block to the next and is dominated by the highest spin kept, which suggests that the degeneracy is a low-$\Lambda$ phenomenon, as the higher-$\Lambda$ runs at the end of this section confirm.\footnote{The mechanism is Section \ref{sec:howfails} in its sharpest form, with one refinement, namely that the certificate has no identity component and its existence is equivalent to the statement that no functional can be strictly positive on the $\e$-sector. An exposing certificate with the same $\e$-row structure appears at the disallowed point $(0.51,\,2.5)$ in the feasibility formulation (Section \ref{sec:exp-twist}), consistent with the observation of Section \ref{sec:diagnosis} that the navigator margin vanishes at allowed and disallowed points alike.}

\subsubsection{The face, and its removal}

Tracing the constraint rows back to functional components makes the face completely explicit. The $M_{\e\e}$ rows of the even blocks are fed only by the six $\e\e\e\e$-equation components, and setting them to zero is six independent conditions (rank $6$ of $6$); the $M_{\s\e}$ rows are fed only by the twelve $\e\s\s\e$- and $\s\s\e\e$-equation components, and $M_{\s\e}\equiv0$ across the even blocks has full rank $12$. The certificate forces $M(x;y)\,e_\e\equiv0$ for every feasible $y$, hence all eighteen conditions:
\beq
	\alpha_{\e\e\e\e} = 0 \ \ (6), \qquad \alpha_{\e\s\s\e} = \alpha_{\s\s\e\e} = 0 \ \ (12).
	\label{eq:face1}
\eeq
Each of them sets a functional component to zero by construction, so no numerically determined quantity has to be promoted to an imposed condition. The reduction is then the moves (R1)+(R2): delete the eighteen $B$-columns, and conjugate each $2\times2$ block to its $\s\s$-corner, keeping only the $\s\s$ constraint rows. Two exactness checks: the $c$-entries on every dropped row vanish identically in the \SDP\ data, so the saturation conditions are homogeneous and the restriction is linear rather than affine; and the congruence is constant, so the bilinear bases are untouched. The result is an \SDP\ with $N=11$ and the same $24$ blocks, now all $1\times1$.

\subsubsection{Iterating: a second reduction step}

Solving the reduced problem confirms that the value is preserved (Table \ref{tab:facepipeline}). Re-running \texttt{ycompute} on it, however, again returns a margin that is zero at the duality gap ($\vareps^*=-2.1\times10^{-31}$ at duality gap ${\sim}10^{-30}$), so one reduction step removed only part of the degeneracy, the situation anticipated in the footnote of Section \ref{sec:certfromycompute}. The second diagnostic solve exposes the residual face even more clearly. The six $\s\e\s\e$-equation components of its optimal functional sit at $10^{-65}$, and with them the block residuals \eqref{eq:saturation} of the fifteen $1\times1$ blocks they feed, against a duality gap of ${\sim}10^{-30}$, which is the pinning of Section \ref{sec:threeuses}; the second exposing certificate is supported entirely on the fifteen $1\times1$ blocks, the $\bZ_2$-odd channel. The second reduction imposes
\beq
	\alpha_{\s\e\s\e}=0 \ \ (6),
	\label{eq:face2}
\eeq
after which the fifteen odd-channel blocks have identically zero data on the remaining five components ($B$-columns and $c$-entries all vanish), i.e.\ their constraints have become the tautology $0\succeq0$. Dropping them leaves an \SDP\ with $N=5$ and nine blocks: the $\s\s$-corner of the external block plus the eight even-channel $\s\s$-corners. The two-parameter mixed-correlator navigator is therefore, on its feasible set, a single-correlator $\langle\s\s\s\s\rangle$ problem presented in mixed-correlator form, with even scalar gap $3$, $\e$ exchanged at $\De_\e$, and the objective $\vec\alpha\cdot\vec V_{0,0}$ and the normalization $\vec\alpha\cdot\vec M_\Sigma=-1$ both restricted to the $\s\s\s\s$ components. In total the face is cut out by setting $24$ functional components to zero; the four crossing equations beyond $\langle\s\s\s\s\rangle$ add no constraints, and their only effect is to place the whole feasible set on the boundary of the cone.

\subsubsection{Preservation of the value and reopening of the margin}

Table \ref{tab:facepipeline} collects the numbers. All navigator and diagnostic solves terminate with \texttt{PrimalDualOptimal}. The navigator value is preserved along the chain to $2\times10^{-23}$ and $6\times10^{-23}$, well inside each solve's own duality gap, so $v_{\mathrm{red}}=\vnav$ at the duality gap of the solves; by Section \ref{sec:sandwich} this is a posteriori evidence that the face was identified correctly and nothing was lost. The final problem satisfies $Y$-Slater, with margins $\vareps^*=6.1\times10^{-8}$ and $\vareps^*_1=5.50\times10^{-2}$, far above the duality gap. The reduction also pays off computationally, as the reduced problem solves $6.6\times$ faster; degeneracy diagnosis thus doubles as formulation debugging.

\begin{table}[htpb]
\centering
\small
\setlength{\tabcolsep}{4pt}
\begin{tabular}{lcccccc}
	\toprule
	system & $N$ & blocks & $\vareps^*_1$ & $\vareps^*$ & $\abs{v-\vnav}$ & iter.\ / time \\
	\midrule
	original            & 29 & 24 & $2.2\times10^{-31}$ & $3.1\times10^{-31}$ & --- & 238 / 179\,s \\
	reduction 1 \eqref{eq:face1} & 11 & 24 & $5.2\times10^{-31}$ & $-2.1\times10^{-31}$ & $1.8\times10^{-23}$ & 172 / 52\,s \\
	reduction 2 \eqref{eq:face2} & 5 & 9 & $5.50\times10^{-2}$ & $6.1\times10^{-8}$ & $5.7\times10^{-23}$ & 152 / 27\,s \\
	\bottomrule
\end{tabular}
\caption{The facial-reduction pipeline on the case-study \SDP\ at the allowed point. $\vareps^*_1$ and $\vareps^*$ are the constant-weight and weighted $Y$-margins (column \eqref{eq:weighted-column} with $\rho=1$ and $\rho=(1+x)^{d_j}$); $\abs{v-\vnav}$ compares the $Y$-objective of each solve to the original's ($\vnav=-7.372480698624455\times10^{-6}$, duality gaps $(3$--$7)\times10^{-21}$); iterations and runtime refer to the navigator solves at duality-gap threshold $10^{-20}$. The margins are from solves to duality gap ${\sim}10^{-30}$.}
\label{tab:facepipeline}
\end{table}

\subsubsection{Lessons}

At $\Lambda=5$ the navigator of this two-parameter formulation is a single-correlator $\langle\s\s\s\s\rangle$ quantity. The three-parameter formulation of Table \ref{tab:diagnosis-case-study} keeps its margin because restoring the OPE-ratio coordinate no longer forces the $\e$-row components to vanish.

Both exposing certificates are confined to a subset of the crossing equations and dominated by the highest kept spin, which suggests a dependence on the truncation order. Appendix \ref{app:skydive-lambda} confirms this: regenerated at $\Lambda=9$ to $19$, the same formulation has a strictly positive $Y$-margin at the same point and Skydive centers natively, so the face is a $\Lambda=5$ truncation artifact. Such degeneracies can appear and disappear as truncation parameters vary.

The twist-gap experiment of Section \ref{sec:exp-twist} supplies a contrasting case, since at the disallowed point $(0.51,\,2.5)$ the certificate has the same $\e$-row structure but the degeneracy survives at $\Lambda=11$ unchanged. The symptom and the type of face are the same in both cases, but the response to raising the truncation differs, and the input does not indicate which case applies.

\section{Feasibility mode and the Slater conditions}
\label{sec:feasibility}

\begin{sloppypar}
	Everything so far was phrased for \emph{optimization} runs: a navigator or OPE-type objective, solved to a tight duality gap. But a large fraction of published bootstrap results were produced in \emph{feasibility mode}: the objective is set to zero, the functional is normalized against the identity contribution, $y\cdot V_{\mathrm{id}}=1$, and SDPB is run in a dedicated termination configuration \cite{Simmons-Duffin:2015qma}. Two such configurations are in use. Most common in modern bootstrap is the \emph{jump mode}: \texttt{detectPrimalFeasibleJump} and \texttt{detectDualFeasibleJump} enabled, \texttt{findPrimalFeasible} and \texttt{findDualFeasible} disabled. Such a run is meant to end in one of two ways: \emph{primal feasible jump detected}, read as ``a positive solution of the truncated crossing equations was exhibited, the point is (at this truncation) allowed,'' or \emph{dual feasible jump detected}, read as ``an excluding functional was exhibited, the point is disallowed.'' The \emph{threshold mode} enables \texttt{--findPrimalFeasible} / \texttt{--findDualFeasible} instead, terminating with \emph{found primal/dual feasible solution} when the corresponding residual crosses its threshold.\footnote{One can also enable any other mixture of these flags. In this section we discuss only the most commonly used sets of termination conditions.} In this section we discuss the implications the Slater conditions have for such runs, the possible failure modes of such classifications, and how the Slater tools can eliminate these failures.
\end{sloppypar}

\subsection{Interpreting termination conditions}
\label{sec:feasverdicts}

In feasibility mode the $Y$-feasible set is the set of normalized excluding functionals: $\{y : y\cdot V_{\mathrm{id}}=1,\ y\cdot V_O \succeq 0\ \forall O\}$, in \SDP\ form $\{y : c - By \succeq 0\}$ after the normalization is eliminated.\footnote{The elimination: pick a component $n_0$ with $(V_{\mathrm{id}})_{n_0}\neq0$, solve $y\cdot V_{\mathrm{id}}=1$ for $y_{n_0}$, and substitute into every constraint. The inhomogeneous part of the substitution generates the vector $c$ (each constraint's share of the identity vector), and the remaining $N-1$ components are the free variables, still denoted $y$.} The primal problem is homogeneous ($b=0$), so its feasible points are positive spectral densities solving the homogeneous equations $B^\top x=0$. Not all of these are crossing solutions. Unpacking the normalization elimination, a feasible density reproduces the identity vector with coefficient proportional to $-\,c\cdot x$, so the feasible rays come in three classes. \emph{Crossing rays} ($c\cdot x<0$) are positive solutions of the truncated crossing equations; \emph{zero-cost rays} ($c\cdot x=0$, Section \ref{sec:howfails}) are positive densities with no identity component at all, i.e.\ exposing certificates (Section \ref{sec:exposing}); and \emph{wrong-sign rays} ($c\cdot x>0$) are positive densities reproducing the identity with the unphysical sign. The distinction is invisible to the feasibility constraints and is carried entirely by the objective. Weak feasibility is guaranteed, since $b=0$ makes $x=0$ feasible. On the $Y$ side nothing is guaranteed, because at every allowed point the functional set is empty and so not even weakly feasible, and the $Y$-Slater condition fails there by construction.

\begin{sloppypar}
\paragraph{Classifiers used in the literature.} A \emph{jump} detector fires when the preceding iteration took a full Newton step on the corresponding side, whatever the current error values (flags \texttt{--detectPrimalFeasibleJump} and \texttt{--detectDualFeasibleJump}). The constraints in question are affine, so a full Newton step lands on them exactly,\footnote{The statement is an identity, because the linearization of an affine map is the map itself, with no Taylor remainder. Writing an affine residual as $r(v)=Av-a$, the Newton direction obeys $A\,dv=-r$, so $r(v+\alpha\,dv)=(1-\alpha)\,r(v)$: partial steps contract affine residuals by exactly $(1-\alpha)$, and a full step reaches zero at the working precision. Of SDPB's KKT equations, the primal and dual feasibility conditions are affine in $(x,X,y,Y)$; only the complementarity condition $XY=\mu I$ is not, and there the full step leaves the second-order remainder $dX\,dY$, which is the only reason the solver needs more than one iteration. A corollary: once an affine residual vanishes, the corresponding Newton rows become homogeneous ($B^\top dx=0$, etc.), so all subsequent steps preserve exact feasibility (in our runs the primal error sits at $10^{-173}$ from the jump iteration onward).}\ and the step-length rule (the step is $\min(\gamma\alpha,1)$, with $\alpha$ the largest step that keeps the cone variable positive semidefinite and $\gamma<1$ SDPB's \texttt{stepLengthReduction}) permits a full step only when the endpoint stays safely positive definite. A jump thus exhibits an iterate that is exactly affine-feasible and strictly inside its cone. A \emph{threshold} detector (\texttt{--findPrimalFeasible} / \texttt{--findDualFeasible}) fires when the affine residual of the current iterate drops below its error threshold; the cone variable is strictly positive definite as always, so this exhibits an approximate version of the same object. Threshold termination is widespread in the published literature. It is the default of the common front ends (PyCFTBoot \cite{Behan:2016dtz}, simpleboot \cite{simpleboot} and the examples shipped with cboot \cite{cboot}), and published feasibility scans have used it alone or in a race with the jump detectors, in which whichever fires first decides; some of these report that the dual jump never fired \cite{Bartlett-Tisdall:2023ghh,Bartlett-Tisdall:2024mbx,Nancarrow:2022wdr} or did not occur in part of the parameter space \cite{Nakayama:2025mrm,Meineri:2025qwz}. Appendix \ref{app:survey} tabulates the termination settings of the works we surveyed. 

Each side's affine residual decreases monotonically, at a rate set by how much of the Newton step its cone admits, and collapses to roundoff at the first full step, the jump. In jump mode the run is a race between the two jumps, and in threshold mode a race between the two thresholds.%
\footnote{SDPB terminates on the first enabled condition that is met, checked in the order: threshold verdicts $\to$ jump verdicts $\to$ iteration/time limits. The jump conditions do not check any error thresholds. Since a full step drives the corresponding residual far below any threshold, such runs can report \emph{found primal/dual feasible solution} even when the underlying event was a jump.}%
\end{sloppypar}

These termination conditions concern strict cone interiors. In jump mode, the endpoint of a full dual step satisfies $(\Tr(A_pY))_p=c-By$ exactly, with $Y\succ0$, a strictly feasible point of the dual set, and the endpoint of a full primal step is a strictly positive density satisfying the homogeneous equations exactly (a crossing solution only if it lies in the $c\cdot x<0$ class). Hence \textbf{a dual (primal) jump can fire only if the dual (primal) feasible set has nonempty interior}. 

In threshold mode the exhibited point is strictly inside its cone but satisfies the affine constraints only to tolerance, and how small that residual can get before the cone obstructs is governed by the distance from the affine subspace to the cone's interior, which is the margin that Slater failure sends to zero. Both sets of termination conditions are affected by Slater condition failure.

\subsection{False classifications in threshold mode}
\label{sec:feascorrupt}

Threshold races can fail in both directions. In an equal-threshold race between the dual and primal error, either side can reach its threshold first, whether the point is allowed or disallowed.  We realized examples of both in the $\s$--$\e$ system, at $\Lambda=5$: a false ``dual feasible'' near the boundary, and a false ``primal feasible'' at the disallowed point $(0.51,\,2.5)$.

\paragraph{A false ``dual feasible'' at an allowed point.} To declare an allowed point disallowed in threshold mode, two things must happen: the dual error must be able to cross its threshold; and it must get there before the primal side reaches its threshold and ends the run. Both can occur close to the boundary of the allowed region, on its allowed side. 

The dual error at an allowed point stalls at the $Y$-margin $\vareps^*$. The dual error is $\|c-By-(\Tr(A_pY))_p\|$ with $Y\succ0$ maintained by construction, so its infimum over everything the solver could ever do is the distance from the affine family $\{c-By\}$ to the PSD cone. That distance is of order $|\vareps^*|$ in the units of the slack weight, since the slack-optimal $y^*$ satisfies $c-By^*+|\vareps^*|\,\widehat Y\succeq0$ and, by optimality, no $y$ does better. We measured this on fifteen allowed points of the $\Lambda=5$ feasibility formulation, on the segment from $(0.5181,\,1.412625)$ to $(0.51,\,2.5)$ at distances from $5.8\times10^{-2}$ (the case-study point itself) down to $10^{-11}$ from the boundary (Table \ref{tab:floor-law}). With all termination thresholds out of reach, the dual error stalls at a level between $0.18$ and $0.45$ of $|\vareps^*|$ across ten decades of margin. A threshold $\tau$ is therefore crossable at all only on the sliver of allowed points with $|\vareps^*|=O(\tau)$.

\begin{table}[htpb]
\centering
\small
\renewcommand{\arraystretch}{1.15}
\begin{tabular}{rrrrr}
\toprule
dist. & $Y$-margin $\vareps^*$ & dual-error stall & stall$/|\vareps^*|$ & $X$-margin $t^*$ \\
\midrule
$5.8\times10^{-2}$ & $-2.71\times10^{-1}$ & $7.3\times10^{-2}$ & $0.27$ & $1.5\times10^{-15}$ \\
$1.1\times10^{-2}$ & $-2.47\times10^{-2}$ & $5.3\times10^{-3}$ & $0.22$ & $5.3\times10^{-16}$ \\
$1.3\times10^{-3}$ & $-2.64\times10^{-3}$ & $5.4\times10^{-4}$ & $0.20$ & $7.4\times10^{-17}$ \\
$5.9\times10^{-5}$ & $-1.72\times10^{-4}$ & $5.2\times10^{-5}$ & $0.30$ & $7.3\times10^{-18}$ \\
$2.3\times10^{-6}$ & $-7.60\times10^{-6}$ & $2.7\times10^{-6}$ & $0.36$ & $3.2\times10^{-19}$ \\
$1.3\times10^{-7}$ & $-4.42\times10^{-7}$ & $1.8\times10^{-7}$ & $0.42$ & $1.9\times10^{-20}$ \\
$1.7\times10^{-9}$ & $-5.85\times10^{-9}$ & $2.6\times10^{-9}$ & $0.45$ & $2.5\times10^{-22}$ \\
$9.7\times10^{-12}$ & $-3.35\times10^{-11}$ & $1.5\times10^{-11}$ & $0.45$ & $1.4\times10^{-24}$ \\
\midrule
$-1.6\times10^{-10}$ (disallowed) & $\approx0$ ($10^{-88}$) & none ($<10^{-80}$) & --- & $\approx0$ ($3\times10^{-46}$) \\
\bottomrule
\end{tabular}
\caption{Margins and the dual-error stall level near the boundary of the allowed region ($\Lambda=5$ feasibility formulation; distance measured along the segment from $(0.5181,\,1.412625)$ to $(0.51,\,2.5)$ in units of its length). $\vareps^*$ from \texttt{ycompute} with the full-degree weight at duality gap ${<}10^{-45}$; stall level = smallest dual error over $400$ iterations of a run with all termination thresholds at $10^{-300}$; $t^*$ from \texttt{xcompute} with $\vareps$-layer $\lambda=10^3$ and floor $t_0=\lambda/(2\Tr\widehat Y)$, at duality gap ${<}10^{-45}$; its positive values are set by the cap (they scale exactly with $\lambda$, Section \ref{sec:xdiag}), and only their sign is a verdict. Eight of fifteen computed points shown. Within $10^{-4}$ of the boundary $t^*=4.2\times10^{-14}|\vareps^*|$ to better than $2\%$ (at $\lambda=10^3$).}
\label{tab:floor-law}
\end{table}

The second gate is the race timing. For this we must also consider the primal error and its $X$-margin $t^*$: the largest $t$ for which a normalized positive homogeneous solution exists with $X\succeq t\,I$, i.e.\ the feasibility problem's own strict-feasibility margin on the density side, mirroring $\vareps^*$ (Section \ref{sec:xdiag}). For a feasibility problem near the boundary, $t^*$ and $\vareps^*$ are related and grow together. Here the size of $t^*$ matters, because at fixed $\lambda$ the capped value measures how deep the cone of positive densities reaches per unit of the weighted trace $\langle X,\widehat Y\rangle$, which is what limits the primal step lengths. All values in Table \ref{tab:floor-law} and in the example below are at $\lambda=10^3$.\footnote{Heuristically, by Gordan's alternative (the strict-inequality relative of Farkas' lemma) the boundary of the allowed region is where $\vareps^*$ rises to zero from below and $t^*$ falls to zero from above. On the disallowed side of the two-parameter formulation both then stay zero (last row of Table \ref{tab:floor-law}), while in the healthy formulations of Table \ref{tab:feas-xslater} $\vareps^*$ turns positive and $t^*$ stays zero. On the allowed side $\vareps^*$ vanishes linearly in the distance to the boundary, and over the range of Table \ref{tab:floor-law} so does $t^*$, so there the two are proportional with problem-dependent prefactors, while away from the boundary they need not track each other. Gordan's alternative concerns finitely many vectors; its conic analogue for positive-semidefinite blocks needs a regularity condition, which is what degenerates on the boundary. Table \ref{tab:floor-law} supports this picture, since within $10^{-4}$ of the boundary $t^*/|\vareps^*|$ is constant to $2\%$ over seven decades, while further out the ratio drifts by a factor seven. The proportionality does not persist all the way to the boundary, however, because at the point of the example below, $6\times10^{-20}$ away on the same segment, $t^*/|\vareps^*|=4\times10^{-20}$ instead of $4\times10^{-14}$.} Near the boundary a small $\abs{\vareps^*}$ thus goes with a small $t^*$, so the crossing solutions lie close to the edge of the cone, Newton steps towards them are cut short by the cone constraints, and the primal side is slow. On the sliver $|\vareps^*|\lesssim\tau$ the dual threshold can then be crossed first, which is how the false exclusion arises.

\paragraph{The example.} Consider the allowed point
\begin{equation*}
	(\De_\s,\De_\e)=(0.5176308413360587319142,\ 1.47560665459297980059),
\end{equation*}
about $6\times10^{-20}$ from the boundary on the segment of Table \ref{tab:floor-law} (much closer than any point listed there). There \texttt{ycompute} attains $\vareps^*=-2.0\times10^{-19}$ (duality gap $4\times10^{-46}$), and a crossing ray (Section \ref{sec:feasverdicts}) with relative objective $c\cdot x/\|x\|_\infty=-9.4\times10^{-20}$ confirms the verdict; the $X$-margin is positive, $t^*=7.8\times10^{-39}$ at $\lambda=10^3$ and duality gap $6\times10^{-46}$ (and $7.8\times10^{-12}$ at $\lambda=10^{30}$). Yet the equal-threshold race at $\tau=10^{-5}$ returns ``found dual feasible solution'', with a ``functional'' of minimum block eigenvalue $-47.6$ at scale $\|y\|_\infty=8.7\times10^{29}$. With the tight thresholds of production runs the affected sliver is correspondingly thinner, but it is never empty.

\paragraph{A false ``primal feasible'' at a disallowed point.} When $Y$-Slater fails, the dual jump has no strictly feasible target and can never fire. The dual error does decrease slowly along the face, without full steps (at a rate set by conditioning), but the primal error decreases faster. At the disallowed point $(0.51,\,2.5)$ it crosses every equal threshold first, by $11$ iterations at $\tau=10^{-5}$ and by about twice that at $\tau=10^{-30}$. The primal side therefore wins the race at every threshold, because the exposing certificate of the face is a zero-cost ray (Section \ref{sec:howfails}) that exists whether or not the point is allowed. The winning run returns this object as ``found primal feasible solution'' at iteration $175$; its terminal $x$ is the exposing certificate itself rather than an approximate crossing solution (it does not include the identity exchange), with relative objective $c\cdot x/\|x\|_\infty=-10^{-66}$ (see also Section \ref{sec:feasworked}).

\paragraph{Conclusion.} Crossing an error threshold shows only that a residual is small, and which side crosses first depends on the two sides' convergence rates, for which no guarantee exists. Declaring feasible the side whose threshold was met first is thus an unreliable heuristic. Jump detections are more trustworthy, although their guarantees are still subtle, and they offer no conclusive answers for the degenerate \SDPs\ that are the subject of this paper; we examine both points next.

\subsection{Reliability of classifications in \emph{jump mode}}
\label{sec:feasjump}

Jump mode is more robust, but its two verdicts are not equally rigorous. A \emph{dual} jump is unconditionally reliable, since its endpoint is an exact, strictly positive excluding functional, the verdict ``disallowed'' follows by weak duality alone, and the exhibited object can be checked independently of how the solver found it. A \emph{primal} jump certifies a strictly positive density solving the homogeneous equations exactly; this does not always mean that no positive functional exists within the search space, so an unconditional interpretation as ``allowed'' is inappropriate. It is correct for only one of the three ray classes of Section \ref{sec:feasverdicts}: a crossing ray ($c\cdot x<0$) proves that no excluding functional exists (below); a zero-cost ray proves nothing, being an exposing certificate available at allowed and disallowed points alike; and a wrong-sign ray can likewise occur at both disallowed and allowed points. A primal jump at a disallowed point requires a strictly positive density there, i.e.\ the $X$-Slater condition at a disallowed point. In the formulations of this paper this condition fails (Section \ref{sec:feasxslater}), so a false primal jump is impossible there. In general it is not excluded, but it would require a density that is strictly positive on every exchanged operator and reproduces the identity with the unphysical sign. A strictly positive density with no identity component at all cannot occur at a disallowed point, because weak duality gives $c\cdot x=\Tr(XY)>0$ for $X\succ0$ and any excluding functional. The possibility of a false primal jump does not call published bootstrap bounds into question. Bounds are obtained by excluding solutions through the existence of positive functionals, and every dual jump correctly certifies the existence of such a functional. At worst a ``false'' primal jump (a step onto a strictly positive ray with $c\cdot x>0$) has caused searches for such a functional to give up where one could have been found, i.e.\ it could have made bounds suboptimal.

\paragraph{The rigorous fix.} At a disallowed point an excluding functional $y$ exists, so every feasible density obeys $0\le x^\top(c-By)=c\cdot x$ by weak duality (the nontrivial direction of Farkas' lemma is not needed), and there is no crossing ray to land on. Hence at a disallowed point no feasible density satisfies
\beq
	c\cdot x\;\le\;-\kappa\,\|x\|_\infty
	\label{eq:hardened-jump}
\eeq
for any $\kappa>0$. Demanding this alongside the primal jump, which we call the \emph{sign-checked primal jump}, upgrades the classifier from a heuristic to a theorem, namely that a sign-checked primal jump can occur only at an allowed point.\footnote{The check requires no additional computation because the relevant quantity is already output by SDPB.}

\paragraph{The race between the jumps.} In a formulation where strictly interior $X$-feasible points exist at a disallowed point (possible in principle, but not realized in any of our examples, Section \ref{sec:feasxslater}), nothing structural prevents a primal jump, and the verdict is decided by a race, namely which side first takes a full Newton step. Throughout this analysis we therefore assume that both Slater conditions hold at the disallowed point, i.e.\ that a strictly positive excluding functional exists and the $X$-feasible set has interior (equivalently, Section \ref{sec:slaterconds}, the set of excluding functionals is bounded). These are the conditions under which the two Newton targets described below exist.

The race has a clear asymmetry, for which we give a heuristic account of the Newton iteration rather than a theorem about predictor--corrector methods. Every iteration of SDPB aims its Newton direction at the central-path point with parameter $\mu_\beta$ (Section \ref{sec:centralpath}), which on each side is the optimum of a \emph{barrier-augmented objective}: the \SDP\ objective of that side plus the target parameter $\mu_\beta$ times a $\log\det$ term that diverges at the boundary of the cone and so keeps the optimum strictly inside it. On the primal side this function is $c\cdot x-\mu_\beta\log\det X$, minimized over the slice $B^\top x=0$; on the dual side, since $b=0$ leaves no objective term at all, it is just the barrier, $\mu_\beta\log\det(c-By)$, maximized over $y$. By the step-length rule, a step of full length is permitted only when the point aimed at lies far enough inside the cone compared to the length of the step being taken.

The Newton direction $(dx,dX,dy,dY)$ solves both affine constraint sets exactly,
\begin{equation*}
	B^\top(x+dx)=0,\qquad
	\sum_p(x+dx)_p A_p = X+dX,\qquad
	\Tr\bigl(A_p(Y+dY)\bigr)+\bigl(B(y+dy)\bigr)_p=c_p,
\end{equation*}
and the complementarity condition to linear order. The endpoint is asked to satisfy $(X+dX)(Y+dY)=\mu_\beta I$ with $\mu_\beta=\beta\mu$, where $\mu=\langle X,Y\rangle/K$ is the current complementarity value, $K$ the dimension of the matrices $X,Y$, and $\beta$ the centering parameter (Section \ref{sec:centralpath}). Typically $\beta<1$, so that each iteration aims for a smaller complementarity than the current one. Dropping the quadratic term $dX\,dY$, this reads
\begin{equation*}
	dX\,Y+X\,dY \;=\; \mu_\beta I-XY.
\end{equation*}
Contracting the third equation with $x+dx$ and using the first two gives $c\cdot(x+dx)=\langle X+dX,\,Y+dY\rangle$; expanding this product and eliminating the cross terms with the trace of the fourth equation leaves
\beq
	c\cdot(x+dx)\;=\;\beta\,\langle X,Y\rangle\;+\;\langle dX,dY\rangle .
	\label{eq:endpoint-objective}
\eeq
The objective at the endpoint of every Newton step is a positive first term, shrinking geometrically over the run, plus a term quadratic in the displacements.\footnote{SDPB's Mehrotra corrector step \cite{Mehrotra:1992} (eq.~(2.31) of \cite{Simmons-Duffin:2015qma}) modifies the right-hand side of the complementarity equation by a further term quadratic in the predictor displacements, shifting \eqref{eq:endpoint-objective} by the same; the conclusions are unchanged.}

While the displacements are small (the convergent regime), the landing value $c\cdot(x+dx)\approx\beta\langle X,Y\rangle$ is positive and slides to zero with the duality gap. A landing point with a large objective of either sign (in particular a solution of large norm with $c\cdot x$ comparable to that norm) would require the quadratic term to be large, hence displacements of the same order; and in this picture displacements grow without bound only when no stationary point exists.

Whether a stationary point exists is decided along feasible rays. Restricted to a feasible ray $x=s\,\hat x$, the primal barrier-augmented objective reads $s\,(c\cdot\hat x)-\mu_\beta K\log s+\mathrm{const}$, with a minimum at $s_*=\mu_\beta K/(c\cdot\hat x)$, provided $c\cdot\hat x>0$.

At a robustly allowed point ($\vareps^*<0$) crossing solutions exist, $c\cdot\hat x<0$.\footnote{For SDPs an empty functional set does not by itself guarantee a crossing ray, since in the weakly infeasible case of Table \ref{tab:feas-verdicts} ($\vareps^*=0$, not attained) the functional set is empty but approached asymptotically, and no crossing ray exists. The statement in the text holds for strictly infeasible functional sets, where the conic Farkas lemma supplies the ray.} Along them both terms decrease without bound and the (convex) barrier-augmented objective has no stationary point, so no central path exists. This absence agrees with Section \ref{sec:centralpath}, where the central path requires both Slater conditions, and at an allowed point the functional set is empty (Section \ref{sec:feasverdicts}), the strongest possible failure of $Y$-Slater. Feasibility mode is built on this dichotomy. At every robustly allowed point the primal problem is unbounded below and its dual is infeasible, and the solver shows this in the usual way. The iterates diverge along a crossing ray, each Newton step still well-defined, its landing value driven negative by the growing quadratic term of \eqref{eq:endpoint-objective}. The full step, once permitted (moving deeper into the cone is never blocked by the step-length rule), lands exactly on the affine slice, and the returned solution is dominated by the crossing-ray direction.

At a disallowed point with a strictly positive excluding functional $Y^*$ every feasible ray has $c\cdot\hat x=\langle \hat X,Y^*\rangle\ge\lambda_{\min}(Y^*)\Tr\hat X>0$, so the stationary point exists, at scale $s_*\propto\mu_\beta$, sliding toward the apex of the cone, $X=0$, as $\mu_\beta\to0$. The iteration converges, displacements shrink, and every landing point has small norm $\|x\|$ and small positive objective $c\cdot x\approx\beta\langle X,Y\rangle$. (These landing points are the wrong-sign rays of Section \ref{sec:feasverdicts}, which prove nothing.)

The dual target does not move, because with no objective term the maximizer of $\log\det(c-By)$ is the same point for every $\mu_\beta$, namely the analytic center of the functional set, its most interior point, at a fixed distance from the cone boundary. It exists because the functional set is bounded and has interior, the two assumptions above, whereas on an unbounded functional set $\log\det(c-By)$ has no maximizer. The dual iterates converge toward this fixed target. Partial steps never fire the detector (they leave the affine residual nonzero, Section \ref{sec:feasverdicts}), but the Newton displacement shrinks as the iterate closes in. Once the displacement is short compared with the target's fixed distance from the boundary, the full step is permitted and the dual jump fires; convergence alone forces it. The primal target, by contrast, slides toward the apex with its distance from the boundary shrinking in proportion to $\mu_\beta$, at the same rate as the displacements chasing it. The full-step condition then compares two quantities of the same order and is decided by constants rather than by convergence, so a full primal step is possible at every iteration but forced at none.

So, in a formulation whose $X$-side has interior at disallowed points, the jump classifier can misclassify. We expect such a misclassification to be rare, because the forced event generically occurs before the accidental one, but we have no example in which the race takes place.

The sign-checked primal jump removes the race from the verdict altogether; in practice one reads $c\cdot x/\|x\|_\infty$ at the jump, which must be negative and stand clear of the iterate's primal residual, i.e.\ of roundoff after a full step (footnote in Section \ref{sec:feasworked}). When neither a dual jump nor a sign-checked primal jump can fire, as on the degenerate formulations of this paper, the sound course is the one stated at the end of Section \ref{sec:feascorrupt}.

\subsection{Failure of the \texorpdfstring{$X$}{X}-Slater condition at disallowed points}
\label{sec:feasxslater}

In feasibility mode the $X$-side is homogeneous, $B^\top x=0$, and $x=0$ is always feasible. Whether the $X$-side is strictly feasible is decided by the reformulation of Slater's theorem in Section \ref{sec:slaterconds}. At a disallowed point the functional set is nonempty and, the objective being zero, every excluding functional is optimal. So $X$-Slater holds if and only if the set of normalized excluding functionals is bounded. That set is unbounded exactly when there is a nonzero functional that is positive semidefinite on all blocks and has vanishing identity component (a recession direction). If that direction is strictly positive on all blocks, pairing it with any feasible density forces the density to vanish, and the $X$-feasible set is $\{0\}$. Conversely, $X$-Slater at a disallowed point requires a density that is strictly positive on every exchanged operator, and by weak duality such a density reproduces the identity with the unphysical sign.

Whether such a recession direction exists is therefore the question, and a sufficient condition is easy to state. Call a functional \emph{unitarity-positive} if it is positive semidefinite on every block of the spectrum with only the unitarity bounds imposed, with no condition on its identity component. If a nonzero unitarity-positive functional $\vec\gamma$ exists, $X$-Slater fails at every disallowed point. Its identity component is non-positive, $\vec\gamma\cdot\vec V_{0,0}=-\sum_{\mathcal O}p^{\mathrm{GFF}}_{\mathcal O}\,\vec\gamma\cdot\vec V_{\mathcal O}\le0$, because the generalized free field with the same external data solves crossing with positive coefficients on operators obeying the unitarity bounds and exchanges the identity.\footnote{The generalized free field exchanges operators of every spin, while the truncated problem imposes positivity only up to a maximal spin; the inequality is exact when $\vec\gamma$ is positive on the spin tail as well, and otherwise holds up to the usual tail estimate.} This is also why such functionals coexist with the generalized free field, which rules out functionals with positive identity component but not these. At a disallowed point with excluding functional $\vec\alpha$ ($\vec\alpha\cdot\vec V_{0,0}=1$, positive on the assumed spectrum), the combination $\vec\gamma-(\vec\gamma\cdot\vec V_{0,0})\,\vec\alpha$ is nonzero, positive on the assumed spectrum, since the unitarity bounds are weaker than any gap assumption, and annihilates the identity. It is a recession direction, the set of normalized excluding functionals is unbounded, and $X$-Slater fails. Nonzero unitarity-positive functionals exist whenever some OPE coefficient admits a finite bound under unitarity alone, since the extremal functional of such a bound is one; the lower bounds on $C_T$ as a function of $\De_\s$ \cite{Poland:2011ey,ElShowk:2012ht} are the standard example, and their functionals remain unitarity-positive when embedded in a multi-correlator system containing the $\langle\s\s\s\s\rangle$ equations, where they act on those equations alone. This is the situation of every feasibility problem in this paper, so the failure is expected at every disallowed point, including those where the $Y$-side is healthy, and it is what the diagnostic finds. The runs marked \enquote{unbounded} in Table \ref{tab:feas-xslater} exhibit the recession direction directly, as a strictly positive functional with vanishing identity component.

The diagnostic of Section \ref{sec:xdiag} settles which case is realized. Table \ref{tab:feas-xslater} shows it on the feasibility-mode \SDPs\ of Section \ref{sec:feasibility}, on the untwisted single-correlator problems of Section \ref{sec:fakepole-exp}, and on the twisted and $\Lambda=11$ variants of Section \ref{sec:exp-twist} (we did not compute $t^*$ for the twisted single-correlator runs or the $\e$--$\e'$ system) and on the three-parameter formulation of Section \ref{sec:casestudy} at four disallowed points. The near-boundary points of Table \ref{tab:floor-law} behave the same way. The $X$-margin vanishes at every disallowed point, including those where the $Y$-side is healthy, and it is positive at the allowed points, shrinking as the boundary is approached. $X$-Slater, which holds in every navigator problem we tested (apart from the deliberately planted failure of Section \ref{sec:xdiag}), thus fails at every disallowed point of the feasibility-mode runs. For the problems marked \enquote{unbounded} in the table (the three-parameter formulation at $(0.5181,\,1.412625)$, the face-reduced problem and the minimal-prefactor single-correlator problem) the $Y$-margin diverges, as it does along a strictly positive functional with vanishing identity component, which is the case in which the $X$-feasible set is $\{0\}$.

\begin{table}[htpb]
\centering
\small
\renewcommand{\arraystretch}{1.2}
\setlength{\tabcolsep}{4pt}
\begin{tabularx}{\textwidth}{@{}>{\raggedright\arraybackslash}X>{\raggedright\arraybackslash}p{2.6cm}>{\raggedright\arraybackslash}p{2.3cm}>{\raggedright\arraybackslash}p{1.9cm}l@{}}
	\toprule
	point / variant & status & $Y$-side & $X$-margin $t^*$ & $X$-Slater \\
	\midrule
	\multicolumn{5}{@{}l}{\emph{two-parameter formulation}} \\
	$(0.5181,\,1.412625)$ & allowed & infeasible & $1.5\times10^{-15}$ & holds \\
	$(0.51763084,\,1.47560625)$ & allowed, $3.7\times10^{-7}$ from the boundary & infeasible & $5.3\times10^{-20}$ & holds \\
	$(0.51762934,\,1.47580743)$ & disallowed, near boundary & degenerate & $\approx0$ & \textbf{fails} \\
	$(0.5170875,\,1.5485469)$, $(0.51405,\,1.9563125)$, $(0.51,\,2.5)$ & disallowed & degenerate & $\approx0$ & \textbf{fails} \\
	$(0.51,\,2.5)$, face-reduced & disallowed & healthy, unbounded$^*$ & $\approx0$ & \textbf{fails} \\
	$(0.51,\,2.5)$: twist $\delta=10^{-6},10^{-1},\frac12,2$; $T$@$3$ + twist $\delta=10^{-6},\frac12$; $\Lambda=11$ with $\delta=0,10^{-6}$ & disallowed & degenerate & $\approx0$ & \textbf{fails} \\
	\multicolumn{5}{@{}l}{\emph{three-parameter formulation, $x_{\mathrm{ope}}=0.5$}} \\
	$(0.5181,\,1.412625)$ & disallowed & healthy, unbounded$^*$ & $\approx0$ & \textbf{fails} \\
	$(0.5177,\,1.30)$, $(0.5177,\,1.55)$, $(0.53,\,1.60)$ & disallowed & healthy & $\approx0$ & \textbf{fails} \\
	$(0.5181,\,1.412625)$, shared prefactor & disallowed & degenerate & $\approx0$ & \textbf{fails} \\
	\multicolumn{5}{@{}l}{\emph{single correlator $\langle\s\s\s\s\rangle$, $\De_\s=0.518$}} \\
	$\De_{\e}\ge1.3$, shared / minimal prefactor & allowed & infeasible & $1.7$ / $18.5$ & holds \\
	$\De_{\e}\ge1.8$, minimal prefactor & disallowed & healthy, unbounded$^*$ & $\approx0$ & \textbf{fails} \\
	$\De_{\e}\ge1.8$, shared prefactor & disallowed & degenerate & $\approx0$ & \textbf{fails} \\
	\bottomrule
\end{tabularx}
\par\smallskip
{\footnotesize\raggedright $^*$The $Y$-margin diverges, as it does along a strictly positive functional with vanishing identity component; the $X$-margin is still zero.\par}
\caption{The $X$-Slater condition in feasibility mode ($\Lambda=5$ unless stated, diagnostic with $\vareps$-layer $\lambda=10^3$ and floor $t_0=\lambda/(2\Tr\widehat Y)$, duality gaps ${\sim}10^{-31}$). $Y$-side, from the $Y$-margin: \enquote{infeasible} $\vareps^*<0$ (no excluding functional), \enquote{degenerate} $\vareps^*=0$ at the duality gap ($Y$-Slater fails), \enquote{healthy} $\vareps^*>0$. Entries $\approx0$ have $\abs{t^*}\le9\times10^{-31}$, within a factor $10$ of the duality gap; their solves take as many iterations as those at the allowed points. Positive entries are set by the cap and only their sign is a verdict (Section \ref{sec:xdiag}). Unlike the navigator problems of this paper, which all have $t^*>0$, these feasibility problems have $t^*\approx0$ at every disallowed point.}
\label{tab:feas-xslater}
\end{table}

In feasibility mode the \enquote{one Slater} row of Table \ref{tab:guarantees} is thus the rule rather than the exception, since at disallowed points of healthy formulations the $Y$-side is strictly feasible and the $X$-side is not. Strong duality holds and the $X$-optimum ($x=0$) is attained, but there is no central path. Two observations from our runs follow: no primal jump occurs at any disallowed point (with $X$-Slater failing the $X$-feasible set has no interior, Section \ref{sec:feasjump}), and \texttt{ycompute} applied to the feasibility form can diverge (the rows marked unbounded in Table \ref{tab:feas-xslater}). The false \enquote{primal feasible} verdict of Section \ref{sec:feasworked} is accordingly a threshold-mode event.

The argument of Section \ref{sec:feasjump} that forces the dual jump assumed a bounded functional set, with an analytic center for the dual iterate to converge to; with $X$-Slater failing, the functional set is unbounded and that target does not exist. Dual jumps nevertheless occur, by the mechanism that produces primal jumps at allowed points. When the functional set is unbounded there is a nonzero functional that is positive semidefinite on all blocks with vanishing identity component, and the barrier $\log\det(c-By)$ increases without bound along it from any strictly feasible point. The dual iterate runs off in that direction; moving deeper into the cone is never blocked by the step-length rule, so a full step, the jump, is taken. We observe this in every disallowed run with a strictly feasible $Y$-side, e.g.\ the minimal-prefactor single-correlator problem of Section \ref{sec:fakepole-exp} jumps at iteration $15$, and so does the face-reduced problem of Section \ref{sec:feasworked}.

\subsection{How to make a rigorous classification}
\label{sec:feasrigorous}
A sharp classification comes from computing the $Y$-margin (see Section \ref{sec:ycompute}): apply \eqref{eq:pmp-Y-slack} to the feasibility-mode constraint set and let $\vareps^*$ be its optimal value. Table \ref{tab:feas-verdicts} lists the four conic outcomes of Section \ref{sec:attainment} together with the verdict and the feasibility-mode behavior that each entails.

\begin{table}[htpb]
	\centering
	\small
	\renewcommand{\arraystretch}{1.3}
	\begin{tabular}{ll>{\raggedright\arraybackslash}p{3.6cm}>{\raggedright\arraybackslash}p{4.0cm}}
		\toprule
		$Y$-margin & conic status & true verdict & feasibility-mode behavior \\
		\midrule
		$\vareps^*>0$ & strictly feasible & disallowed, robustly & dual jump fires in all our runs (Section \ref{sec:feasxslater})\\
		$\vareps^*=0$, attained & weakly feasible & disallowed; functional set on a face & \textbf{no dual jump can fire}; threshold mode risks a false conclusion \\
		$\vareps^*=0$, not attained & weakly infeasible & allowed; functional set empty, approached asymptotically & \textbf{no dual jump can fire}; threshold mode risks a false conclusion \\
		$\vareps^*<0$ & strictly infeasible & allowed, robustly & primal jump possible; threshold mode can err for $|\vareps^*|\lesssim\tau$ (Section \ref{sec:feascorrupt}) \\
		\bottomrule
	\end{tabular}
	\caption{The four conic outcomes of the feasibility question, resolved by the $Y$-margin $\vareps^*$ of the functional set, and the corresponding behavior of the feasibility-mode termination logic.}
	\label{tab:feas-verdicts}
\end{table}

The two middle rows are the weakly feasible and the weakly infeasible case of Section \ref{sec:attainment}, which the computed value $\vareps^*\approx0$ cannot tell apart; the attainment check of that section does. It passes in all navigator formulations and at the allowed points of our feasibility formulations. At the disallowed points of the feasibility formulations it cannot pass, because the recession directions it looks for exist there (they are the $X$-Slater failure of Section \ref{sec:feasxslater}); a vanishing margin is instead resolved by facial reduction, whose reduced problem returns a strictly positive margin (Section \ref{sec:feasworked}).

The diagnostic presupposes nothing about the feasibility of the set it examines, since the $Y$-margin problem's own feasible set is never empty (a sufficiently negative $\vareps$ is always feasible), and $\vareps^*<0$ is its report that the functional set is empty. When the margin does vanish, the facial-reduction pipeline of Section \ref{sec:pmpfr} applies unchanged, because the face is a property of the constraint data and the reduction never touches the (zero) objective. On the reduced problem the functional set regains interior, the dual jump detector is live again, and \texttt{ycompute} returns a nonzero margin of the appropriate sign. Section \ref{sec:feasworked} carries this out.

\paragraph{Incompatibility of the guarantees with the feasibility flags.} The guarantees of Section \ref{sec:guarantees} are not a repair for feasibility-mode runs. \texttt{yguarantee} relaxes every constraint by $|\vareps|\,\rho_j(x)I$ ($\vareps\le0$), so its functional set is strictly feasible at every point, allowed or not. The dual detectors then find a feasible point everywhere, and a run with the feasibility flags reports ``disallowed'' at allowed points as well. The information is instead in the optimal value, $\max\lambda\vareps\le0$, which is $0$ at a disallowed point and strictly negative at a strictly infeasible allowed one (the weakly infeasible case gives $0$, as for \texttt{ycompute}). This value is the $Y$-margin cut off at zero, read against the duality gap like any margin, and \texttt{ycompute} gives the same classification with more information (a positive margin at a robustly disallowed point, and the exposing certificate when it vanishes). The same argument applies to the $X$-side. \texttt{xguarantee} makes the $X$-side strictly feasible at every point, so a primal jump can occur at every point, and only the sign of the full objective $c\cdot x+\nu\eta$ (with $\nu$ above the trace of some excluding functional) would carry information. For feasibility questions we therefore use \texttt{ycompute} as the classifier, or pose the question as a navigator problem, whose sign is the verdict and to which the guarantees apply as intended.

\subsection{Case Study I revisited in feasibility mode}
\label{sec:feasworked}

We now carry out the program of Section \ref{sec:feasrigorous} on the $\Lambda=5$ formulation of Case Study I (Section \ref{sec:casestudy}). We convert the navigator PMP to the classic feasibility formulation (identity normalization $y\cdot V_{\mathrm{id}}=1$, zero objective) and run SDPB with all four feasibility flags enabled (the configuration of Table 1 of \cite{Simmons-Duffin:2015qma}). Per the priority chain of Section \ref{sec:feasverdicts} the terminations below carry the threshold labels. We run at two points: the case-study point $(\De_\s,\De_\e)=(0.5181,\,1.412625)$ of Section \ref{sec:casestudy}, and a point far outside the allowed region, $(0.51,\,2.5)$, whose exclusion the functional recovered below certifies directly. The outcome is shown in Table \ref{tab:feasmode-outcome}.
\begin{table}[htpb]
\centering
\renewcommand{\arraystretch}{1.25}
\begin{tabular}{lllll}
\toprule
point & truth & flags verdict & iterations & \texttt{ycompute} $\vareps^*$ \\
\midrule
$(0.5181,\,1.412625)$ & allowed & \trueverdict{``primal feasible''} & $39$ & $-0.271$ \\
$(0.51,\,2.5)$ & disallowed & \falseverdict{``primal feasible''} & $175$ & $-5.1\times10^{-31}$ \\
\bottomrule
\end{tabular}
\caption{Feasibility-mode runs at an allowed and a disallowed point of the $\Lambda=5$ Case Study I formulation (all four feasibility flags). The flags return the same verdict at both points (the correct one in green, the false one in red); the \texttt{ycompute} margin tells the trustworthy verdict from the false one.}
\label{tab:feasmode-outcome}
\end{table}

At the allowed point the verdict is correct and the $Y$-margin is strictly negative, so everything is healthy. No normalized excluding functional comes within $|\vareps^*|=0.27$ (in the units of the slack column) of existing, so ``allowed'' holds with a quantified margin. At the disallowed point the run terminates cleanly, after four and a half times as many iterations, with the wrong verdict. The $Y$-margin is zero at the duality gap, so every excluding functional lies on a face and the primal detector wins the race by the mechanism of Section \ref{sec:feascorrupt}. The verdict string does not distinguish this run from the healthy one above it; the step lengths and the relative objective do, as we now show.

\paragraph{Anatomy of the two ``primal feasible'' verdicts.} The two identical verdicts arise in different ways, and the difference is visible in the solver's own output. At the allowed point the run ends in a primal jump. At iteration $39$ a full primal step ($\text{step length}=1$) lands the iterate on its affine subspace, and the terminal $x$ is a crossing ray (Section \ref{sec:feasverdicts}), with $c\cdot x=-4.4\times10^{25}$ at scale $\|x\|_\infty=1.7\times10^{26}$, i.e.\ a relative objective $c\cdot x/\|x\|_\infty$ of $-0.25$. This sign proves that no excluding functional exists (Section \ref{sec:feasjump}). 

At the disallowed point no full primal step ever occurs. The primal error instead decreases slowly and falls below the \texttt{primalError} threshold at iteration $175$, with $c\cdot x=-3.4\times10^{-29}$, i.e.\ zero at the primal residual of the run.\footnote{In feasibility mode the dual objective vanishes identically ($b=0$), so SDPB's reported duality gap is $|c\cdot x|$ itself and provides no independent scale. The primal residual provides such a scale, since for an iterate whose $X$ differs from $\sum_px_pA_p$ by a residual $R$, weak duality gives $c\cdot x=\Tr(XY)-\Tr(RY)\ge-\lVert R\rVert\,\Tr Y$ for every excluding functional, so a negative $c\cdot x$ of the order of the primal error times the size of a functional is compatible with exclusion. Here $c\cdot x\approx-57\times\texttt{primalError}$ over the last seventy iterations, both shrinking by the same factor per step, so the value is zero at that scale; divided by the largest component of $x$ it is $-10^{-66}$. After a primal jump the primal error sits at roundoff, so the margin $\kappa$ of the sign-checked primal jump \eqref{eq:hardened-jump} need only stand clear of roundoff; at the allowed point the relative objective is $-0.25$ against a primal error at the working precision.} This $x$ is an exposing certificate ($B^\top x\approx0$, $c\cdot x\approx0$) supported on the $2\times2$ blocks that carry the face rather than a crossing solution; the solver has converged onto the certificate of Section \ref{sec:certfromycompute}. Its relative objective, $-10^{-66}$, is zero at the scale of the run's primal residual and keeps shrinking with it. At $b=0$ the point $x=0$ is always (weakly) feasible, so ``primal feasible'' is vacuous unless the objective is strictly negative at the solution's own scale. Demanding the sign-checked primal jump \eqref{eq:hardened-jump} of Section \ref{sec:feasjump} avoids such misclassifications.

\begin{figure}[t]
\centering
\includegraphics[width=\textwidth]{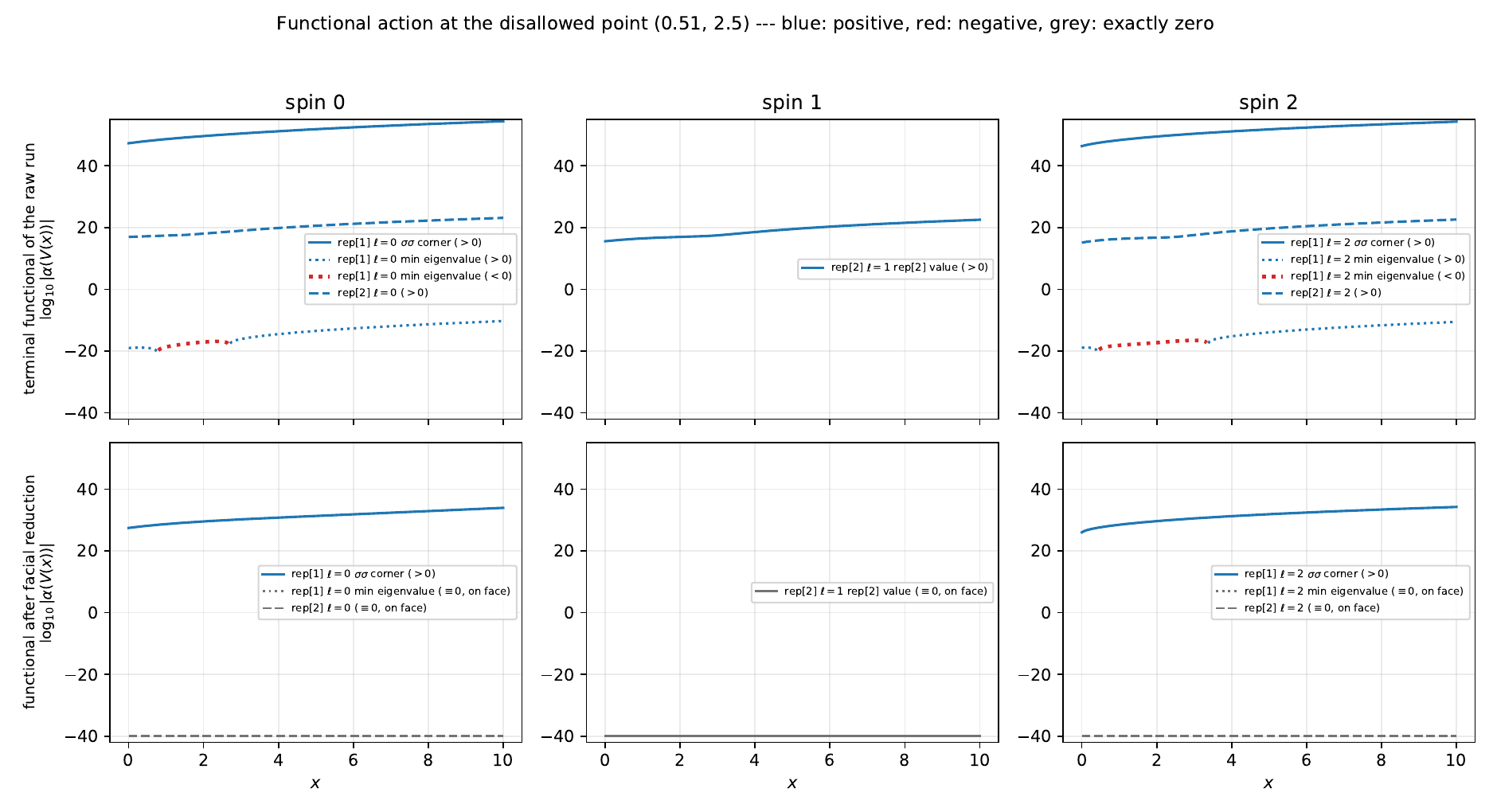}
\caption{Functional actions $\alpha(V(x))$ in polynomial units at the disallowed point $(0.51,\,2.5)$, for spins $0,1,2$ (columns), evaluated by exact-degree interpolation at $60$-digit precision (\texttt{rep[1]}: the $\bZ_2$-even $2\times2$ blocks, \texttt{rep[2]}: the $\bZ_2$-odd blocks). Blue: positive; red: negative; grey: exactly zero. \emph{Top row}: the terminal functional of the raw feasibility-mode run. Its $\s\s$ corners and $\bZ_2$-odd actions are large and positive, but the minimum eigenvalue of the $2\times2$ blocks (dotted) changes sign, being negative on $19\%$ (spin $0$) and $29\%$ (spin $2$) of the plotted range, so the functional is not a valid excluding one, and its nonzero $\s\e$ entries place it off the face. \emph{Bottom row}: the functional recovered from the facially reduced problem, lifted back to the original: every entry the face forces to vanish is exactly zero and the surviving $\s\s$ corners are strictly positive with a wide margin.}
\label{fig:feasfunctionals}
\end{figure}

\paragraph{The repair.} The facial-reduction pipeline of Section \ref{sec:pmpfr}, run as in Section \ref{sec:faceworked}, removes the same face here. The same $24$ functional components are forced to vanish, now in three certificate rounds, $6+12+6$, instead of two, because the feasibility form eliminates a different component for its normalization and exposes the face in a different order. What remains is the $\langle\s\s\s\s\rangle$ problem on the $\s\s$ corners, every other block entry being exactly zero. On this reduced problem the same four feasibility flags return the correct ``dual feasible'' in $15$ iterations, and the recovered functional is strictly positive on every block, with a wide margin (Figure \ref{fig:feasfunctionals}, bottom row, against the indefinite terminal functional of the false run in the top row); it proves that the original ``primal feasible'' verdict was false. Its \texttt{ycompute} margin is unbounded above, because strictly positive functional directions with vanishing identity component exist, the most robust form of ``disallowed''.

\section{Case Study II: Slater failure as a debugging tool for a wrong-spin sector}
\label{sec:faceworked2}

Case Study I (Section \ref{sec:casestudy}) diagnosed and removed a degeneracy that was intrinsic to a formulation. This section demonstrates the same toolbox catching a different type of degeneracy, namely a mistake in a bootstrap setup. We deliberately plant a realistic error in the healthy three-parameter system of Section \ref{sec:casestudy} and watch what each instrument reports. The mistaken problem solves without complaint and returns a plausible-looking navigator value, which answers a different question than the one posed. The Slater diagnostics flag it immediately; restricting the problem to subsets of sectors then localizes it, and the face it creates explains its effect on the value.

\subsection{The mistake}

In the mixed $(\s\s\s\s,\s\s\e\e,\e\e\e\e)$ system the $\bZ_2$-odd channel enters the crossing equations with two sign structures,
\beq
	V_\pm(\De,\ell) \;=\; \bigl(0,\ 0,\ v_1,\ \pm v_2,\ \mp v_3\bigr),
	\label{eq:vpm-ising}
\eeq
where $v_{1,2,3}$ are the $(\De,\ell)$-dependent contributions of a $\bZ_2$-odd operator to the three mixed crossing equations: the two structures share the first entry and differ by the sign of the last two. The sign tracks the $(-1)^\ell$ accompanying the interchange $\lambda_{\s\e\mathcal{O}}\leftrightarrow\lambda_{\e\s\mathcal{O}}$, so that even-spin operators contribute $V_+$ and odd-spin operators $V_-$. Our ``\texttt{wrongSpins}'' setup imposes positivity of $\alpha(V_+)$ at all spins $\ell\le14$, i.e.\ it adds the even-spin sign structure at the odd spins $1,3,\dots,13$ on top of the correct sectors. The mistake is a single configuration error in a sector list; mixed systems carry many (channel, spin-parity, sign) triples, each of which must be paired correctly.\footnote{This failure mode is not specific to the Ising system. We first encountered it in a larger mixed-correlator production setup where the analogous wrong-parity entry had propagated unnoticed from one config file to the next; the structure of the failure and its remedy were the same as in what follows.} The mistaken and correct problems are solved side by side at the three-parameter case-study point $(\De_\s,\De_\e,x_{\mathrm{ope}})=(0.5181,\,1.412625,\,0.5)$ of Section \ref{sec:casestudy} (allowed in the two-parameter system, but disallowed in the healthy three-parameter one) with identical numerics; the reproduction directory of \cite{PolyBoot} (family F11) contains both launchers.

\subsection{Symptoms}

Both the \texttt{wrongSpins} and the healthy navigator solves terminate \texttt{PrimalDualOptimal} at duality gap $10^{-20}$. The value of the former is nevertheless wrong, and both instruments sensitive to the $Y$-Slater condition report a problem:
\begin{itemize}
	\item \emph{The value is wrong.} The healthy three-parameter navigator at this point is $v=+5.8788\times10^{-4}$; the \texttt{wrongSpins} problem returns $v=-7.3725\times10^{-6}$, two orders of magnitude smaller in magnitude and of the opposite sign, so the mistake turns a disallowed point into a falsely allowed one. A user who never sees the healthy number has no way to notice from the value alone.
	\item \emph{\texttt{ycompute} detects the failure.} The $Y$-margin drops from a healthy $\vareps^*=+2.4\times10^{-8}$ to $\vareps^*=-4.0\times10^{-31}$, zero at the duality gap of the solve, so $Y$-Slater fails.
	\item \emph{Skydive stalls.} We use a centering target of $10^{-15}$ here, tighter than the $10^{-10}$ of Section \ref{sec:casestudy}. Under identical solver settings and flags, pure centering steps at fixed $\mu$ (the endpoint $\beta=1$ of the centering parameter) converge at once on the healthy problem, in two steps with unit step length, $R=7\times10^{-4}$ then $1.1\times10^{-17}$. On \texttt{wrongSpins} the dual step is clipped to $0.36$--$0.89$ and the residual stalls near $10^{-11}$ for the $156$ centering steps of the run. At the default target $10^{-10}$ this run would still pass centering; the stall shows only once the target is tightened, whereas the $Y$-margin flags the problem at any setting.
\end{itemize}

\subsection{The geometry of the failure}

Demanding more positivity breaks the problem even though each ingredient alone is innocuous, as the $Y$-margin of the problem restricted to the indicated $\bZ_2$-odd families shows (all other blocks retained):
\beq
\begin{array}{lc}
	\text{$V_+$ at even spins (correct)} & \vareps^* = +\infty\ \text{(unbounded)} \\
	\text{$V_+$ at odd spins (the mistake, alone)} & \vareps^* = +\infty\ \text{(unbounded)} \\
	\text{$V_+$ at even}\ \cup\ \text{$V_-$ at odd (correct pairing)} & \vareps^* = +2.7\times10^{-8}\ ^{\dagger} \\
	\text{$V_+$ at even}\ \cup\ \text{$V_-$ at odd}\ \cup\ \text{$V_+$ at odd (\texttt{wrongSpins})} & \vareps^* = -4.0\times10^{-31} .
\end{array}
\label{eq:wrongspins-margins}
\eeq
($^\dagger$From a separately generated copy of the correct problem, to be compared with $2.4\times10^{-8}$ above.) Each family alone admits a strictly positive functional (rows 1--2), and so does the correct pairing (row 3); adding the $V_+$ family at odd spins to the correct problem (row 4) leaves none. The degeneracy is the general mechanism of Section \ref{sec:howfails}, made explicit, and it rests on two facts. First, exact linear dependence among the sampled constraint vectors is automatic, because the $\bZ_2$-odd families of constraint vectors touch only three equations, so all $560$ sampled rows live in an $18$-dimensional space, and rank counting guarantees a large space of exact linear dependencies among the truncated input data, at any truncation order. Whether the problem degenerates is then decided by orientation alone: by Gordan's alternative, either a strictly positive functional exists, or one of those dependencies has all-nonnegative coefficients, i.e.\ it is an exposing certificate. Second, the normalized sample vectors are strongly collinear (a familiar fact underlying the ill-conditioning of bootstrap \SDPs), and along the sampled family the orientation is tied to the $(-1)^\ell$ sign structure. The physical assignment $\{V_+@\text{even},\,V_-@\text{odd}\}$ places every generator on one side, so that no nonnegative combination of the generators vanishes and the $\bZ_2$-odd sector constrains the functional instead of forcing parts of it to zero. Adding $V_+$ at the odd spins, where $V_-$ is already present, covers both orientations, and a nonzero nonnegative combination of the generators then vanishes.

Once an exposing certificate $\sum_i\hat x_iV_i=0$ ($\hat x_i\ge0$) exists, every feasible functional obeys $\alpha(V_i)=0$ exactly on its support, so the whole $Y$-feasible set lies on a face, the corresponding primal masses become zero-cost rays (Section \ref{sec:howfails}), the analytic center ceases to exist, and the centering stall of the symptoms above follows by the mechanism of Section \ref{sec:centralpath}.

The face also explains the value, $v_{\texttt{wrongSpins}}=-7.3724806986244537\times10^{-6}$, which agrees to fifteen digits with the two-parameter navigator value $-7.372480698624455\times10^{-6}$ of Table \ref{tab:facepipeline}. The forced conditions on the support of the certificate remove the $\s\e$-sector freedom that the OPE-ratio coordinate had added, so the planted mistake demotes the three-parameter problem to the two-parameter one, and the solver reports the optimum of a smaller problem than the one the user believes they posed.

\section{The folklore twist-gap remedy and its domain of validity}
\label{sec:twistgap}

Bootstrap folklore holds that poorly conditioned solves, and feasibility-mode runs that fail to produce a clean primal or dual feasible jump, can be repaired by imposing a small \emph{twist gap}: starting the spinning continua slightly above the unitarity bound, at $\De_{\mathrm{unitarity}}+\delta$.\footnote{A small twist gap, $10^{-7}$ to $10^{-6}$, is imposed in e.g.\ \cite{Chester:2019ifh,Erramilli:2022kgp,Liu:2023elz,Kousvos:2024dlz,DiPietro:2025ozw,Chang:2024whx,Chester:2020iyt}; \cite{Chester:2019ifh,Erramilli:2022kgp,DiPietro:2025ozw} give numerical stability as the motive, \cite{Chester:2019ifh} reports that without it no dual feasible jump occurred, and \cite{Liu:2020tpf} imposes $10^{-4}$ because without it the solver showed ``a steady decay of \texttt{dualError} without vanishing.''} In this section we identify a simple and, we believe, widespread source of $Y$-Slater failure, exact by construction, that a twist gap does cure. It lies in the convention used to factor the positive prefactor out of the conformal blocks. The \emph{minimal prefactor} strips only the poles the block actually has; the \emph{shared prefactor} strips one common list of poles for every channel, including poles whose residues vanish for equal external dimensions. We call the latter \emph{spurious poles}. The shared prefactor plants exact zero rows at the endpoints of unitarity-gapped sectors. We work out the mechanism in elementary terms, confirm experimentally that it removes the Slater margin, and give the fix, the minimal prefactor. A second endpoint mechanism is genuine rather than conventional. It occurs in channels whose OPE basis mixes equal and unequal pairs of operators, no exact reduction within the PMP class removes it, and there a twist gap is the right cure. Finally we show experimentally that a twist gap does not cure $Y$-Slater failures that are not localized at the endpoint, so the folklore has a precise domain of validity.

\subsection{A common mistake: the shared prefactor}
\label{sec:fakepoles}

SDPB imposes positivity of a polynomial rather than of a conformal block. Blocks enter in the rational-approximation form $g_{\De,\ell}\approx r_*^{\De}\,P(\De)/Q(\De)$, with the denominator $Q(\De)=\prod_i(\De-p_i)$ collecting the poles of the block in $\De$ \cite{Kos:2013tga,Simmons-Duffin:2015qma}, and every functional constraint $\alpha(V(\De))\succeq0$ on a sector $\De\ge\De_{\mathrm{gap}}$ has the positive scalar $r_*^{\De}/Q(\De)$ as a common factor of all matrix entries. The passage to the PMP divides this prefactor out, so that ``$g\succeq0$'' becomes ``$P\succeq0$''. Since the prefactor is positive on the open constraint region, the step never changes the feasible set. It can change strict feasibility, however, and the relevant condition is on the numerator rather than on the prefactor, namely that the polynomial matrix handed to SDPB must not be forced to vanish anywhere on the \emph{closed} constraint region, endpoint included. Whether the prefactor has a pole there is by itself irrelevant. A pole that every entry of the block actually has at the endpoint is harmless, since the numerator there is the (nonzero) residue (for blocks that mix entries with and without the pole, see Section \ref{sec:mixedentry}). Our minimal-prefactor control problem has this situation in all of its $\s\e$-channel spinning blocks, whose block pole sits at the unitarity bound $x=0$. A pole that is not in the block is harmful, since cancelling it forces a matching zero into the numerator.

The pole locations come in three families \cite{Kos:2013tga,Kos:2014bka}: Type I at $\De=1-\ell-k$, Type II at $\De=D/2-k$, and Type III at $\De=\ell+D-1-k$, with $k=1,2,\dots$ for types I and II and $k=1,\dots,\ell$ for type III. For equal external dimensions ($\De_{12}=0$ or $\De_{34}=0$) the residues of the odd-$k$ poles of types I and III vanish identically, so these spurious poles are absent from the block and the minimal prefactor omits them. It is nevertheless common practice to use the shared prefactor for every channel of a mixed system,\footnote{The convention is built into the standard block generator \texttt{scalar\_blocks} \cite{scalar_blocks}. As of its current release (v1.0.1) the pole set it declares is computed from the dimension, the spin and the truncation order alone, independently of $\De_{12}$ and $\De_{34}$, so for equal external dimensions the zero-residue families are retained and the compensating zeros are built into the numerator tables. The residue formulas carry the factors $(\frac{1-k+\De_{12}}{2})_k(\frac{1-k+\De_{34}}{2})_k$, which vanish for odd $k$ as soon as $\De_{12}=0$ \emph{or} $\De_{34}=0$, while the declared pole list is unchanged. A minimal per-channel set can only be supplied by hand, through the \texttt{--poles} option; the pole at the unitarity bound is not affected by the pole-shifting of \cite{Kos:2013tga}.} writing the same block as
\beq
	g \;\propto\; \frac{P(\De)}{Q(\De)} \;=\; \frac{(\De-\De_*)\,P(\De)}{(\De-\De_*)\,Q(\De)}
	\label{eq:matched-pair}
\eeq
with a matched zero--pole pair at each spurious pole $\De_*$. As a rational function of $\De$ nothing changes, which is why the convention looks harmless. Two of the spurious poles, however, are not at generic points. The leading spurious pole of type III sits at $\De_*=\ell+D-2$, the unitarity bound of a spin-$\ell$ sector. That bound is the endpoint $x=0$, in the shifted variable $x=\De-\De_{\mathrm{gap}}$, of every spinning ($\ell\ge1$) sector gapped at unitarity in a channel with $\De_{12}=0$ or $\De_{34}=0$.\footnote{The leading spurious pole of type I for spin $0$ sits at $\De=0$, the identity operator, so the identity vector becomes the $0/0$ limit of the rational form; this is a separate hazard.}

\subsection{The vacuous constraint at a matched endpoint zero}

Consider what the divided-through constraint says at the endpoint. With the shared prefactor the polynomial handed to SDPB is $\tilde P(x)=x\,P(x)$, up to factors positive on the closed region, namely the same polynomial as with the minimal prefactor times an overall factor of $x$ in every matrix entry, since the prefactor is common to all of them. At $x=0$ the constraint matrix is therefore identically zero, and the constraint reads $\alpha(0)\succeq0$, which holds for every functional $\alpha$ and carries no information. Because the zero is inserted to cancel a pole that the function does not have, it is exact and does not go away at higher precision.

A one-line toy example isolates the phenomenon. The function $f(x)=x+1$ is positive on $x\ge0$ with margin $1$ at the endpoint. Rewritten as $x(x+1)/x$, the same function has a numerator $x(x+1)$ that is still nonnegative on $x\ge0$ but whose value at $x=0$ is now $0$ instead of $1$. The rewriting replaced ``the margin of the function at the endpoint'' by ``the margin of the numerator at the endpoint'', and the inserted zero sets the latter to exactly zero.

Now compare the two conventions as constraint sets on the functional:
\begin{equation*}
	P(x)\succeq0\quad\forall x\ge0
	\qquad\text{versus}\qquad
	x\,P(x)\succeq0\quad\forall x\ge0 .
\end{equation*}
These define the same feasible set, because $xP\succeq0$ on the half-line forces $P\succeq0$ for $x>0$ and continuity extends this to $x=0$. No verdict and no bound changes, so the mistake is invisible at the level of feasibility and can go unnoticed; strict feasibility, however, is lost completely. The $Y$-Slater condition asks for a functional with strictly positive slack on every constraint; at the endpoint, $\alpha(\tilde P(0))=\alpha(0)=0$ for every $\alpha$, as an identity of the data, so the $Y$-margin of Section \ref{sec:ycompute} obeys $0\ge\vareps\,\rho$ at that row, forcing $\vareps^*\le0$. In the discretized \SDP\ the statement is more concrete. In all seven affected blocks of the example below, \texttt{pmp2sdp} places the first sample point at the prefactor pole $x=0$,\footnote{\texttt{pmp2sdp} places a sample point at every pole within $10^{-10}$ of the endpoint and evaluates the prefactor there with the pole distance floored at $10^{-16}$; the scaling is large but finite, and the sampled numerator is the string $0$, so the row is exactly zero.} so the constraint data acquire zero rows. A zero diagonal row is by itself an exposing certificate supported on a single sample point, its most degenerate form (Section \ref{sec:exposing}). The feasible set is thus unchanged, but every feasible functional has zero slack at the endpoint of every affected sector, and the $Y$-margin is zero.

\subsection{Experimental confirmation}\label{sec:fakepole-exp}

We regenerate the healthy three-parameter system of Section \ref{sec:casestudy} at the three-parameter case-study point $(\De_\s,\De_\e,x_{\mathrm{ope}})=(0.5181,\,1.412625,\,0.5)$ in both conventions with identical numerics ($\Lambda=5$): a control with the minimal prefactor and a variant with the shared prefactor. Only the $\bZ_2$-even sectors differ between the two: the $\s\e$-channel blocks have $\De_{12}\neq0$, so none of their residues vanish and their pole sets are identical in either convention. The reproduction directory of \cite{PolyBoot} lists both launchers under the run family F12.

\begin{itemize}
	\item \emph{Same navigator value, slower convergence.} Both problems terminate at duality gap $10^{-20}$ with the verdict \texttt{PrimalDualOptimal}, and the navigator values agree. The agreement realizes numerically the ``same feasible set'' statement above. The way the solver approaches this value, however, differs. With no margin left, no dual jumps occur (primal jumps occur in both, since the $X$-margin is healthy in both formulations), and this slows the progress considerably.\footnote{Similarly, in feasibility mode the shared-prefactor variant never achieves a dual jump and terminates at the iteration limit ($500$ iterations) with no verdict, although at fixed $x_{\mathrm{ope}}=0.5$ this point is disallowed (Table \ref{tab:feas-xslater}).}
	\item \emph{Detection of the failure by \texttt{ycompute}.} At the same point, the $Y$-margin drops from $\vareps^*=2.4\times10^{-8}$ to $\vareps^*=-1.4\times10^{-32}$ (at duality gap ${\sim}10^{-31}$), i.e.\ to zero at the duality gap of the solve. The prefactor convention alone removes the $Y$-Slater margin.
\end{itemize}

The mechanism is not specific to mixed systems, and it does not need a navigator. We repeated the comparison on the simplest bootstrap \SDP: the single-correlator $\langle\s\s\s\s\rangle$ feasibility problem at $\Lambda=5$, $\De_\s=0.518$, with a gap $\De_\e\ge\De_{\mathrm{gap}}$ in the scalar sector and every spinning sector starting at unitarity, blocks generated with the minimal and with the shared prefactor. At the allowed point $\De_{\mathrm{gap}}=1.3$ both conventions return a strictly negative margin and the same verdict. At the disallowed point $\De_{\mathrm{gap}}=1.8$ the minimal-prefactor problem is as healthy as a feasibility problem can be, with a $Y$-margin unbounded above and a jump-mode run that ends with a dual jump at iteration $15$. The shared-prefactor problem instead has $\vareps^*=-9\times10^{-93}$ at duality gap $9\times10^{-46}$, a $Y$-margin that is zero at the duality gap. Its jump-mode run never makes the dual jump; the dual error stalls at $2\times10^{-13}$, and after $340$ iterations the solver aborts with a Cholesky failure of $Y$. By the mechanism of the preceding subsections this failure is not specific to our example; any single-correlator study of identical scalars that uses shared-prefactor tables and imposes no twist gap carries it at every disallowed point.

The $X$-side output of the \texttt{ycompute} solve confirms the predicted localization block by block. In the shared-prefactor problem the certificate's mass sits on the seven even-spin $\bZ_2$-even blocks (spins $2,4,\dots,14$), i.e.\ the $\bZ_2$-even spinning sectors gapped at unitarity whose prefactor acquired the endpoint pole. It is concentrated at the bottom edge of the sampled continuum and lies many orders of magnitude above every unaffected block, including all $\s\e$-channel blocks, whose pole sets the mistake does not touch. In the minimal-prefactor control problem all blocks carry mass of order one, spread across interior sample points. In the navigator solve itself the same endpoint rows carry the divergent masses (the terminal $x$ reaches $4.1\times10^{35}$ in each of the seven affected blocks). These divergent rows are the zero-cost rays of Section \ref{sec:howfails}, with the same signature as the face of Case Study I (Section \ref{sec:faceworked}), here planted by hand.

\subsection{Facial reduction from the diagnostic output}
\label{sec:fakepolefr}

Suppose one did not know the mechanism of the preceding subsections and saw only that $\vareps^*$ is zero at the duality gap. The facial-reduction pipeline of Section \ref{sec:pmpfr}, run mechanically on the \texttt{ycompute} output, recovers both the cause and the cure. Each step below uses only the diagnostic data in hand at that point.

\begin{enumerate}
	\item \emph{Detect and localize.} $\vareps^*=-1.4\times10^{-32}$ at duality gap ${\sim}10^{-31}$, so a face exists. The converged $x$ of the \texttt{ycompute} solve is the exposing certificate (Section \ref{sec:exposing}), and its support names the face as the seven even-spin $\bZ_2$-even blocks (spins $2$--$14$), with the mass concentrated on the endpoint sample $x=0$ of each. In the generic case the certificate spreads over many rows and blocks; here it is, per block, a point mass on a single row.
	\item \emph{Identify.} The next step is always to ask which linear combination of the supported rows vanishes. With a single supported row per block the simplest candidate is that each supported row is itself the zero vector. Checking the constraint data directly confirms that each supported row is exactly zero. The block-residual scan \eqref{eq:saturation} does not flag these blocks, because only one sample row per block is affected and $r_j$ is dominated by the healthy rows (the first limitation noted in Section \ref{sec:threeuses}). The certificate finds the face instead. The identification step is also where the endpoint type separates from the generic type, since in Sections \ref{sec:casestudy} and \ref{sec:faceworked2} the supported rows are individually nonzero and only a nontrivial positive combination of them vanishes.
	\item \emph{From zero row to reduction move.} A zero row at the sample $x=0$ means every polynomial entry of the block shares a root there, at the boundary of the constraint region. The same block's prefactor, read off from the input, has a pole at the same point. A common numerator root coinciding with a denominator pole is a matched pair, a removable singularity of the rational function the block represents. The reduction move that stays in the PMP class and removes the face at every truncation and sample choice is to cancel the pair, dividing all entries of the block by $x$ and deleting the pole at $0$.\footnote{Deleting the vacuous sampled row from the discretized \SDP\ also restores strict feasibility for the given discretization, but only the cancellation is an exact reduction within the PMP class, and only it names the cause.} In monomial coordinates the division is a one-line edit of the problem file: drop the constant coefficient (which the generator wrote as the string $0$) and remove the pole; no arithmetic is needed. Here all rows at that sample vanish. If all rows but the unequal-pair diagonal ones vanish, the block mixes equal and unequal pairs and division is not available (Section \ref{sec:mixedentry}).
	\item \emph{Reduce and audit.} We perform this edit on the seven blocks named by the certificate in the shared-prefactor problem and re-solve. The navigator value is preserved to $2\times10^{-20}$, i.e.\ up to the duality gap, as the same-feasible-set argument requires, since the reduced problem is the same problem, now correctly presented. The margin is restored to $\vareps^*=2.38201\times10^{-8}$, agreeing with the minimal-prefactor control ($2.38202\times10^{-8}$) to five digits; the dual jump reappears (iteration $65$, dual error $10^{-174}$); and the zero-cost rays are gone ($\max_j\|x_j\|_\infty$ drops from $4.1\times10^{35}$ to $9.9\times10^{2}$).
\end{enumerate}

The edit removed only the single pole the certificate pointed at; the reduced problem still carries every other spurious pole of the shared prefactor (they sit strictly below the gap), and its margin nevertheless matches the minimal-prefactor control. Only the pole at the endpoint matters, as the mechanism predicts, and the pipeline, run without knowledge of the cause, performs the minimal repair.

Unlike the value-level degeneracies of the earlier case studies, which Section \ref{sec:exposing} shows are invisible in the PMP data, this one is visible in the input, as constraint rows of zero norm. A static check before any solve, \emph{no diagonal constraint row vanishes identically}, catches it, and also the mixed-channel mechanism of Section \ref{sec:mixedentry}. The pipeline does not need to know which kind it has found, because the kinds separate at steps 2 and 3, where the certificate is either a point mass on zero rows (all rows of the block here; only the equal-pair rows in Section \ref{sec:mixedentry}) or a nontrivial combination of nonzero rows spread over the continuum, as in Section \ref{sec:exp-twist}, where no twist gap cures it.

\subsection{The fix, and the origin of the folklore}
\label{sec:minimalprefactor}

The minimal prefactor, which strips only the poles the block actually has and drops the spurious ones, whose residues vanish for equal external dimensions, fixes the failure. This fix is the common special case of the requirement of Section \ref{sec:fakepoles} that no convention force the numerator handed to SDPB to vanish on the closed constraint region. The control of the experiment above uses this fix; with the same physics and numerics, its margin is $2.4\times10^{-8}$ instead of zero.

A twist gap also cures this failure. Raising the sector gap to unitarity${}+\delta$ moves the endpoint past the planted zero, whose location is pinned at the unitarity bound; the zero row leaves the constraint region, and the margin returns. We checked this on the single-correlator problem of Section \ref{sec:fakepole-exp} at the disallowed point. With the shared prefactor and a twist gap $\delta=10^{-6}$, $10^{-3}$ or $10^{-1}$ in every spinning sector, the jump-mode run ends with a dual jump at iteration $13$--$14$ and the $Y$-margin is unbounded above, as for the minimal prefactor, while without the twist the dual jump never occurs. We propose this as the mechanistic origin of the folklore. A pipeline built on shared-prefactor tables carries an exact, endpoint-localized $Y$-Slater failure in every spinning sector gapped at unitarity with $\De_{12}=0$ or $\De_{34}=0$, i.e.\ in all spinning sectors of a single-correlator study of identical scalars, wherever a feasible functional exists, i.e.\ at every disallowed point in feasibility mode. The twist gap has been the visible knob that removes it. Conversely, since Section \ref{sec:exp-twist} shows that a twist gap cures only endpoint-localized degeneracies, the fact that a small twist gap was found to help numerically in the setups of \cite{Chester:2019ifh,Erramilli:2022kgp,DiPietro:2025ozw}, and in \cite{Chester:2019ifh} to restore the dual feasible jumps, indicates that an endpoint-planted zero was present there, from the shared prefactor or, in the multi-operator systems among them, possibly from the genuine mechanism of Section \ref{sec:mixedentry}. For the shared-prefactor zero, however, the twist gap is inferior as a cure on two counts. It modifies the physical assumptions, since it excludes operators in $[\De_{\mathrm{unitarity}},\,\De_{\mathrm{unitarity}}+\delta)$ and removes the stress tensor from a twisted spin-$2$ sector unless it is restored by hand as an isolated operator, and it treats a symptom whose cause can be removed from the input data.

\subsection{Channels mixing equal and unequal pairs, where a twist gap is the right cure}
\label{sec:mixedentry}

The endpoint zero of the preceding subsections is an artifact of a convention, and the minimal prefactor removes it. A second mechanism also creates this type of degeneracy. It arises whenever one positivity matrix contains both OPE pairs of equal operators and pairs of unequal ones, and the continuum of that channel starts at the unitarity bound, for spin $\ell\ge1$. In this case there is no easy way to avoid the degeneracy, and the twist gap is a practical solution. However, if one does not want to change the physical assumptions one should instead use our guarantee transforms, which in the examples we studied return the navigator value of the original problem.

\paragraph{Mechanism.} The leading type-III pole at $\De=\ell+D-2$, the endpoint $x=0$ of such a sector, is genuine for the entries built from an unequal pair, since its residue is proportional to $\De_{12}\De_{34}\neq0$. It is spurious for the entries built from an equal pair, and for mixed entries with $\De_{12}=0$ or $\De_{34}=0$. SDPB requires a single scalar prefactor per matrix block, so even the minimal prefactor of the block must contain the pole, and clearing it multiplies every entry in the equal-pair rows and columns by $x$, once. With the rows ordered (equal, unequal),
\beq
	P(x)=\begin{pmatrix} x\,a(x) & x\,b(x)\\ x\,b(x)^{\mathsf T} & c(x)\end{pmatrix},\qquad c(0)=R\neq0,
	\label{eq:mixed-endpoint}
\eeq
so at the endpoint the block is $\mathrm{diag}(0,R)$ and the equal-pair direction vanishes for every functional. Unlike in the shared-prefactor case, only the equal-pair rows and columns of the block vanish there, while the block as a whole does not. Dividing out the prefactor again leaves the feasible set unchanged (continuity gives $a(0;y)\succeq0$ and $c(0;y)\succeq0$ separately), and the $Y$-Slater condition is again lost, as an exact statement. Physically, the pole reflects the null descendant of a spin-$\ell$ operator at the unitarity bound. A conserved current cannot couple to two scalars of different dimension, so the unequal-pair OPE coefficient must vanish there like $\sqrt{\De-\ell-D+2}$ (see below); pairs of equal operators carry no such pole.

The structure occurs whenever a representation appears, at a common spin parity, both in some $\mathcal O_i\times\mathcal O_i$ and in some $\mathcal O_j\times\mathcal O_k$ with $j\neq k$ and $\De_j\neq\De_k$. The simplest case is two scalars in the same representation, such as the $\e$--$\e'$ system, whose even-spin channel has the basis $(\e\e,\e\e',\e'\e')$. It does not occur in the $\s$--$\e$ system of our case studies, where $\bZ_2$ separates the equal pairs ($\s\s$, $\e\e$) from the unequal one ($\s\e$). With a global symmetry the structure appears in other ways. Some examples: in the $O(2)$ system of a vector $v$, a singlet $s$ and a charge-$2$ operator $t$, charge $2$ appears in $v\times v$ and in $s\times t$ (adding a charge-$4$ operator $q$ gives a second instance, charge $4$ in $t\times t$ and $s\times q$); for the $S_3$ doublet $\phi$ and a singlet $\e$ of the Potts model the doublet channel contains $\phi\times\phi$ and $\e\times\phi$; and for $O(N)$ with $N\ge3$, with a singlet $s$ and a rank-two operator $t$, the rank-two channel contains $t\times t$ and $s\times t$, since the rank-two representation lies in its own symmetric square.

\paragraph{Detection by the diagnostic.} We test the mechanism in its simplest setting, two scalars $\e$ and $\e'$ in the same representation, with the crossing equations of the five correlators $\langle\e\e\e\e\rangle,\dots,\langle\e'\e'\e'\e'\rangle$ generated with \texttt{autoboot} \cite{Go:2019lke}: seven equations, an even-spin channel with OPE basis $(\e\e,\e\e',\e'\e')$ and an odd-spin channel containing only $\e\e'$. We take the feasibility problem (identity normalization, zero objective) at $\Lambda=5$ with $(\De_\e,\De_{\e'})=(1,\,1.5)$, a free external OPE matrix, the next scalar $\e''$ gapped at $\De_{\e''}$, and all spinning continua at unitarity, and diagnose it with \texttt{ycompute}. These values have no physical significance and serve only to produce a test problem with an allowed and a disallowed point. A static scan of the \SDP\ finds the predicted pattern, where in each of the seven even-spin blocks (spins $2$ to $14$) every entry vanishes at the endpoint sample $x=0$ except the unequal-pair diagonal entry $(\e\e',\e\e')$. At $\De_{\e''}=3$ and $5$ the point is allowed, no excluding functional exists, and the $Y$-margin is strictly negative with or without a twist. The zero entries only add constraints $0\succeq\vareps\,\rho(0)$, which bound $\vareps^*$ from above by zero and are satisfied by any negative margin, so at an allowed point the degeneracy is invisible. At $\De_{\e''}=8$ the point is disallowed and the $Y$-margin is zero at the duality gap ($\vareps^*=-3.4\times10^{-41}$ at duality gap $3.4\times10^{-41}$). The converged certificate sits entirely where the mechanism says, at the endpoint sample on the equal-pair diagonal rows of the seven even-spin blocks, with masses growing with spin from $6\times10^{13}$ to $3\times10^{39}$, more than twenty orders of magnitude above every other row. With a twist gap $\delta=10^{-3}$ on the continua, and the stress tensor kept as an isolated operator at $\De=3$ with its Ward-identity OPE direction ($\lambda_{\e\e T}:\lambda_{\e\e' T}:\lambda_{\e'\e' T}=\De_\e:0:\De_{\e'}$), no sample point lies at the pole any more, and the $Y$-margin of the same point is unbounded above. The functional set has thus regained its interior, as for the minimal-prefactor problem of Section \ref{sec:fakepole-exp}.

\paragraph{No exact reduction within the PMP class.} The pipeline of Section \ref{sec:pmpfr} stops at the reduction move, because none of its moves applies. Dividing the block by $x$, the fix of Section \ref{sec:fakepolefr}, makes the unequal-pair entries singular. The congruence of (R3) divides the equal-pair corner by $x^2$, but the contact is of first order. In fact no rational reduction exists at all. For small $x<0$ the block $P(x)$ has one negative eigenvalue for each equal-pair row ($x\,a(x)\prec0$ for $a(0)\succ0$) and one positive eigenvalue for each unequal-pair row. A rational congruence $U(x)^{\mathsf T}P(x)U(x)$, optionally divided by a scalar function positive on $(0,\infty)$ such as $x$, keeps the inertia mixed on a punctured neighborhood of $0$ (a scalar that changes sign at $0$ only exchanges the two counts), so it cannot be positive definite at $x=0$. The balanced, strictly positive form needs a half-integer power,
\begin{equation*}
	\begin{pmatrix} x^{-1/2} & 0\\ 0 & 1\end{pmatrix}P(x)\begin{pmatrix} x^{-1/2} & 0\\ 0 & 1\end{pmatrix}
	=\begin{pmatrix} a & \sqrt{x}\,b\\ \sqrt{x}\,b^{\mathsf T} & c\end{pmatrix},
\end{equation*}
which is the same constraint written with the unequal-pair OPE coefficients normalized by $\sqrt{\De-\ell-D+2}$, the natural normalization of a coupling that switches off as the operator becomes conserved, but it is not polynomial. The substitution $x=s^2$ makes it polynomial, but it doubles the degree, needs an extra congruence by $1+s$ on the equal-pair rows to keep them at full degree (Consequence (b) of Appendix \ref{app:faithful}), and turns the prefactor $r_*^{x}$ into $r_*^{s^2}$, for which \texttt{pmp2sdp} cannot generate sample points and bilinear bases; they would have to be supplied by hand. At the level of the \SDP\ the exact reduction is trivial, since the forced zeros are one direction of the Gram matrix $Y_1$ per equal-pair row (in a bilinear basis whose only element nonzero at $x=0$ is the constant, that row times that element), and deleting these directions removes the face. Within the PMP class this requires bilinear bases that differ between the rows of one block, which SDPB does not support. The diagnostic therefore detects and localizes the face, and names its exact reduction, but that reduction is not available to SDPB. A valid lower bound within the class nevertheless exists, by restricting the functionals beyond the face (Section \ref{sec:sandwich}). Imposing $a(0)=0$, linear conditions on $y$, makes the contact of second order (the off-diagonal entries $x\,b$ already vanish to first order), after which (R3) applies. It may be weaker than $\vnav$, since the true face is larger.\footnote{Row-dependent bilinear bases would be a modest extension of SDPB: the positivity condition and the degree are unchanged, only the bookkeeping of the Gram blocks is.}

\paragraph{Remedies.} Three options remain.
\begin{itemize}
	\item \emph{A twist gap on the affected channels.} Starting the continuum at unitarity${}+\delta$ turns the endpoint block into $\bigl(\begin{smallmatrix}\delta a&\delta b\\\delta b^{\mathsf T}&c\end{smallmatrix}\bigr)$, whose smallest eigenvalue is ${\approx}\,\delta\,\lambda_{\min}(a(0))$ for small $\delta$, so the margin is restored, linearly in $\delta$. The physical assumption it adds is mild. The only spinning operators exactly at the unitarity bound are conserved currents (the stress tensor, and the currents of any continuous global symmetry). These must be kept as isolated operators, and they couple only to equal pairs. Every other spinning operator lies strictly above the bound, by an amount set by the theory. In the 3d Ising model the smallest excess, at spin $4$, is $0.023$ \cite{Simmons-Duffin:2016wlq}, so $\delta=10^{-3}$, the value used in the test above, is safe there; in theories with nearly conserved higher-spin currents, such as $O(N)$ models at large $N$, where the excess is of order $1/N$, $\delta$ must be chosen below the smallest anomalous dimension one wants to keep. Smaller values also work, since the margin shrinks linearly with $\delta$ but stays positive (for the shared-prefactor zero we tested down to $\delta=10^{-6}$, Section \ref{sec:minimalprefactor}). The twist can be restricted to the channels that mix equal and unequal pairs.
	\item \emph{The $Y$-guarantee.} For navigator problems, \texttt{yguarantee} restores the $Y$-Slater condition without changing the physical assumptions, and returns the navigator value exactly above its penalty threshold, which is finite whenever the $X$-optimum is attained (Section \ref{sec:twosided}). It cannot be combined with feasibility-mode flags, whose verdicts it makes meaningless (Section \ref{sec:feasrigorous}), so feasibility questions met by this mechanism are better posed as navigator problems. We tested this on the navigator form of the $\e$--$\e'$ problem at the same two points. In navigator form the face is present at both, the allowed one included, and \texttt{ycompute} returns a margin that is zero at the duality gap at each; \texttt{yguarantee} returns the navigator value, to all digits at the duality gap, with its check $x_c>0$ passing at every penalty from $\lambda=10^2$ to $10^{10}$, and \texttt{xyguarantee} returns the same value with both checks passing (the reproduction directory of \cite{PolyBoot}, family F18). This is the one degeneracy in this paper that facial reduction cannot remove within the PMP class, and the guarantee covers it without touching the spectrum assumptions.
	\item \emph{Row-dependent bilinear bases} would give the exact reduction at unchanged degree, and would make the twist unnecessary.
\end{itemize}
The twist gap is thus the genuine cure in this situation and it was correctly applied in several of the setups cited above, such as the $O(2)$ and $O(3)$ systems of \cite{Chester:2019ifh,Chester:2020iyt}, where some channel combines an equal and an unequal pair. Still, if one does not want to alter the physical assumptions the guarantee is the better choice: it returns the navigator value of the original problem.

\subsection{Degeneracies beyond the reach of a twist gap}
\label{sec:exp-twist}

If the twist-gap remedy works by restoring the $Y$-Slater condition, our diagnostic should detect this, and should equally tell us when it cannot work. The failure of the preceding subsections was localized at the sector endpoints; nothing guarantees that property in general. We therefore test the remedy on a $Y$-Slater failure of a different origin, the disallowed point $(0.51,\,2.5)$ of Section \ref{sec:feasworked}, whose feasibility-mode run approaches the exposing certificate and returns the false ``primal feasible'' verdict. We regenerate the identical \SDP\ with every sector gapped at unitarity${}+\delta$ (the explicitly gapped scalar channels are untouched), for a ladder of twists $\delta=10^{-6},\ 10^{-1},\ \tfrac12,\ 2$. Since the twisted spin-$2$ sector starts above $\De=3$, these formulations also remove the stress tensor from the spectrum; we therefore also test a variant with an isolated $T$ at $\De=3$ below the twisted spin-$2$ continuum, at $\delta=10^{-6}$ and $\tfrac12$. Table \ref{tab:twist} shows the outcome.

\begin{table}[htpb]
\centering
\small
\renewcommand{\arraystretch}{1.15}
\begin{tabular}{lcccc}
\toprule
variant & flags verdict & iterations & $c\cdot x/\|x\|_\infty$ & \texttt{ycompute} $\vareps^*$ \\
\midrule
$\delta=0$ (Table \ref{tab:feasmode-outcome}) & \falseverdict{``primal feasible''} & $175$ & $-1.2\times10^{-66}$ & $-5.1\times10^{-31}$ \\
$\delta=10^{-6}$ & \falseverdict{``primal feasible''} & $175$ & $-1.9\times10^{-66}$ & $-7.9\times10^{-31}$ \\
$\delta=10^{-1}$ & \falseverdict{``primal feasible''} & $175$ & $-1.1\times10^{-66}$ & $-7.3\times10^{-31}$ \\
$\delta=\tfrac12$ & \falseverdict{``primal feasible''} & $175$ & $-2.0\times10^{-67}$ & $-7.7\times10^{-31}$ \\
$\delta=2$ & \falseverdict{``primal feasible''} & $189$ & $-2.0\times10^{-70}$ & $-6.2\times10^{-31}$ \\
$T$@$3$ iso.\ $+\ \delta=10^{-6}$ & \falseverdict{``primal feasible''} & $175$ & $-6.7\times10^{-85}$ & $-8.2\times10^{-31}$ \\
$T$@$3$ iso.\ $+\ \delta=\tfrac12$ & \falseverdict{``primal feasible''} & $173$ & $-1.5\times10^{-85}$ & $-8.0\times10^{-31}$ \\
\bottomrule
\end{tabular}
\caption{The twist-gap experiment at the disallowed point $(0.51,\,2.5)$: the feasibility-mode run of Table \ref{tab:feasmode-outcome} regenerated with every spinning continuum starting at unitarity${}+\delta$, and, in the last two rows, with the stress tensor restored as an isolated operator at $\De=3$ below the twisted spin-$2$ continuum (margins measured at duality gap ${\sim}10^{-30}$). None of the deformations changes the outcome: the false verdict appears at essentially the same iteration, the terminal $x$ is always the exposing certificate (zero relative objective), and the $Y$-margin remains zero at the duality gap even at $\delta=2$. False verdicts in red.}
\label{tab:twist}
\end{table}

The twist gap does not cure this failure at any of the values of $\delta$ we tried. All runs terminate with the same false verdict at essentially the same iteration, and the terminal $x$ is again an exposing certificate. For $\delta=10^{-6}$ and $10^{-1}$ its support pattern is identical to the untwisted one (the nine face-carrying blocks at $\|x\|_\infty\sim10^{37}$, everything else at least $26$ orders down), with per-block magnitudes shifted by only a few percent; at $\delta=\tfrac12$ and $2$ its scale grows ($\|x\|_\infty\sim10^{38}$ and $10^{40}$). With the stress tensor isolated, the certificate puts its largest weight on the stress-tensor block ($\|x\|_\infty\sim10^{56}$); restoring $T$ does not cure the degeneracy but is absorbed into the certificate. The same holds at $\Lambda=11$, where regenerating the untwisted and $\delta=10^{-6}$ formulations reproduces the outcome, with false ``primal feasible'' verdicts at iterations $178$ and $176$, relative objectives ${\sim}-10^{-68}$, and margins that are zero at the duality gap ($-8.9\times10^{-31}$ and $-9.4\times10^{-31}$ at duality gap ${\sim}10^{-30}$). The face is of the same type as in Section \ref{sec:faceworked}. This certificate lives in the $\e\e$-corners of the nine $2\times2$ blocks, a later certificate round forces the $\bZ_2$-odd components to zero as well, and what remains is the $\langle\s\s\s\s\rangle$ problem, which excludes this point. Unlike at the allowed point of Section \ref{sec:faceworked}, however, the face persists at $\Lambda=11$.

The support of the certificate explains this result, since a twist gap can only remove a degeneracy localized at the bottom of the continuum. Both endpoint failures of this section, the shared-prefactor one of Section \ref{sec:fakepoles} (planted by a convention) and the mixed-channel one of Section \ref{sec:mixedentry} (forced by kinematics), are of this type, and for both the twist is a cure. The degeneracy here is not of that kind, because the exposing certificate has support throughout the sampled continuum of the $\e\e$-corner components, with relative weights $0.06$--$1$ across all sample points of every face block, while the $\s\s$-corner rows carry ${\sim}10^{-65}$ of that weight. A twist removes the constraints for $\De\in[\De_{\mathrm{unitarity}},\De_{\mathrm{unitarity}}+\delta)$, which carry part of this certificate, so the old combination is no longer null; but a certificate spread over the whole continuum can be rebuilt on the remaining range, and the experiment shows that at this point it is, at every $\delta$ we tried. In practice, then, whether a twist gap can help depends on where the exposing certificate is supported, which the $X$-side output of a single \texttt{ycompute} solve shows. Here it indicates that the remedy is unlikely to help, and the experiment confirms this.

\paragraph{A planted contact at infinity.} The third location a contact can have, $x=\infty$, we plant in the healthy three-parameter navigator of Section \ref{sec:fakepole-exp}, at the same allowed point and $\Lambda=5$, by the mechanism of the published example of Section \ref{sec:mixedcurrent}: a $\De$-dependent normalization of one OPE coefficient. We multiply the $\e\e$ row and column of the spin-$2$ $\bZ_2$-even block by $1+x$, i.e.\ conjugate the block by $\mathrm{diag}(1,\,1+x)$. Since $1+x>0$ on the half-line this changes neither the feasible set nor the value, but the entry degrees become $(26,27;27,28)$ in place of $(26,26;26,26)$: the block's declared degree is $28$ and the $\s\s$ row has a deficit of two, so by Consequence (b) of Appendix \ref{app:faithful} its Gram representation is degenerate at $x=\infty$. Table \ref{tab:degmismatch} shows the three problems, with all poles of the prefactor kept so that the degrees are exact.\footnote{With the reduced-prefactor interpolation of \texttt{pmp2sdp} used elsewhere in this paper (eight poles kept) the value is $5.88\times10^{-4}$ and the deficit is no longer exact at the \SDP\ level, because interpolating the deficient entries at the block's sample points produces a spurious leading coefficient. The margin then comes out at $9\times10^{-19}$, twelve orders above the duality gap and ten below the healthy value, with the same certificate: the fourth mechanism of Table \ref{tab:mixedcurrent} in miniature.} The navigator value is unchanged to the duality gap and the solve needs only ten more iterations, so the run itself reports nothing. Its dual step, however, never reaches full length, and the dual error stalls at $5\times10^{-61}$ where the control's collapses to $10^{-149}$ at the dual jump of iteration $100$; the $Y$-margin drops from $2.4\times10^{-8}$ to zero at the duality gap; and the exposing certificate is carried entirely by the modified block (every other block at least $10^{9}$ down), where in the sampled Gram representation it is the rank-one matrix on the highest basis element of the $\s\s$ row, i.e.\ the leading coefficient of that row is forced to vanish. This is the contact data of (R$\infty$), read off the solve as Section \ref{sec:pmpfr} prescribes, and no twist gap can touch it, since the sector's endpoint plays no role. The move itself, the congruence by $\mathrm{diag}(1+x,\,1)$ which raises the $\s\s$ row to the declared degree, restores the margin to its control value to all printed digits: on this block the repaired constraint is $(1+x)^2$ times the original, and the full-degree weight $\rho_j=(1+x)^{d_j}$ absorbs the factor, so the two margin problems coincide. With this, each of the three contact locations of Section \ref{sec:pmpfr} has been produced, diagnosed and removed on the Ising system: the endpoint (Sections \ref{sec:fakepoles} and \ref{sec:mixedentry}), the interior (Section \ref{sec:faceworked} and the point above) and infinity.

\begin{table}[htpb]
\centering
\small
\renewcommand{\arraystretch}{1.15}
\setlength{\tabcolsep}{4pt}
\begin{tabularx}{\textwidth}{@{}>{\raggedright\arraybackslash}p{2.8cm}cccc>{\raggedright\arraybackslash}X@{}}
\toprule
problem & entry degrees & $\vnav$ & dual jump & \texttt{ycompute} $\vareps^*$ & certificate \\
\midrule
control & $(26,26;26,26)$ & $9.5477\times10^{-4}$ & it.\ $100$ & $2.38\times10^{-8}$ & -- \\
planted, $\mathrm{diag}(1,1+x)$ & $(26,27;27,28)$ & same to $4\times10^{-21}$ & none & $-5\times10^{-34}$ & spin-$2$ block, $\s\s$ row, leading coefficient \\
repaired, $\mathrm{diag}(1+x,1)$ & $(28,28;28,28)$ & same to $7\times10^{-21}$ & it.\ $101$ & $2.38\times10^{-8}$ & -- \\
\bottomrule
\end{tabularx}
\caption{A degree mismatch planted in the spin-$2$ $\bZ_2$-even block of the healthy three-parameter navigator at $(0.5181,\,1.412625,\,0.5)$, $\Lambda=5$, all prefactor poles kept; navigator solves to duality gap $10^{-20}$ at $640$ bits, margins to ${\sim}10^{-30}$ (the planted margin is zero at its duality gap of $3\times10^{-31}$). Entry degrees are listed as $(\s\s,\s\e;\e\s,\e\e)$; \enquote{dual jump} is the iteration of the first full dual step of the navigator solve. The planted problem has the same value and feasible set as the control, but its Gram representation is degenerate at $x=\infty$; the (R$\infty$) congruence removes the contact and the margin returns to the control's value exactly.}
\label{tab:degmismatch}
\end{table}

\section[A published degenerate example: the mixed current--scalar bootstrap]{A published degenerate example:\\ the mixed current--scalar bootstrap}
\label{sec:mixedcurrent}
As a test on a published setup with spinning external operators we revisited the 3d mixed current--scalar bootstrap of \cite{Reehorst:2019pzi}: a charged scalar $\phi$ and the conserved current $J$ of an $O(2)$ global symmetry, with the five correlators $\langle\phi\bar\phi\phi\bar\phi\rangle$, $\langle J\phi J\bar\phi\rangle$, $\langle JJ\phi\bar\phi\rangle$, $\langle\phi JJ\bar\phi\rangle$ and $\langle JJJJ\rangle$. The published formulation contains four degeneracies, listed in Table \ref{tab:mixedcurrent}, which prevent full primal and dual Newton steps. Each comes from a choice in the formulation, and each is removed by one of the moves of Section \ref{sec:pmpmoves} or by a different choice of equations. We describe them only as far as needed to place them in the classification of this paper.

\begin{table}[htpb]
\centering
\small
\renewcommand{\arraystretch}{1.2}
\begin{tabularx}{\textwidth}{>{\raggedright\arraybackslash}p{4.0cm}>{\raggedright\arraybackslash}X>{\raggedright\arraybackslash}X}
\toprule
degeneracy & kind & resolution \\
\midrule
redundant $\langle JJJJ\rangle$ equations & exact linear relation among the functional components at $\Lambda=13$ (none at $\Lambda\le5$); $B$ loses column rank (Section \ref{sec:feasrigorous}) & remove the redundant direction, (R1), or an independent equation set \\
zeros planted by the normalization of conserved $\langle JJ\mathcal O\rangle$ structures & forced contact at finite $x$: a numerator zero planted by a convention, as in Section \ref{sec:fakepoles} & lowest-degree normalization; scalar (R3), as in Section \ref{sec:fakepolefr} \\
unequal polynomial degrees within one matrix block & forced contact at $x=\infty$ (Appendix \ref{app:faithful}, Consequence (b)), at every point and every $\Lambda$ & degree-matching congruence; (R$\infty$) with contact data known in advance \\
spurious leading coefficients (roundoff, block truncation) & the same contact at $x=\infty$, at the numerical level & analytic prediction of the true degrees; exact truncation \\
\bottomrule
\end{tabularx}
\caption{The degeneracies of the published mixed current--scalar setup \cite{Reehorst:2019pzi}, their kind in the classification of this paper, and their resolution. The fourth was hidden behind the third and appeared only after it had been removed.}
\label{tab:mixedcurrent}
\end{table}

\paragraph{Redundant equations.} The published $\langle JJJJ\rangle$ system includes equations on the crossing line $u=v$ that conservation of $J$, together with regularity at the crossing point, makes consequences of the bulk equations. Whether an exact linear relation among the truncated functional components survives the derivative truncation depends on $\Lambda$: at $\Lambda=3$ and $5$ the truncated set has full rank, while at $\Lambda=13$ exactly one exact relation survives among the $176$ $\langle JJJJ\rangle$ components, and it annihilates every constraint vector, the normalization and the objective. This degeneracy is a $B$-degeneracy rather than a Slater failure: it is not a face of the cone; instead the optimal functional is undetermined along a line, the matrix $B$ loses column rank, and SDPB's Newton system, which assumes full column rank, becomes singular (full column rank is also the first ingredient of the attainment check of Section \ref{sec:feasrigorous}). The degeneracy is removed by (R1), i.e.\ by dropping the redundant direction, or, as we did, by an independent set of equations (the conformal-frame bulk equations plus the point equation), which is not equivalent at equal $\Lambda$ and moves the bound slightly. Redundancy must be judged on the full system rather than correlator by correlator, because two line equations of $\langle JJ\phi\bar\phi\rangle$ coincide on the charge-$0$ blocks but are distinguished by the charge-$1$ blocks of $\langle\phi JJ\bar\phi\rangle$, and both must be kept.

\paragraph{Planted zeros.} Imposing conservation on $\langle JJ\mathcal O\rangle$ fixes the three-point structures up to a normalization, which is a polynomial in $\De$. The published normalization carried a factor $\De-3$ in the parity-odd spin-$0$ structure, so that $1\times1$ block was proportional to $(\De-3)^2$, a double zero inside the positivity region at which nothing physical happens. This planted zero is the mechanism of Section \ref{sec:fakepoles}, a numerator zero that the physics does not require, and its cure is the division of Section \ref{sec:fakepolefr} without a pole to delete, namely dividing the factor out and lowering the declared degree of the $1\times1$ block by two (the scalar case of (R3), with the degree lowered instead of restored by $(1+x)^2$). We adopted this as a rule, normalizing every conserved structure by the lowest-degree coefficients without common factors.

\paragraph{Unequal degrees within one block.} A neutral parity-even operator of even spin $\ell\ge2$ appears in $\phi\times\bar\phi$ with one OPE coefficient and in $J\times J$ with two conserved structures, so its positivity constraint is a $3\times3$ block. The conserved structures carry $\De$-dependent coefficients of different degree, and after the common prefactor is divided out the entries of the block have degrees $d-a_i-a_j$, with $d$ the largest entry degree (the degree SDPB assigns to the block) and row deficits $a=(2,0,1)$ (and $(1,0)$ for the $2\times2$ spin-$0$ block). As a positivity condition at finite $x$ nothing is wrong with this, but for SDPB it is fatal, because one degree $d$ and one bilinear basis serve all rows of a block, so a diagonal entry of degree below $d$ forces the Gram element of its top basis function to vanish, and with it the whole row and column of $Y$, for every feasible functional. The forced vanishing is the contact at $x=\infty$ of Appendix \ref{app:faithful}, Consequence (b), with physical content. It is invisible pointwise, since the constant-weight margin $\vareps^*_1$ can be positive and only the full-degree weight \eqref{eq:slack-weight} detects it. It is present at every external point and every $\Lambda$, and neither precision nor a twist gap helps, because the contact is not at the endpoint (Section \ref{sec:exp-twist}). The contact is removed by the move (R$\infty$) with the contact data read off the PMP, namely the congruence $\widetilde M=U\,M\,U$ of the block $M(x;y)$ with $U(x)=\mathrm{diag}\bigl((1+x)^{a_i}\bigr)$, one factor $1+x$ per unit of deficit (the inverse of the $U$ of Section \ref{sec:pmpmoves}), which gives every diagonal entry the degree $d$, is exact for $x\ge0$, and amounts to measuring the $JJ$ OPE coefficients in $\De$-dependent units. With it, solves that had stalled at the same iterate at two working precisions converge. The mechanism needs a matrix block whose rows grow at different rates, which is why it did not occur in the scalar systems of the earlier sections. The stress-tensor bootstrap of \cite{Chang:2024whx} met it in a first choice of conserved structures that became linearly dependent at $\De=\infty$, so that no dual jump ever occurred; the cure adopted there, structures normalized to grow at the same rate as $\De\to\infty$, is the degree matching above in another guise. \cite{Nancarrow:2022wdr} met the same mechanism in a spin-system bootstrap, mismatched highest powers forcing an off-diagonal entry to vanish exactly so that the dual jump ``will never work,'' and cured it by a fixed $\epsilon x^{d}$ shift of the diagonal, which fails in their two-variable setup. It is prevented by a simple static check, that in every block each diagonal entry reaches the declared degree. Section \ref{sec:exp-twist} reproduces this mechanism on the Ising system by planting a row deficit in one block of the healthy navigator, and shows the certificate and the repair.

\paragraph{Spurious leading coefficients.} The congruence needs the true degrees, and the computed polynomials overstate them, because exact large-$\De$ cancellations between structures leave top coefficients at roundoff level, or at the level of the finite-order block truncation, many orders below their neighbours. For the Gram representation an entry whose top coefficient is noise is an entry of lower degree, so this is the previous degeneracy again, now in every entry that carries $\langle JJJJ\rangle$ equations. A tolerance cannot separate the truncation-level coefficients reliably; we predict the degree of every entry analytically from the large-$\De$ asymptotics of the blocks and truncate the numerators there, checking on every run that the dropped coefficients are negligible.

\paragraph{The remaining degeneracy and the published verdicts.} With these four resolutions the $\Lambda=13$ problem satisfies both Slater conditions. The remaining degeneracy is not a formulation defect. At $\Lambda\le5$ the optimal functional acts only on the scalar correlator, an exposing certificate forces every component on the current equations to vanish, and the bound equals the scalar-only bound, the multi-correlator mechanism of Section \ref{sec:howfails}; by $\Lambda=13$ the mixed system is active. A remaining slowness of the dual side with every pole kept is due to conditioning rather than degeneracy, and is removed by pole reduction. The third degeneracy bears on the feasibility verdicts of Section \ref{sec:feasibility}. Since it is present at every point and every $\Lambda$, no feasible functional ever had a positive-definite Gram matrix $Y$ in the published formulation (the $Y$-Slater condition failed exactly), so by Section \ref{sec:feasverdicts} no dual jump could fire. The allowed regions of \cite{Reehorst:2019pzi}, which are feasibility scans (the $C_T$ and $C_J$ bounds are optimization runs and are not affected), were accordingly decided in the threshold mode of Section \ref{sec:feascorrupt}, with both error thresholds enabled, and with dual jump detection enabled although it could not have fired. That mode is not a rigorous verdict (Section \ref{sec:feascorrupt}), and it is slow by construction, because the dual error creeps down the face instead of collapsing in one step. With the degeneracies removed, the jumps of Section \ref{sec:feasjump} are available again.

\section{Conclusions}
\label{sec:conclusions}

The convergence guarantees behind every numerical bootstrap computation rest on the Slater conditions, and bootstrap SDPs often violate them in practice. The violations are not confined to exceptional points, since in our standard two-parameter mixed-correlator navigator at $\Lambda=5$ they occur at every tested point, allowed or disallowed. At higher $\Lambda$ ($9$ to $19$) it disappears at the tested allowed point, while at a disallowed point of the feasibility formulation it persists at $\Lambda=11$. Whether a given formulation violates a Slater condition at a given point and truncation is a property of the solution rather than of its construction, and can change under a shift of the external point or a change in the truncation parameter $\Lambda$. Since it cannot be predicted a priori it must be measured, and a degeneracy that lives in the solution cannot be removed once and for all before a scan. This is the case for the guarantees over facial reduction: one transformation of the input makes every point non-degenerate by construction, uniformly across a scan or a Skydive run, with the check of the penalty read from each solve. Facial reduction is the route when the cause is visible in the input, or when a feasibility-mode question must be answered by the flags. Our toolbox makes this failure mode routine to detect and cheap to correct. Figure \ref{fig:workflow} summarizes the procedure. The diagnostics \texttt{ycompute} and \texttt{xcompute} each measure one Slater margin within one SDPB call. The guarantees \texttt{yguarantee} and \texttt{xguarantee} enforce the two Slater conditions by construction. These give rigorous bounds on the original optimum for every penalty, and coincide with it once the penalty is inactive, which each run reports itself ($x_c>0$, respectively $Y_\eta>0$).

The combined guarantee, with a large enough trace bound, converges in every case and reports from the same run whether its value is an upper bound, a lower bound or the navigator value itself. Only when an optimum is not attained does the missing end need facial reduction: the lower end, from the $Y$-side, when the $X$-optimum is not attained (the case relevant to our examples), and the upper end, from the $X$-side, in the mirror case. For all but one of the Slater degeneracies we met the reduction can be carried out exactly within the PMP class (for the published setup of Section \ref{sec:mixedcurrent}, up to the block-approximation error of its truncated tables); for the exception (Section \ref{sec:mixedentry}) the guarantee transforms already return the navigator value and no facial reduction is required.

The Skydive case study is the first explicit root-cause diagnosis of the centering-stall failure mode anticipated in Section 5.3.2 of \cite{Liu:2023elz}. On the same problems on which Skydive stalls, classical SDPB optimization often converges despite the Slater failures. The cause of the stall is a $Y$-margin $\vareps^*$ that is zero at the duality gap, and its signature a centering residual that does not converge while the duality gap reaches its target. Our transformation reopens the margin and turns a solver that cannot take a single step into one that centers and moves.

In feasibility mode a threshold race between the two errors can misclassify in both directions, and the verdict strings do not distinguish a genuine jump from a stalled approach to an exposing certificate (Section \ref{sec:feascorrupt}); a verdict is reliable only from a dual jump (disallowed) or from a sign-checked primal jump, a primal jump with a strictly negative $c\cdot x$ at the solution's own scale (allowed, Section \ref{sec:feasjump}). At disallowed points the $X$-Slater condition fails in every feasibility problem we tested, so a $Y$-Slater failure there leaves jump mode without a verdict; the guarantees cannot repair this, since they make the flags' verdicts meaningless (Section \ref{sec:feasrigorous}), and the failure is instead removed by facial reduction, or by posing the question as a navigator problem. The mixed current--scalar bootstrap of Section \ref{sec:mixedcurrent} shows how this plays out in practice. One of its degeneracies was present at every point, so no dual jump could fire and its allowed regions were decided by threshold races. For feasibility questions we recommend \texttt{ycompute} itself as the classifier: its sign, read against the duality gap of the solve, is the verdict (a margin that is zero at the gap calls for the attainment check of Section \ref{sec:attainment} or for facial reduction), its magnitude the margin, and its $X$-side output the exposing certificate when the margin vanishes.

Finally, a simple source of degeneracy deserves attention in practice. The shared prefactor, the convention built into \texttt{scalar\_blocks}, plants an exact zero at the unitarity bound of every spinning sector gapped at unitarity in channels with equal external dimensions; values are unchanged, the $Y$-Slater condition is destroyed, and a standard single-correlator feasibility run then makes no dual jump at any disallowed point. The minimal prefactor removes this at no cost, and a static check of the input (no identically vanishing diagonal row and, for spinning systems, every diagonal entry reaching its block's declared degree, Section \ref{sec:mixedcurrent}) catches it. The same check catches the one genuine endpoint degeneracy we know of, in channels that mix equal and unequal operator pairs, where the cure is a small twist gap on those channels with conserved currents kept as isolated operators. Whether a twist gap can help at all is read off the support of the exposing certificate. A twist gap cures endpoint-localized degeneracies and, in our experiments, not those supported throughout the continuum.

Two recommendations for practice follow. First, feasibility questions should be run in jump mode, with the threshold flags disabled, and the termination settings should be reported: the survey of Appendix \ref{app:survey} finds that threshold termination entered the literature through the race settings of the SDPB manual and through front-end defaults, and that most published feasibility results are silent on the settings that decided them. Second, block generators and front ends should adopt the minimal prefactor and the static checks of Figure \ref{fig:workflow} as defaults, so that the planted endpoint zero, and the twist gap that compensates for it, disappear from bootstrap practice without any user needing the tools of this paper.

The census of Appendix \ref{app:survey} shows the same failures, and hand-made versions of the same cures, in the S-matrix and quantum-mechanical bootstraps; the transforms of this paper act on the PMP and do not depend on the solver. The tools introduced here make Slater failures in SDPB and Skydive detectable, and in every case we met easy to remove.

\section*{Acknowledgements}
We thank Prof.\ Balt van Rees for useful discussions about Slater conditions and Skydive. The calculations in this work were performed on the CREATE HPC cluster~\cite{CREATE} at King's College London. The work of MR is funded by UK Research and Innovation (UKRI) under the UK government's Horizon Europe funding guarantee [grant number EP/X042618/1]. We thank Prof.\ David Simmons-Duffin for arranging access to the Caltech Resnick High Performance Computing Center, although in the end it was not needed for the computations reported here. We thank Claude and ChatGPT for their tireless assistance in all phases of the research.

\appendix

\section{The \texorpdfstring{$Y$}{Y}-margin at the \texorpdfstring{\SDP}{SDP} level: conic margins and unboundedness}
\label{app:Y-slack-PMP-to-SDP}

To translate \eqref{eq:pmp-Y-slack} into an \SDP, we use the PMP$\to$\SDP\ conversion rules of \cite{Simmons-Duffin:2015qma}. Writing the $(r,s)$ entry of $M_n^j(x)$ as a scalar polynomial $P^n_{j,rs}(x)$, and letting $p=(j,r,s,k)$ run over blocks, matrix entries, and sample points $x_k$, the \SDP\ constraints take the standard affine form
\beq
	\Tr(A_pY)+\sum_{n=1}^N B_{p,n}y_n=c_p\qquad\forall p,\qquad Y\succeq 0,
	\label{eq:sdp-affine}
\eeq
with $B_{(j,r,s,k),\,n}=-P^n_{j,rs}(x_k)$ and $c_{(j,r,s,k)}=P^0_{j,rs}(x_k)$. The slack deformation appends $M^j_{N+1}=-\rho_j(x)\,I$, whose translation is the extra column
\beq
	B_{(j,r,s,k),\,N+1}=\rho_j(x_k)\,\delta_{rs},
	\label{eq:slack-column}
\eeq
(times the reduced sample scalings in SDPB's normalization, which makes it the column $w$ of \eqref{eq:weighted-column}, $\Tr(A_p\widehat Y)$ for the Gram matrix $\widehat Y$ representing $\rho_j(x)I$) so the deformed constraints read
\beq
	\Tr(A_pY)+\sum_{n=1}^N B_{p,n}y_n+\vareps\,\rho_j(x_k)\,\delta_{rs}=c_p .
	\label{eq:sdp-slack-constraints}
\eeq
The slack term can be rewritten as a trace against a fixed positive-semidefinite direction $\widehat Y$ chosen so that $\Tr(A_{(j,r,s,k)}\widehat Y)=\rho_j(x_k)\,\delta_{rs}$; for the full-degree weight \eqref{eq:slack-weight} such a $\widehat Y$ can be taken positive \emph{definite} (this is the interior-of-the-cone property of the weight, Section \ref{sec:ycompute}).\footnote{Alternatively one can start at the \SDP\ level with a clean $Z-\vareps I\succeq0$; pushed back through the translation this deforms the PMP by $-\vareps\, S_j(x)\,I$ with $S_j(x)$ nonnegative sums of squares, a benign reweighting of the same construction.} After the change of variables $Z:=Y+\vareps\,\widehat Y$ the constraints become the original affine constraints while the cone constraint becomes $Z-\vareps\,\widehat Y\succeq0$, so the $Y$-margin problem is
\beq
	\text{maximize }\vareps
	\quad\text{subject to}\quad
	\Tr(A_pZ)+\textstyle\sum_n B_{p,n}y_n=c_p,\qquad
	Z-\vareps\,\widehat Y\succeq 0.
	\label{eq:sdp-slack-coneform}
\eeq
In this form $\vareps^*$ is manifestly the largest \emph{conic margin} of the feasible set in the direction $\widehat Y$, since for fixed feasible $(Z,y)$ the largest admissible $\vareps$ is the smallest generalized eigenvalue of the pair $(Z,\widehat Y)$. This form makes the sign classification of Section \ref{sec:ycompute} immediate.

The same form also makes precise the relation between the polynomial-level margin $\vareps$ and the isotropic Gram margin, i.e.\ the largest $t$ with $Z\succeq t\,I$ for some feasible $Z$, the quantity one would naively call ``the'' $Y$-Slater margin. With the full-degree weight \eqref{eq:slack-weight} the direction $\widehat Y$ is positive \emph{definite}, and for fixed $Z$ the largest admissible $\vareps$ is the smallest eigenvalue of $\widehat Y^{-1/2}Z\,\widehat Y^{-1/2}$, bounded on both sides by Rayleigh quotients (as written for $\lambda_{\min}(Z)\ge0$; for $\lambda_{\min}(Z)<0$ the two bounds exchange):
\begin{equation*}
	\frac{\lambda_{\min}(Z)}{\lambda_{\max}(\widehat Y)}
	\;\le\;\vareps\;\le\;
	\frac{\lambda_{\min}(Z)}{\lambda_{\min}(\widehat Y)} .
\end{equation*}
Maximizing over the feasible set, $\vareps^*$ and $\tau^*_Y:=\sup\{\lambda_{\min}(Z):(Z,y)\ \text{feasible}\}$ therefore always have the same sign (the exactness statement of Appendix \ref{app:faithful}, Consequence (c)), and comparable magnitude when $\widehat Y$ is well conditioned. The ratio between them is the quantitative content of the units remark of Section \ref{sec:ycompute}, namely that $\vareps^*$ measures the same distance to the cone boundary, in units set by the weight, the bilinear basis, and the sample scalings. For the constant weight the equivalence holds in one direction only, because $\widehat Y$ is then singular; $Y$-Slater still forces a positive pointwise margin, but not conversely. The toy example of Section \ref{sec:ycompute} has pointwise margin $1$ with Gram margin $0$, and the magnitudes are not comparable.

\paragraph{Unboundedness.} The $Y$-margin problem is unbounded if there exists a recession direction that increases $\vareps$: a vector $\Delta y$, a matrix $\Delta Y\succeq0$ and $k>0$ with
\beq
	\Tr(A_p \Delta Y) + (B\,\Delta y)_p = -\,k\,\Tr(A_p \widehat Y)\qquad\forall p ,
	\label{eq:recession}
\eeq
in which case $(y,Y,\vareps)\to(y+\alpha\Delta y,\ Y+\alpha \Delta Y,\ \vareps+k\alpha)$ preserves feasibility for all $\alpha\ge0$. For linear programs the converse also holds; for SDPs it need not, since a non-polyhedral set can admit an unbounded linear objective without an improving recession direction (e.g.\ $\max t$ over $\{(s,t):s\ge t^2\}$). An unbounded $\vareps^*$ therefore shows only that the feasible set is unbounded, typically along such a direction. (Without the $B\Delta y$ term no such direction could exist, since $\sum_{p\in\mathrm{diag}}A_p=I$ for SDPB~3.1 bases, and $X_0\succ0$ otherwise, pairs positively with $\Delta Y+k\widehat Y\succ0$.) The PMP-level counterpart is a direction $\Delta y$ with $\sum_n(\Delta y)_n M^j_n(x)\succeq\omega\,\rho_j(x) I$ for some $\omega>0$, as stated in Section \ref{sec:ycompute}. The cutoff $\vareps\le0$ of the $Y$-guarantee removes the issue.

\section{Faithfulness of the PMP-to-\texorpdfstring{\SDP}{SDP} translation}
\label{app:faithful}\label{sec:faithful}

The translation itself might manufacture degeneracy, by mapping a $y$ that is strictly feasible at the polynomial level to the boundary of the Gram cone, so that the Slater failure we diagnose would be an artifact of the rewrite rather than a property of the bootstrap problem. Section \ref{sec:weightedslack} already exhibited one instance: the toy block $p(x)=1$ declared at degree $2$, strictly positive at every point of the half-line, whose unique Gram representation is nonetheless singular. There the rewrite itself creates the degeneracy, through the vanishing leading coefficient. In this appendix we show that this is the only mechanism by which it can. At finite $x$ the rewrite is faithful, and any degeneracy it creates is located at $x=\infty$ and visible in the leading coefficients at the declared degrees.

Compactify the half-line. For each block define
\beq
	\widehat M^j(\zeta;y) := (1-\zeta)^{d_j}\, M^j\Bigl(\tfrac{\zeta}{1-\zeta};\, y\Bigr),
	\qquad \zeta\in[0,1],
	\label{eq:compactified}
\eeq
a polynomial in $\zeta$ of degree $\le d_j$ whose value at $\zeta=1$ is the leading coefficient matrix of the block. By the identity $(1-\zeta)^{d}(1+x)^{d}=1$, the compactification maps the full-degree weight $\rho=(1+x)^{d}$ of \eqref{eq:slack-weight} to the constant weight in $\zeta$, so weighted margins on the half-line are uniform margins on $[0,1]$.

\paragraph{Lemma (faithfulness).} With SDPB's translation ($d_j+1$ distinct sample points per block, bilinear bases of exact degrees $\lfloor d_j/2\rfloor$ and $\lfloor (d_j-1)/2\rfloor$), where for pole-reduced \SDPs\ $M^j$ denotes the reduced PMP of the footnote in Section \ref{sec:ycompute}, i.e.\ the degree-$d_j$ interpolant of the block times $\chi_j/\tilde\chi_j$, and $d_j$ its degree:
\begin{enumerate}
	\item $y$ is \SDP-feasible if and only if $\widehat M^j(\zeta;y)\succeq0$ on $[0,1]$ for every $j$;
	\item $y$ admits a strictly feasible $Y\succ0$ if and only if $\widehat M^j(\zeta;y)\succ0$ on the \emph{closed} interval $[0,1]$ for every $j$.
\end{enumerate}
For (1): positive semidefiniteness on $[0,1)$ at $\zeta<1$ is pointwise positivity on the half-line, extended to $\zeta=1$ by continuity, and the existence of a positive-semidefinite Gram pair at exactly these degrees is the degree-exact Markov--Lukács parameterization of matrix polynomials positive semidefinite on a half-line \cite{GeninHachezNesterovVanDooren:2003}, the classical theorem that a polynomial matrix positive semidefinite on $[0,\infty)$ is $\sigma_1+x\sigma_2$ with $\sigma_i$ sums of squares of fixed degree. The degree exactness makes SDPB's rewrite an equivalence rather than a relaxation. Sampling adds nothing: evaluation at $d_j+1$ distinct points is a linear bijection on the coefficient space, and the sample scalings are positive row rescalings. For (2), one direction: if $\widehat M\succ0$ on the compact interval then $\widehat M\succeq\vareps I$, i.e.\ $M\succeq\vareps(1+x)^{d}I$ on $[0,\infty)$; subtracting a small multiple of $(\sigma_1(x)+x\,\sigma_2(x))I$, with $\sigma_i$ the sums of squares of the bilinear basis elements (positive on the half-line, of exact total degree $d$), leaves a matrix polynomial that is still positive semidefinite there, hence Gram-representable by (1). The total Gram matrix is then a small multiple of the identity plus a positive-semidefinite matrix, i.e.\ positive definite. The other direction: $Y\succ0$ gives $M(x;y)\succeq\lambda_{\min}(Y)\,(\sigma_1+x\sigma_2)\,I\succeq c\,(1+x)^{d} I$ for some $c>0$, i.e.\ $\widehat M\succeq cI$.

\paragraph{Consequence (a): at finite $x$ the rewrite is faithful.} Call a point $\bar\zeta$ of $[0,1]$ where $\widehat M^j(\bar\zeta;y)$ is singular for every feasible $y$ a \emph{forced contact}. Every forced contact at $\bar\zeta<1$ is a pointwise, polynomial-level degeneracy of the PMP. The translation can neither create nor destroy these. If the naive pointwise reading sees strict feasibility and the \SDP\ does not, the discrepancy is always located at $\zeta=1$. All of this is relative to the PMP handed to \texttt{pmp2sdp}; for pole-reduced \SDPs\ that is the reduced PMP, and the pole reduction itself, the interpolation of $P\chi/\tilde\chi$, is a separate approximation that the Lemma does not cover.

\paragraph{Consequence (b): rewrite-induced degeneracy at $\zeta=1$, from causes as simple as degree bookkeeping.} Consider the block $\mathrm{diag}(1+x^2,\,1)$, declared degree $2$. Pointwise it is $\succeq I$ uniformly on the half-line, but a positive-definite Gram matrix would imply $M\succeq c(1+x)^2 I$, contradicting the constant $(2,2)$ entry, so $Y$-Slater fails because one diagonal entry has submaximal degree. The same mechanism operates when the declared block degree exceeds the maximal degree actually attained (padding), and when the leading coefficients cancel throughout the $y$-span. These degeneracies are removed by the move (R$\infty$) of Section \ref{sec:pmpmoves} or, in the bookkeeping cases, by tightening the declared degree.

\paragraph{Consequence (c): the two slack weights measure the two notions of strictness.} The constant-weight $Y$-margin $\vareps^*_1$ ($\rho_j=1$) is the pointwise margin, and the full-degree weight \eqref{eq:slack-weight} of Section \ref{sec:ycompute} gives the compactified margin,
\begin{equation*}
	\vareps^*_1=\sup_y\,\inf_{x\ge0}\,\lambda_{\min}M(x;y),
	\qquad
	\vareps^*=\sup_y\,\inf_{\zeta\in[0,1]}\,\lambda_{\min}\widehat M(\zeta;y),
\end{equation*}
and by the Lemma the sign classification of the latter is exact: $\vareps^*>0$ if and only if $Y$-Slater holds. Comparing the two therefore separates the mechanisms. A positive constant-weight $Y$-margin with vanishing full-degree $Y$-margin is a translation-level degeneracy at infinity, while a vanishing constant-weight $Y$-margin is a pointwise degeneracy: a forced contact at finite $x$ (the case study of Section \ref{sec:casestudy} is of this kind) or a forced decay of $\lambda_{\min}M(x;y)$ as $x\to\infty$.\footnote{This faithfulness is special to univariate problems on a half-line or interval: the degree-exact Markov--Lukács decomposition has no multivariate analogue, and a multivariate PMP generalization would face a sum-of-squares gap on top of everything discussed here.}

In summary, the claim we set out to substantiate holds in a sharpened form: \emph{the degeneracy of a bootstrap \SDP\ is always a forced contact of the compactified block polynomials $\widehat M^j$ on $[0,1]$}. The minimal face is described by these contacts (their locations, null directions and orders), which are polynomial-level data located at finite $x$ or at $x=\infty$ and nowhere else.\footnote{Forced contacts come in three kinds, matching the reduction moves of Section \ref{sec:pmpmoves}: isolated contacts at finite $\bar x$ (R3), contacts at $\zeta=1$ (R$\infty$), and continua of contact, which force a polynomial identity, i.e.\ a null vector $v(x)$ depending polynomially on $x$. The constant-$v$ case is (R2), and the general case requires a polynomial-vector congruence that we do not spell out here.}

\section{Alternative \texorpdfstring{$X$}{X}-side constructions}
\label{app:xslack-elim}

The identity-based diagnostic and guarantee of Sections \ref{sec:xdiag} and \ref{sec:xguarantee} are the recommended $X$-side tools. This appendix collects alternatives: a guarantee that relaxes the linear constraints instead of the cone (Appendix \ref{app:bslack}), general reference directions $X_0\neq I$ for \SDPs\ whose bilinear bases do not make \eqref{eq:identity-rep} hold (Appendix \ref{sec:X0idea}), the lessons from implementations that fail (Appendix \ref{sec:implessons}), and the dual of the $X$-slack problem (Appendix \ref{app:duals}).

\subsection{The \texorpdfstring{$B$}{B}-slack guarantee}
\label{app:bslack}
Instead of relaxing the cone, one can relax the linear constraints. The cone part of the $X$-problem admits a strictly feasible reference $X_0=\sum_px^{(0)}_pA_p\succ0$ (Appendix \ref{sec:X0idea}; \texttt{xguarantee\_bslack} uses $x^{(0)}=\mathbf 1$ on all rows); only $B^\top x=b$ obstructs. Add a $1\times1$ block variable $s\ge0$ whose $B$-row is $b-d_{\mathrm{all}}$, $d_{\mathrm{all}}:=B^\top x^{(0)}$,
\beq
	B^\top x + (b-d_{\mathrm{all}})\,s = b ,
	\label{eq:bslack-constraint}
\eeq
so that $s=0$ is the original constraint while at $s=1$ it becomes $B^\top x=B^\top x^{(0)}$, which the reference satisfies. Keeping the original navigator $c$ and adding the penalty $c_s=\lambda_X$ gives the guarantee:
\begin{itemize}
	\item $X$-Slater: witnessed explicitly by $(x^{(0)}, s=1)$, since $B^\top x^{(0)} + (b-d_{\mathrm{all}})\cdot 1 = b$ and $X_0\succ0$.
	\item $Y$-Slater: from applying \texttt{yguarantee} to the output. In this order the slack column also enters the new $1\times1$ block, whose $Y$-constraint becomes $\lambda_X-(b-d_{\mathrm{all}})^\top y-\vareps\ge0$, so that $\vareps\to-\infty$ keeps every block strictly feasible. At $\vareps=0$ the new block adds only the half-space $(b-d_{\mathrm{all}})^\top y \le \lambda_X$ to the $Y$-side; up to off-diagonal contributions and the $O(1)$ vector $b$, this is a trace bound, as explained in Section \ref{sec:xguarantee}.
	\item Objective: $c^\top x + \lambda_X s$ on the $X$-side, $b^\top y + \lambda\vareps$ on the $Y$-side; for $\lambda,\lambda_X$ large both slacks are driven to zero and both sides approach $\vnav$. Unlike the $Y$-guarantee alone, the upper bound \eqref{eq:yg-upper} is conditional here, because the half-space restricts the $Y$-side, so that while it is binding the value can fall below $\vnav$ (Table \ref{tab:lambdaX-calibration}). Once the half-space is verified to be non-binding at the optimum, it can be dropped without changing the optimum (the problem is convex), and \eqref{eq:yg-upper} holds again; the calibration protocol below is this check.
\end{itemize}
The two modifications barely interact. The $B$-slack adds a block, i.e.\ a row of $B$ and an entry of $c$; the $Y$-guarantee then adds a column to $B$, with a single entry in that block, and an entry to $b$. There is no elimination and no pivot arithmetic on existing entries, and the original $c$ is intact (Appendix \ref{sec:implessons}).

\paragraph{Calibrating $\lambda_X$.} The trace interpretation of Section \ref{sec:xguarantee} explains the calibration. The $Y$-side half-space of the $B$-slack is $(b-d_{\mathrm{all}})^\top y\le\lambda_X$, with $d_{\mathrm{all}}=B^\top\mathbf 1$ the sum of $B$ over all constraint rows. Its entries reach $10^{17}$--$10^{19}$ in an illustrative test \SDP, while $b$ is of order one, so the half-space is, up to off-diagonal contributions, a bound on the trace of $Y$ (the exact half-space also contains the constant $c_{\mathrm{all}}$ and signed off-diagonal terms). The non-binding condition
\beq
	\lambda_X \;\ge\; \bigl|(b-d_{\mathrm{all}})^\top y^*_{\mathrm{nav}}\bigr|
	\label{eq:lambdaX-condition}
\eeq
therefore lands at $\lambda_X\gtrsim10^{16}$, heuristically the trace of the optimal Gram matrix in SDPB's scaled units, and below this threshold the optimum settles at a wrong value; Table \ref{tab:lambdaX-calibration} shows an example.
\begin{table}[htpb]
\centering
\begin{tabular}{r|c}
	\toprule
	$\lambda_X$ & $v_{\mathrm{xyg\text{-}bs}}$ \\
	\midrule
	$10^{3}$  & $-1.538 \times 10^{-3}$ \\
	$10^{12}$ & $-1.537 \times 10^{-3}$ \\
	$10^{15}$ & $-1.149 \times 10^{-3}$ \\
	$10^{17}$ & $-4.94 \times 10^{-5}$ \\
	$10^{20}$ & $-4.94 \times 10^{-5}$ \\
	\bottomrule
\end{tabular}
\caption{Calibrating $\lambda_X$: the doubly guaranteed value $v_{\mathrm{xyg\text{-}bs}}$ (\texttt{xguarantee\_bslack} followed by \texttt{yguarantee} with $\lambda=10^3$) as a function of the penalty $\lambda_X$ on a test \SDP. Below the binding threshold \eqref{eq:lambdaX-condition} the half-space constraint is active and the optimum settles at a wrong value; above it the value locks onto the true $\vnav=-4.94\times10^{-5}$.}
\label{tab:lambdaX-calibration}
\end{table}
The value is stable at the wrong place for nine orders of magnitude before crossing the binding threshold and locking on to the true value $\vnav=-4.94\times10^{-5}$. A too-small $\lambda_X$ is, however, detectable from the solve itself, with no reference value needed. Below threshold the half-space constraint is binding at the optimum, $(b-d_{\mathrm{all}})^\top y^*=\lambda_X-\bar\vareps$, i.e.\ $\lambda_X$ up to the slack, with a relaxation slack $s^*\sim10^{-19}$ on the $X$ side (against $10^{-50}$ above threshold). Since the slack column enters the new block, the binding half-space also pulls $\bar\vareps$ away from zero ($\approx-5\times10^{-7}$), so that $x_c$ vanishes ($\sim10^{-27}$) as well. Above threshold, $(b-d_{\mathrm{all}})^\top y^*=6.3\times10^{15}$ lies below $\lambda_X$ (by a factor $16$ at $\lambda_X=10^{17}$, growing with $\lambda_X$) and $x_c=17.1$. The practical protocol is therefore a posteriori: solve, then check that $\abs{(b-d_{\mathrm{all}})^\top y^*}$ sits well below $\lambda_X$ and that the cutoff multiplier $x_c$ of the $\vareps$-layer is positive \eqref{eq:xc}; if either check fails, raise the corresponding penalty and re-solve. The identity-based guarantee of Section \ref{sec:xguarantee} makes the same check exact, since its half-space is $\Tr Y\le\nu$.

Replacing the navigator objective by $c\to0$ and $c_s=1$ turns the construction into a \enquote{phase-one} problem whose optimum $s^*$ is the smallest relaxation needed for feasibility. It is zero whenever the original $X$-problem is feasible, strictly or not, so it does not diagnose the $X$-Slater condition. On the planted $X$-Slater failure of Section \ref{sec:xdiag} it returns $s^*\approx10^{-31}$, just as on the healthy problem. The diagnostic of Section \ref{sec:xdiag} replaces it.

\subsection{General reference directions \texorpdfstring{$X_0$}{X0}}
\label{sec:X0idea}
When the sampled bilinear bases are not orthonormal for the sampled measure (\SDPs\ written by SDPB versions before 3.1, or with user-supplied bases), the diagonal-row sum \eqref{eq:identity-rep} is not the identity, but any positive definite
\beq
X_0 = \sum_{p=1}^{P} x^{(0)}_p\,A_p \succ 0
\label{eq:X0-def}
\eeq
can take its place: the $X$-slack problem becomes
\beq
\label{eq:DS-X0}
\begin{aligned}
	\text{maximize}\quad & t \\
	\text{subject to}\quad & X = \sum_{p} x_p A_p, \qquad
	B^\top x = b, \qquad
	X - t\,X_0 \succeq 0,
\end{aligned}
\eeq
whose optimum is positive exactly when $X$-Slater holds, and the shift $z=x-t\,x^{(0)}$ implements it as in Section \ref{sec:xdiag}. The reference need not satisfy $B^\top x^{(0)}=b$, and since the $A_p$ depend only on the discretization data (block sizes, degrees, sample points, scalings and bilinear bases), one reference serves every member of a family of \SDPs\ that shares these data. Finding one is a linear-algebra problem. \emph{Row sums:} $x^{(0)}_p=1$ on the diagonal rows gives $X_0=\bigoplus_j (V_jV_j^{\mathsf T})\otimes I$, with the columns of $V_j$ the sampled bilinear basis vectors of block $j$; this is positive definite whenever those vectors span each block, which is exactly the condition for $\mathrm{span}\{A_p\}$ to contain any positive definite matrix, and it equals $I$ for the default bases of SDPB~3.1 and later. Setting $x^{(0)}_p=1$ on all rows, as the $B$-slack guarantee does, adds the off-diagonal matrix units $\frac12(E_{rs}+E_{sr})$ and gives $(V_jV_j^{\mathsf T})\otimes\frac12(I+J)$, with $J$ the all-ones matrix, also positive definite. \texttt{xcompute} offers both references (option \texttt{--x0}). \emph{Auxiliary \SDP:} maximizing a component of $B^\top x$ subject to $\sum_p x_pA_p\succeq0$ and a trace normalization cannot find a reference where row sums fail, but can improve the conditioning of the $t$-direction.

\subsection{Implementation lessons}
\label{sec:implessons}

Several natural implementations of the $X$-side constructions fail, and the reasons shaped the block-based designs of the main text.
\begin{itemize}
	\item \emph{Modifying $c$ endangers $Y$-Slater; modifying $b$ does not.} Changing $b$ changes only the $Y$-objective, while changing $c$ changes the right-hand side of every $Y$-constraint. Eliminating the slack variable $t$ of \eqref{eq:DS-X0} by Gaussian elimination on a pivot row of $B^\top x+t\,B^\top x^{(0)}=b$ is algebraically valid, and even stays in the PMP class, but it replaces $c$ by a column of $B$; in every case we tested the transformed problem lost its $Y$-Slater condition and SDPB stalled on the diagnostic itself. More basically, the $Y$-side of any $X$-slack problem is homogeneous up to a normalization (Appendix \ref{app:duals}), so its Slater condition is not automatic; the $\vareps$-layer of \texttt{xcompute} secures it.
	\item \emph{Every new variable needs its own block.} Treating $t$ as an extra $X$-variable with zero constraint matrix produces a zero row and column in the Schur complement (the matrix SDPB factorizes, by Cholesky, at every iteration, before the $B$ constraints are included); the factorization aborts although the full Newton system is regular. A $1\times1$ block for $t$ avoids this.
	\item \emph{Relaxing one linear equation does not restore $X$-Slater.} A slack that relaxes a single equation of $B^\top x=b$, instead of shifting $X$, leaves the other equations to be satisfied exactly by a strictly feasible point, which is the original difficulty.
	\item \emph{Safe deformations of the $Y$-feasible set are half-spaces rather than hyperplanes.} A relaxation variable without a sign constraint, as left behind by the elimination above, is dual to an equality on the $Y$-side. Relaxing $X\succeq0$ to $X\succeq-\eta X_0$ with a free, penalized $\eta$ confines $Y$ to the hyperplane $\Tr(X_0Y)=\nu$ (and to $Y=0$ at zero penalty), which is biased for every $\nu$. Giving $\eta$ its own $1\times1$ block, $\eta\ge0$, turns the equality into the inequality $\Tr(X_0Y)\le\nu$, which for $X_0=I$ is \texttt{xguarantee}, an exact penalty. All tools of this paper are built from such sign-constrained, penalized variables.
\end{itemize}
Finally, the $X$-slack problem itself is unbounded if some $\Delta x$ with $B^\top\Delta x=0$ has $\sum_p(\Delta x)_pA_p\succeq\Delta t\,X_0$ for a $\Delta t>0$. Such a direction is the strongest form of $X$-Slater and is common in navigators (the healthy navigator of Section \ref{sec:xdiag} has $t^*=+\infty$); the $\vareps$-layer of \texttt{xcompute} caps it.

\subsection{The dual of the \texorpdfstring{$X$}{X}-slack problem}
\label{app:duals}

For reference we record the Lagrangian dual of the $X$-slack problem \eqref{eq:DS}: $t^*=-\max\{b^\top y\}$ over
\beq
\label{eq:dual-DS-max}
	Y \succeq 0, \qquad
	\Tr(Y) = 1, \qquad
	B y + \bigl(\Tr(A_p Y)\bigr)_{p=1}^P = 0 ,
\eeq
a homogeneous $Y$-problem with a trace normalization, which removes the scaling freedom of the feasible cone and with it the recession directions that would otherwise drive $t\to+\infty$. By \eqref{eq:trace-linear} with $c=0$, $\Tr Y=-h\cdot y$, so the normalization is a single linear constraint on the functional and the dual is itself a PMP. The dual also makes the meaning of $t^*$ explicit: $t^*>0$ exactly when every nonzero homogeneous positive functional has $b\cdot y<0$, which is the boundedness criterion of Section \ref{sec:slaterconds}.

\section{Skydive runs in detail}
\label{app:skydive}

This appendix collects the run-level data behind the case study of Section \ref{sec:casestudy}: the iteration trajectories of the raw and the guaranteed runs, a comparison of guaranteed and raw Skydive runs at an order where both converge, and the dependence of the $\Lambda=5$ degeneracy on the truncation order.

\subsection{Trajectories of the raw and the guaranteed runs}
\label{app:skydive-traj}

Table \ref{tab:control-trajectory} is the trajectory of the raw two-parameter run at the allowed point of Section \ref{sec:symptom}, and Table \ref{tab:endtoend} the first Skydive iteration with and without the \texttt{yguarantee} transform from both starting points (Section \ref{sec:cure-isolated}).

\begin{table}[htpb]
\centering
\begin{tabular}{rccc}
	\toprule
	iter & $\mu$ & duality gap & centering $R$-err \\
	\midrule
	40  & $6.1\times10^{27}$ & $1.00$ & $1.65\times10^{30}$ \\
	200 & $1.4\times10^{-2}$ & $1.00$ & $5.13\times10^{19}$ \\
	240 & $2.0\times10^{-6}$ & $3.8\times10^{-4}$ & $1.80\times10^{19}$ \quad$\leftarrow$ local min \\
	320 & $2.4\times10^{-6}$ & $4.5\times10^{-4}$ & $6.95\times10^{24}$ \\
	400 & $2.7\times10^{-6}$ & $4.9\times10^{-4}$ & $4.23\times10^{30}$ \\
	500 & $2.7\times10^{-6}$ & $4.8\times10^{-4}$ & $8.84\times10^{36}$ \quad$\leftarrow$ maxIterations \\
	\bottomrule
\end{tabular}
\caption{Control run (no guarantee) at the allowed point: the duality gap converges to its $10^{-3}$ target while the centering residual diverges by ${\sim}17$ orders of magnitude. No navigator step is ever taken.}
\label{tab:control-trajectory}
\end{table}

\begin{table}[htpb]
\centering
\renewcommand{\arraystretch}{1.25}
\small
\begin{tabularx}{\textwidth}{l>{\raggedright\arraybackslash}X>{\raggedright\arraybackslash}X}
	\toprule
		\multicolumn{3}{c}{\emph{allowed starting point} $(0.5181,\,1.412625)$}\\
		\midrule
		& control & guaranteed ($\lambda=10^3$) \\
		duality gap & $4.8\times10^{-4}$ (fine) & $9.1\times10^{-4}$ (fine) \\
		centering $R$-err & diverges: dips to $1.8\times10^{19}$, climbs to $8.8\times10^{36}$ & $3.1\times10^{38}$ $\to$ $1.8\times10^{14}$ $\to$ $5.7\times10^3$ $\to$ $1.4\times10^{-7}$ $\to$ $\mathbf{8.7\times10^{-13}}$ \\
		reaches $10^{-10}$? & no & \textbf{yes} \\
		ext.\ step size & $0$ (no step ever) & $\mathbf{8.1\times10^{-3}}$ (real step) \\
	termination & \texttt{maxIterations} & \texttt{updateSDP} \\
	\midrule
	\multicolumn{3}{c}{\emph{disallowed starting point} $(0.51,\,2.5)$}\\
	\midrule
	& control & guaranteed ($\lambda=10^3$) \,/\, ($\lambda=10^{10}$) \\
		duality gap & $4.4\times10^{-4}$ (fine) & $6.2\times10^{-4}$ \,/\, $8.9\times10^{-4}$ (fine) \\
		centering $R$-err & diverges $\to5.9\times10^{33}$ & $\to\mathbf{4.5\times10^{-14}}$ \,/\, $\to\mathbf{2.0\times10^{-13}}$ \\
		reaches $10^{-10}$? & no & \textbf{yes} \,/\, \textbf{yes} \\
		ext.\ step size & $0$ (no step ever) & $\mathbf{4.7\times10^{-3}}$ \,/\, $\mathbf{2.7\times10^{-3}}$ (real steps) \\
		termination & \texttt{maxIterations} & \texttt{updateSDP} \,/\, \texttt{updateSDP}\\
		\bottomrule
\end{tabularx}
\caption{First Skydive iteration (centering plus one external step) with the \texttt{yguarantee} transform applied to the base and all derivative \SDPs, started from an allowed and from a disallowed point. In both cases the two-parameter raw run never takes a navigator step (it exhausts its $500$ inner iterations), while the guaranteed runs center to threshold and progress, $139$--$147$ inner iterations each; at the allowed point the guaranteed $R$-err is listed at iterations $40$, $80$, $120$, $138$ and after the final centering step. Termination \texttt{updateSDP}: centering reached its threshold and an external step was taken; \texttt{maxIterations}: the inner iteration budget ran out. At the disallowed point we also ran $\lambda=10^{10}$, above the threshold $L_*$ of Section \ref{sec:exp-survey}; $\lambda=10^3$ is below it there, which biases the value but not the centering. Classical SDPB optimization converges despite the $Y$-Slater violation. The raw runs do not depend on the slack column; the guaranteed runs use the column \eqref{eq:weighted-column}.}
\label{tab:endtoend}
\end{table}

\subsection{Comparing a guaranteed and a raw Skydive run}
\label{sec:skydive-ab}
When the original problem has no $Y$-margin the comparison is simple and discrete, since the \texttt{yguarantee} run converges while the raw run fails at the first point (Section \ref{sec:casestudy}). Here we compare what happens when the $Y$-Slater condition already holds and both runs converge. We use three otherwise identical Skydive runs on the three-parameter $(\De_\s,\De_\e,x_{\mathrm{ope}})$ Ising navigator at $\Lambda=11$: one without the \texttt{yguarantee} transform, and two with it, at $\lambda=10^3$ and $10^{10}$. These are single runs, so the differences between them are indicative only, and we take all three terminal points to lie in the same flat basin. Table \ref{tab:skydive-ab} collects the data. The raw and the $\lambda=10^3$ trajectories end within $3\times10^{-3}$ of each other in parameter space, where the navigator is flat at the $10^{-6}$ level, at nearly the same cost: four extra external steps and $13\%$ more inner iterations, at the same solver time per inner iteration. The $\lambda=10^{10}$ run ends about $0.02$ away, at a lower navigator value, and costs more, with $16\%$ more external steps and $33\%$ more inner iterations than the raw run.

\begin{table}[htpb]
\centering
\renewcommand{\arraystretch}{1.15}
\begin{tabular}{lccc}
\toprule
 & raw & \multicolumn{2}{c}{guaranteed} \\
 & & $\lambda=10^3$ & $\lambda=10^{10}$ \\
\midrule
external steps to $|\nabla \vnav|<10^{-4}$ & $61$ & $65$ & $71$ \\
inner (\texttt{dynamical\_sdp}) iterations$^*$ & $2083$ & $2356$ & $2767$ \\
inner iterations per external step & $34.1$ & $36.2$ & $39.0$ \\
solver time per inner iteration & $0.100\,$s & $0.097\,$s & $0.081\,$s \\
terminal navigator & $1.21\times10^{-6}$ & $1.30\times10^{-6}$ & $6.55\times10^{-7}$ \\
terminal $|\nabla \vnav|$ & $9.6\times10^{-5}$ & $9.8\times10^{-5}$ & $9.2\times10^{-5}$ \\
terminal $\De_\s$ & $0.53172$ & $0.53118$ & $0.53655$ \\
terminal $\De_\e$ & $1.48983$ & $1.48762$ & $1.50852$ \\
terminal $x_{\mathrm{ope}}$ & $1.52452$ & $1.52296$ & $1.53599$ \\
\bottomrule
\end{tabular}
\caption{Comparison of three otherwise identical Skydive runs on the three-parameter $\Lambda=11$ Ising navigator, without (raw) and with (guaranteed, $\lambda=10^3$ and $10^{10}$, column \eqref{eq:weighted-column}) the \texttt{yguarantee} transform, all from the disallowed start $(0.52,\,1.5,\,1.1)$ with gradient goal $10^{-4}$. $^*$Summed over the final solve at each external point, from the solver's own iteration logs, for all three runs.}
\label{tab:skydive-ab}
\end{table}

\paragraph{The transform's bias against the other error scales.} The terminal navigator values of the first two runs differ by $8\times10^{-8}$, but the runs also end at different points of a basin in which the navigator varies at the $10^{-6}$ level, so this difference does not isolate the transform's bias. We therefore audited the $\lambda=10^3$ run at its own terminal point: tight solves (duality gap ${\sim}6\times10^{-31}$) of the raw navigator \SDP\ and of its $Y$-guaranteed versions at $\lambda=10^3$, $10^6$ and $10^{10}$ return the same optimum to all $24$ printed digits, with terminal slacks $\bar\vareps=-7.3\times10^{-37}$, $-6.1\times10^{-40}$ and $-7.2\times10^{-44}$. The cutoff multiplier confirms this in each run: $x_c=998.93$ at $\lambda=10^3$, and $x_c=\lambda$ to the printed digits at $10^6$ and $10^{10}$, i.e.\ $\langle X,\widehat Y\rangle\approx1.07$ independently of $\lambda$ (this problem satisfies $Y$-Slater, so its $X$-optimum is attained and the optimal set is bounded), and hence $L_*\le1.07$. The value $\lambda=10^3$ is far above $L_*$, and the bias \eqref{eq:yg-bias} of the guaranteed search vanishes. The difference between the trajectories is the ordinary path dependence of a BFGS search in a nearly flat basin; the third run, at $\lambda=10^{10}$, also far above $L_*$, ends at yet another point.

\subsection{Dependence on the truncation order}
\label{app:skydive-lambda}

The exposing certificates of Section \ref{sec:faceworked} are confined to a subset of the crossing equations and dominated by the highest kept spin, which suggests a dependence on the truncation order. Regenerating the same formulation at $\Lambda=9,11,15,19$ and rerunning \texttt{ycompute} at the same allowed point $(0.5181,\,1.412625)$ gives strictly positive $Y$-margins $\vareps^* = 6.4\times10^{-15},\ 1.7\times10^{-16},\ 1.6\times10^{-20},\ 8.4\times10^{-25}$ respectively, each $16$--$26$ orders of magnitude above the duality gap of ${\sim}5\times10^{-41}$. The values are not comparable across $\Lambda$, since the units of the slack column change with the block degrees (Section \ref{sec:ycompute}). Skydive's centering succeeds natively at all four orders. The face is therefore a $\Lambda=5$ truncation artifact. Such degeneracies can appear and disappear as truncation parameters vary.

\section{Termination settings in the published literature}
\label{app:survey}

Section \ref{sec:feasverdicts} distinguishes four ways of ending a feasibility-mode run: by a \emph{threshold} on the primal or dual error (\texttt{--findPrimalFeasible}, \texttt{--findDualFeasible}), by a \emph{jump} (\texttt{--detectPrimalFeasibleJump}, \texttt{--detectDualFeasibleJump}), by a \emph{race} of all four, or by a \emph{hybrid} with a threshold on one side and a jump on the other. This appendix records which published works used which. The candidate list is every paper that INSPIRE lists as citing the SDPB paper \cite{Simmons-Duffin:2015qma}, its successor \cite{Landry:2019qug}, the navigator paper \cite{Reehorst:2021ykw} or the Skydive paper \cite{Liu:2023elz}: 436 arXiv papers as of 28 October 2026, across the conformal, S-matrix, quantum-mechanical and matrix bootstraps. We searched the source of each for the option names and for the vocabulary of termination (feasible jump, step length, primal and dual error, hot start, instability) and read every hit in context, together with parameter tables, ancillary notebooks and the public code where the paper defers to it; a work that cites another paper's parameter table inherits that paper's classification. Table \ref{tab:survey} lists the works that report feasibility verdicts and state, or inherit, their termination flags, grouped by mode, each labelled by a few words of its title, together with the works that report trouble with jumps or convergence.

Two facts about the defaults are needed to read the table. First, no released version of SDPB, from 1.0 to 3.1 \cite{SDPB-repo}, enables any of the four flags by default, and both thresholds default to $10^{-30}$; a paper that switches on the two threshold flags and keeps ``default values for the other parameters'' therefore runs in threshold mode. Second, the common front ends enable the threshold flags on their own: PyCFTBoot \cite{Behan:2016dtz} passes \texttt{--findPrimalFeasible --findDualFeasible} unconditionally and reads its verdict from the threshold messages, simpleboot \cite{simpleboot} passes the same two flags in its default option string and its result parsers recognize only the threshold messages, and the examples shipped with cboot \cite{cboot} pass the same two flags. A user of these front ends who does not override the options is in threshold mode without having chosen it. One further fact bears on the primal threshold specifically. SDPB~1.x computed the primal error without the term $|b-B^{\mathsf T}x|$, an omission corrected in SDPB~2.0 \cite{Landry:2019qug}, so threshold-based ``found primal feasible solution'' verdicts issued by SDPB~1.x rest on an incomplete residual; \cite{Go:2019lke} relate their observation that primal-threshold verdicts are unreliable under hot starts to this.

{\footnotesize
\setlength{\tabcolsep}{4pt}
\captionsetup{justification=raggedright,singlelinecheck=false}
\begin{longtable}{@{}>{\raggedright\arraybackslash}p{2.7cm}>{\raggedright\arraybackslash}p{1.6cm}>{\raggedright\arraybackslash}p{1.1cm}>{\raggedright\arraybackslash}p{1.1cm}>{\raggedright\arraybackslash}p{2.45cm}>{\raggedright\arraybackslash}p{4.9cm}@{}}
\caption{SDPB termination settings of published feasibility-mode bootstrap works, grouped by termination mode (Section \ref{sec:feasverdicts}). The flag columns are \texttt{findPrimalFeasible}/\texttt{findDualFeasible} and \texttt{detectPrimalFeasibleJump}/\texttt{detectDualFeasibleJump}: ``on/on'' means both enabled, ``as [$n$]'' means the settings are inherited by citation from reference [$n$], a superscript f means the flag is supplied by the front end (PyCFTBoot, simpleboot or cboot) rather than stated in the paper, and a dash means not stated. Thresholds are primal/dual, a single value when both are equal. The survey file with verbatim quotations and source locations is an ancillary file to this submission and is also included with the reproduction material.}
\label{tab:survey}\\
\toprule
Reference & Stated in & find P/D & jump P/D & Thresholds & Remarks \\
\midrule
\endfirsthead
\toprule
Reference & Stated in & find P/D & jump P/D & Thresholds & Remarks \\
\midrule
\endhead
\bottomrule
\endlastfoot
\multicolumn{6}{@{}l}{\emph{Race: thresholds and jumps, whichever fires first}}\\*[2pt]
SDPB manual \cite{Simmons-Duffin:2015qma} & Table 1 & on/on & on/on & $10^{-30}$--$10^{-75}$ (by $\Lambda$) & The settings that later works inherit. \\
$O(N)$ archipelago \cite{Kos:2015mba} & appendix table & on/on & on/on & $10^{-30}$--$10^{-70}$ (by $\Lambda$) & \\
3d fermions \cite{Iliesiu:2015qra} & appendix table & on/on & on/on & $10^{-25}$--$10^{-100}$ / $10^{-25}$--$10^{-40}$ & \\
Accidental symmetries \cite{Chester:2015qca} & text & \multicolumn{2}{l}{as \cite{Simmons-Duffin:2015qma}} & & ``The parameters specified in the first column of Table 1 in the SDPB manual.'' \\
Precision islands \cite{Kos:2016ysd} & App.\ A & \multicolumn{2}{l}{as \cite{Kos:2015mba}} & & \\
TASI lectures \cite{Simmons-Duffin:2016gjk} & footnote & \multicolumn{2}{l}{as \cite{Simmons-Duffin:2015qma}} & & \\
Continuous global symmetry \cite{Dymarsky:2017xzb} & appendix table & on/on & on/on & $10^{-30}$--$10^{-70}$ (by $\Lambda$) & \\
Mixed current--scalar \cite{Reehorst:2019pzi} & text & \multicolumn{2}{l}{as \cite{Dymarsky:2017xzb}} & & No parameter table of its own; see Section \ref{sec:mixedcurrent}. \\
2d Ising mixed \cite{delaFuente:2019hbl} & App.\ A & on/on & on/on & $10^{-75}$ / $10^{-30}$ & Gap runs; flags of the OPE runs not listed. \\
Coulomb and Higgs branch \cite{Gimenez-Grau:2020jrx} & App.\ B & on/on & on/on & -- & Reports that in practice every run ended with a jump. \\
Weizmann lectures \cite{Chester:2019wfx} & text & on/on & on/on & $10^{-10}$ / $10^{-5}$ & Recommends all four flags ``to rule out points as fast as possible.'' \\
Traceless symmetric $O(N)$ \cite{Reehorst:2020phk} & Table 2 & on/on & --/on & $10^{-60}$ & Primal jump flag not listed. \\
Points and lines \cite{Meineri:2025qwz} & appendix table & on/on & on/on & $10^{-30}$ & Reports a region where instabilities prevent any jump. \\
Duality-inspired fusion rule \cite{Nakayama:2025mrm} & appendix & on/on & on/on & $10^{-60}$ / $10^{-10}$ & Threshold flags added because the dual jump does not occur in part of the parameter space; hot start disabled. \\
3d conformal manifold \cite{Baggio:2017mas} & text & as \cite{Simmons-Duffin:2015qma} & as \cite{Simmons-Duffin:2015qma} & as \cite{Simmons-Duffin:2015qma} & ``The first column of Table 1 in the SDPB manual.'' \\
4d fermions \cite{Karateev:2019pvw} & appendix table & on/on & on/on & $10^{-30}$ / $10^{-20}$ & Dimension bounds; the OPE bounds have all four off. \\
$U(m)\times U(n)$ \cite{Kousvos:2022ewl} & text & on/on\textsuperscript{f} & on/on & -- & PyCFTBoot with the two jump flags added; the threshold flags come from the front end. \\
3-state Potts \cite{Chester:2022hzt} & table & on/on\textsuperscript{f} & on/on & $10^{-60}$ & simpleboot; feasibility column. Navigator column has the jumps off. \\
Non-abelian currents \cite{He:2023ewx} & text & on/on\textsuperscript{f} & on/on & $10^{-25}$ / $10^{-20}$ & simpleboot with both jump flags added. \\
Replica Potts \cite{Kousvos:2024dlz} & table & on/on\textsuperscript{f} & on/on & $10^{-60}$ & simpleboot with both jump flags on. \\
Redundancy channels \cite{Kousvos:2025ext} & text & on/on\textsuperscript{f} & on/on & -- & simpleboot with both jump flags on. \\
\midrule
\multicolumn{6}{@{}l}{\emph{Threshold only}}\\*[2pt]
PyCFTBoot \cite{Behan:2016dtz} & text and code & on/on & off/off & $10^{-30}$ / $10^{-15}$ & Front-end default; ``allowed CFTs return \emph{found primal feasible solution} well before this.'' \\
3d stress tensor \cite{Dymarsky:2017yzx} & table and text & on/on & off/off & $10^{-30}$ & Exclusion plot only; the bound runs have all four flags off. \\
Hypercubic \cite{Stergiou:2018gjj} & text & on/on & off/off & $10^{-30}$ / $10^{-20}$ & ``Default values for other parameters.'' \\
Cubic mixed \cite{Kousvos:2018rhl} & text & on/on & off/off & $10^{-30}$ / $10^{-20}$ & Same settings. \\
MN and tetragonal \cite{Stergiou:2019dcv} & text & on/on & off/off & $10^{-30}$ / $10^{-20}$ & Same settings. \\
Minimal 3d SCFT \cite{Atanasov:2018kqw} & App.\ B and code & on/on & off/off & $10^{-120}$ / $10^{-10}$ & Flags from the public code of the paper. \\
Multi-fixed-point \cite{Dowens:2020cua} & text & on/on & off/off & $10^{-30}$ / $10^{-15}$ & ``A dual feasible solution is eventually found, in which case we say that it is disallowed.'' \\
$U(1)$ atlas \cite{Bartlett-Tisdall:2024mbx} & App.\ B & -- & -- & $10^{-30}$ / $10^{-20}$ & Flags not printed. Dual jumps never observed; exclusions read from the creeping dual error, allowed points from primal jumps. \\
Pure gravity in AdS$_3$ \cite{Bae:2016yna} & text & --/on & -- & -- / $10^{-200}$ & Terminates when a functional is found ``to sufficient precision set to $10^{-200}$''; failure to find one is called inconclusive. \\
Ising in varying dimension \cite{Cappelli:2018vir} & table & on/on & off/off & $10^{-30}$ & \\
$O(m)\times O(n)$ \cite{Henriksson:2020fqi} & text & on/on & off/off & $10^{-30}$ / $10^{-15}$ & ``Default values are used for other parameters of the solver.'' \\
Free partition functions \cite{Afkhami-Jeddi:2020ezh} & table & off/on & off/off & $10^{-100}$ & Dual threshold only; the primal side is never detected. \\
Bootstrap deformations \cite{Afkhami-Jeddi:2021iuw} & ancillary notebook & on/on & off/off & $10^{-30}$ & Verdict read from the threshold messages. \\
$\epsilon$-continuation of Ising \cite{Henriksson:2022gpa} & table & on/on & -- & $10^{-30}$ & Island checks; reports a primal-side numerical instability. \\
Front-end users \cite{Gowdigere:2018lxz,Rong:2020gbi,Nakayama:2023wrx,Nakayama:2024iiw} & -- & on/on\textsuperscript{f} & off/off\textsuperscript{f} & -- & No settings printed; PyCFTBoot, simpleboot or cboot with their defaults. \\
\midrule
\multicolumn{6}{@{}l}{\emph{Hybrid: dual threshold, primal jump}}\\*[2pt]
autoboot \cite{Go:2019lke} & \S4.3 & off/on & on/off & -- & Found primal-threshold verdicts unreliable under hot starts. \\
qboot thesis \cite{Go:2020ahx} & text & off/on & on/off & -- & Same recommendation; also reports runs in which $\mu$ ``mysteriously starts to increase'' while both step lengths become very small. \\
$SU(N)$ adjoint \cite{Manenti:2021elk} & table & on/on & --/on & $10^{-60}$ & Primal jump flag not listed. \\
$\calN=4$ mixed correlators \cite{Bissi:2020jve} & table & on/on & --/on & -- & Primal jump flag not listed; SDPB~1. \\
Boundary QED \cite{Bartlett-Tisdall:2023ghh} & text and table & -- & -- & $10^{-30}$ / $10^{-20}$ & As \cite{Bartlett-Tisdall:2024mbx}: ``never able to see dual jumps,'' exclusions from the creeping dual error, allowed points from primal jumps. \\
\midrule
\multicolumn{6}{@{}l}{\emph{Jumps only, thresholds out of reach}}\\*[2pt]
$O(2)$ model \cite{Chester:2019ifh} & Table 5 & off/off & on/on & $10^{-200}$ & Explicit recommendation: ``only primal/dual jumps are good reasons to terminate.'' Dual jumps absent without a small twist gap (Section \ref{sec:twistgap}). \\
$O(3)$ model \cite{Chester:2020iyt} & Table 4 & -- & -- & $10^{-200}$ & Flags not printed; methods of \cite{Chester:2019ifh}. \\
$N_f=4$ QED$_3$ \cite{Albayrak:2021xtd} & appendix table & off/off & on/on & $10^{-200}$ & \\
Simplest DQCP \cite{Chester:2025uxb} & Table V & off/off & on/on & $10^{-60}$ & \\
Product functionals \cite{Ghosh:2026dse} & appendix & off/off & on/on & $10^{-200}$ & Thresholds set low ``to avoid premature classification near the boundary.'' \\
GNY archipelago \cite{Erramilli:2020rlr} & table & off/off & on/on & $10^{-200}$ & \\
Hyperbolic surfaces \cite{Bonifacio:2021aqf} & table & off/off & on/on & $10^{-100}$ & \\
Hyperbolic manifolds \cite{Bonifacio:2021msa} & ancillary notebook & off/off & on/on & $10^{-100}$ & Discusses that threshold-limited verdicts have no controlled error. \\
Abelian currents in 4d \cite{Karateev:2025sjw} & table & off/off & on/on & $10^{-30}$ & \\
Stress-tensor cartography \cite{Erramilli:2026ykj} & table & off/off & on/on & $10^{-200}$ & Higher precision needed for part of one scan. \\
\midrule
\multicolumn{6}{@{}l}{\emph{Flags not stated}}\\*[2pt]
3d Ising stress tensor \cite{Chang:2024whx} & text & -- & -- & -- & Jump-oriented workflow; reports that a first setup never produced a dual jump, traced to three-point structures linearly dependent at some $\De$ (including $\De=\infty$) and cured by a basis independent at every $\De$. \\
Endpoint bootstrap \cite{Lanzetta:2025xfw} & text & -- & -- & -- & Points ``ruled out'' when no allowed value is found within 75 iterations, attributed to hot-starting. \\
Quantum spin systems \cite{Nancarrow:2022wdr} & text & -- & -- & -- & The optimal functional is exactly semidefinite, so the dual jump detector ``will never work''; cured by an $\epsilon$-shift of the diagonal (a $Y$-margin in our terms), which fails in their two-variable setup. \\
Geometry bootstrap \cite{Bonifacio:2020xoc} & text & -- & -- & -- & ``It was necessary to disallow the program from terminating due to detection of a primal feasible solution.'' \\
Level repulsion in $\calN=4$ \cite{Chester:2023ehi} & table & -- & -- & $10^{-30}$ & Fine grids with nearly redundant constraints cause SDPB ``to stall, with the step sizes going to zero.'' \\
ANEC bootstrap \cite{Manenti:2019kbl} & text & -- & -- & -- & Infeasibility decided by \texttt{maxComplementarity}; the authors note such exclusions are not rigorous. \\
Confinement in AdS \cite{DiPietro:2025ozw} & text & -- & -- & -- & A modified SDPB that ``recovers from a `stalling' phase by aggressively increasing the $\beta$ parameter''; hot starts in optimization mode ``usually impossible.'' \\
Non-identical scalars in 4d \cite{Ferretti:2026iij} & text & -- & -- & $10^{-30}$ & S-matrix bootstrap. Near-zero eigenvalues ``can lead to instabilities where the algorithm never manages to achieve a dual jump''; after a diagonal rescaling ``the SDPB solver always has both primal and dual jumps.'' \\
Heterotic string \cite{Chiang:2026ubv} & text & -- & -- & -- & S-matrix bootstrap. Leading coefficients that positivity forces ``to exactly vanish for every functional'' are solved for exactly and the functional dimension reduced before SDPB ``to prevent numerical instability'': facial reduction by hand. Infeasibility read from \texttt{maxComplementarity exceeded}. \\
\end{longtable}
}

The census behind the table is in the ancillary survey file. Of the 436 papers, about 160 cite SDPB without running it (reviews, analytic work, and bootstraps solved with SDPA-GMP, MOSEK, CVXPY or custom solvers, which is the rule in the quantum-mechanical and matrix bootstrap), about 125 run it only in optimization mode, where termination is by duality gap and none of the four flags enters (every S-matrix bootstrap paper that runs SDPB is in this group), and about 95 report feasibility verdicts without stating the settings that decided them, against some 50 that state or inherit them. The table supports the statements of Section \ref{sec:feasverdicts} in a specific form: threshold termination entered the literature through the race settings of the SDPB manual and through the front-end defaults, and where a work explains why it kept a threshold, the reason is that in some region a dual jump did not occur.

The degeneracies of this paper are not confined to the conformal bootstrap. In the S-matrix bootstrap, \cite{Chiang:2026ubv} find leading coefficients that positivity forces to vanish for every functional and reduce the functional space by hand before calling SDPB, \cite{Ferretti:2026iij} recover the dual jumps by a rescaling, and \cite{Correia:2022dyp} verify Slater's condition by constructing an interior point. In the matrix and quantum-mechanical bootstrap, \cite{Lin:2025srf} trace their numerical instability to ``the lack of strict feasibility in the primal form of the SDP,'' \cite{Lawrence:2024mnj} identify the affine constraints implied by the inequalities before running an interior-point method, which is facial reduction, \cite{Nancarrow:2022wdr} shift the diagonal by an $\epsilon$, which is the $Y$-margin of Section \ref{sec:ycompute} used as a cure at fixed $\vareps$, \cite{Li:2025tub} find a navigator that is ``always negative'' at double precision and relax the criterion by a tolerance, and \cite{Kull:2022wof} obtain a rigorous bound from an unconverged iterate by shifting the solver's dual variable to strict feasibility.

\bibliographystyle{utphys}
\bibliography{refs}

\end{document}

%% file: shorthand.tex
\def\vareps{\varepsilon}
\def\De{\Delta}
\def\s{\sigma}
\def\e{\epsilon}

\def\bR{\mathbb{R}}
\def\bS{\mathbb{S}}
\def\bZ{\mathbb{Z}}
\def\calF{{\cal F}}
\def\calN{{\cal N}}

\def\calR{{\cal R}}

\newcommand{\beq}{\begin{equation}}
\newcommand{\eeq}{\end{equation}}

\def\ge{\geqslant}
\def\le{\leqslant}

\newcommand{\Tr}{\operatorname{Tr}}
\newcommand{\SDP}{\ensuremath{\overline{\mathrm{SDP}}}}
\newcommand{\SDPs}{\SDP{}s}
\newcommand{\vnav}{v_{\mathrm{Nav}}}
\DeclarePairedDelimiter{\abs}{\lvert}{\rvert}

%% file: main.bbl
\providecommand{\href}[2]{#2}\begingroup\raggedright\begin{thebibliography}{100}

\bibitem{Poland:2018epd}
D.~Poland, S.~Rychkov, and A.~Vichi, ``{The Conformal Bootstrap: Theory,
  Numerical Techniques, and Applications},''
  \href{http://dx.doi.org/10.1103/RevModPhys.91.015002}{{\em Rev. Mod. Phys.}
  {\bfseries 91} (2019) 015002},
  \href{http://arxiv.org/abs/1805.04405}{{\ttfamily arXiv:1805.04405
  [hep-th]}}.

\bibitem{Poland:2022qrs}
D.~Poland and D.~Simmons-Duffin, ``{Snowmass White Paper: The Numerical
  Conformal Bootstrap},'' in {\em {Snowmass 2021}}.
\newblock 3, 2022.
\newblock \href{http://arxiv.org/abs/2203.08117}{{\ttfamily arXiv:2203.08117
  [hep-th]}}.

\bibitem{Rychkov:2023wsd}
S.~Rychkov and N.~Su, ``{New developments in the numerical conformal
  bootstrap},'' \href{http://dx.doi.org/10.1103/RevModPhys.96.045004}{{\em Rev.
  Mod. Phys.} {\bfseries 96} no.~4, (2024) 045004},
  \href{http://arxiv.org/abs/2311.15844}{{\ttfamily arXiv:2311.15844
  [hep-th]}}.

\bibitem{Rattazzi:2008pe}
R.~Rattazzi, V.~S. Rychkov, E.~Tonni, and A.~Vichi, ``{Bounding scalar operator
  dimensions in 4D CFT},''
  \href{http://dx.doi.org/10.1088/1126-6708/2008/12/031}{{\em JHEP} {\bfseries
  12} (2008) 031}, \href{http://arxiv.org/abs/0807.0004}{{\ttfamily
  arXiv:0807.0004 [hep-th]}}.

\bibitem{ElShowk:2012ht}
S.~El-Showk, M.~F. Paulos, D.~Poland, S.~Rychkov, D.~Simmons-Duffin, and
  A.~Vichi, ``{Solving the 3D Ising Model with the Conformal Bootstrap},''
  \href{http://dx.doi.org/10.1103/PhysRevD.86.025022}{{\em Phys. Rev. D}
  {\bfseries 86} (2012) 025022},
  \href{http://arxiv.org/abs/1203.6064}{{\ttfamily arXiv:1203.6064 [hep-th]}}.

\bibitem{Kos:2014bka}
F.~Kos, D.~Poland, and D.~Simmons-Duffin, ``{Bootstrapping Mixed Correlators in
  the 3D Ising Model},'' \href{http://dx.doi.org/10.1007/JHEP11(2014)109}{{\em
  JHEP} {\bfseries 11} (2014) 109},
  \href{http://arxiv.org/abs/1406.4858}{{\ttfamily arXiv:1406.4858 [hep-th]}}.

\bibitem{Kos:2016ysd}
F.~Kos, D.~Poland, D.~Simmons-Duffin, and A.~Vichi, ``{Precision Islands in the
  Ising and $O(N)$ Models},''
  \href{http://dx.doi.org/10.1007/JHEP08(2016)036}{{\em JHEP} {\bfseries 08}
  (2016) 036}, \href{http://arxiv.org/abs/1603.04436}{{\ttfamily
  arXiv:1603.04436 [hep-th]}}.

\bibitem{Simmons-Duffin:2015qma}
D.~Simmons-Duffin, ``{A Semidefinite Program Solver for the Conformal
  Bootstrap},'' \href{http://dx.doi.org/10.1007/JHEP06(2015)174}{{\em JHEP}
  {\bfseries 06} (2015) 174}, \href{http://arxiv.org/abs/1502.02033}{{\ttfamily
  arXiv:1502.02033 [hep-th]}}.

\bibitem{Landry:2019qug}
W.~Landry and D.~Simmons-Duffin, ``{Scaling the semidefinite program solver
  SDPB},'' \href{http://arxiv.org/abs/1909.09745}{{\ttfamily arXiv:1909.09745
  [hep-th]}}.

\bibitem{Reehorst:2021ykw}
M.~Reehorst, S.~Rychkov, D.~Simmons-Duffin, B.~Sirois, N.~Su, and B.~C. van
  Rees, ``{Navigator Function for the Conformal Bootstrap},''
  \href{http://dx.doi.org/10.21468/SciPostPhys.11.3.072}{{\em SciPost Phys.}
  {\bfseries 11} (2021) 072}, \href{http://arxiv.org/abs/2104.09518}{{\ttfamily
  arXiv:2104.09518 [hep-th]}}.

\bibitem{Liu:2023elz}
A.~Liu, D.~Simmons-Duffin, N.~Su, and B.~C. van Rees, ``{Skydiving to Bootstrap
  Islands},'' \href{http://dx.doi.org/10.1007/JHEP11(2025)003}{{\em JHEP}
  {\bfseries 11} (2025) 003}, \href{http://arxiv.org/abs/2307.13046}{{\ttfamily
  arXiv:2307.13046 [hep-th]}}.

\bibitem{Chester:2019ifh}
S.~M. Chester, W.~Landry, J.~Liu, D.~Poland, D.~Simmons-Duffin, N.~Su, and
  A.~Vichi, ``{Carving out OPE space and precise $O(2)$ model critical
  exponents},'' \href{http://dx.doi.org/10.1007/JHEP06(2020)142}{{\em JHEP}
  {\bfseries 06} (2020) 142}, \href{http://arxiv.org/abs/1912.03324}{{\ttfamily
  arXiv:1912.03324 [hep-th]}}.

\bibitem{Bartlett-Tisdall:2023ghh}
S.~Bartlett-Tisdall, C.~P. Herzog, and V.~Schaub, ``{Bootstrapping boundary
  QED. Part I},'' \href{http://dx.doi.org/10.1007/JHEP05(2024)235}{{\em JHEP}
  {\bfseries 05} (2024) 235}, \href{http://arxiv.org/abs/2312.07692}{{\ttfamily
  arXiv:2312.07692 [hep-th]}}.

\bibitem{Bartlett-Tisdall:2024mbx}
S.~Bartlett-Tisdall, C.~P. Herzog, and V.~Schaub, ``{An Atlas for 3d conformal
  field theories with a U(1) global symmetry},''
  \href{http://dx.doi.org/10.1007/JHEP06(2025)237}{{\em JHEP} {\bfseries 06}
  (2025) 237}, \href{http://arxiv.org/abs/2412.01608}{{\ttfamily
  arXiv:2412.01608 [hep-th]}}.

\bibitem{Nancarrow:2022wdr}
C.~O. Nancarrow and Y.~Xin, ``{Bootstrapping the gap in quantum spin
  systems},'' \href{http://dx.doi.org/10.1007/JHEP08(2023)052}{{\em JHEP}
  {\bfseries 08} (2023) 052}, \href{http://arxiv.org/abs/2211.03819}{{\ttfamily
  arXiv:2211.03819 [hep-th]}}.

\bibitem{Ferretti:2026iij}
G.~Ferretti, D.~Karateev, A.~Piazza, and M.~Serone, ``{Bounds on scattering
  amplitudes of non-identical scalar particles in 4d},''
  \href{http://arxiv.org/abs/2609.09282}{{\ttfamily arXiv:2609.09282
  [hep-th]}}.

\bibitem{Meineri:2025qwz}
M.~Meineri and B.~Radhakrishnan, ``{The bootstrap of points and lines},''
  \href{http://dx.doi.org/10.1007/JHEP04(2026)040}{{\em JHEP} {\bfseries 04}
  (2026) 040}, \href{http://arxiv.org/abs/2506.17382}{{\ttfamily
  arXiv:2506.17382 [hep-th]}}.

\bibitem{Nakayama:2025mrm}
Y.~Nakayama and T.~Onagi, ``{Conformal bootstrap with duality-inspired fusion
  rule},'' \href{http://dx.doi.org/10.1103/h2t4-bb53}{{\em Phys. Rev. D}
  {\bfseries 113} no.~10, (2026) L101701},
  \href{http://arxiv.org/abs/2511.00386}{{\ttfamily arXiv:2511.00386
  [hep-th]}}.

\bibitem{Go:2019lke}
M.~Go and Y.~Tachikawa, ``{autoboot: A generator of bootstrap equations with
  global symmetry},'' \href{http://dx.doi.org/10.1007/JHEP06(2019)084}{{\em
  JHEP} {\bfseries 06} (2019) 084},
  \href{http://arxiv.org/abs/1903.10522}{{\ttfamily arXiv:1903.10522
  [hep-th]}}.

\bibitem{Chiang:2026ubv}
L.-Y. Chiang, Y.-t. Huang, and H.-C. Weng, ``{Cornering the Heterotic
  String},'' \href{http://arxiv.org/abs/2610.12318}{{\ttfamily arXiv:2610.12318
  [hep-th]}}.

\bibitem{Lin:2025srf}
H.~W. Lin and Z.~Zheng, ``{High-Precision Bootstrap of Multimatrix Quantum
  Mechanics},'' \href{http://dx.doi.org/10.1103/cyq8-4sd7}{{\em Phys. Rev.
  Lett.} {\bfseries 136} no.~18, (2026) 181603},
  \href{http://arxiv.org/abs/2507.21007}{{\ttfamily arXiv:2507.21007
  [hep-th]}}.

\bibitem{Lawrence:2024mnj}
S.~Lawrence, B.~McPeak, and D.~Neill, ``{Bootstrapping time-evolution in
  quantum mechanics},'' \href{http://arxiv.org/abs/2412.08721}{{\ttfamily
  arXiv:2412.08721 [hep-th]}}.

\bibitem{BorweinWolkowicz:1981}
J.~M. Borwein and H.~Wolkowicz, ``{Facial reduction for a cone-convex
  programming problem},''
  \href{http://dx.doi.org/10.1017/S1446788700017250}{{\em J. Austral. Math.
  Soc. Ser. A} {\bfseries 30} (1981) 369--380}.

\bibitem{BorweinWolkowicz:1981b}
J.~M. Borwein and H.~Wolkowicz, ``{Regularizing the abstract convex program},''
  \href{http://dx.doi.org/10.1016/0022-247X(81)90138-4}{{\em J. Math. Anal.
  Appl.} {\bfseries 83} (1981) 495--530}.

\bibitem{WakiMuramatsu:2013}
H.~Waki and M.~Muramatsu, ``{Facial Reduction Algorithms for Conic Optimization
  Problems},'' \href{http://dx.doi.org/10.1007/s10957-012-0219-y}{{\em J.
  Optim. Theory Appl.} {\bfseries 158} (2013) 188--215}.

\bibitem{Pataki:2013}
G.~Pataki, \href{http://dx.doi.org/10.1007/978-1-4614-7621-4_28}{``{Strong
  Duality in Conic Linear Programming: Facial Reduction and Extended Duals},''}
  in {\em Computational and Analytical Mathematics}, vol.~50 of {\em Springer
  Proceedings in Mathematics \& Statistics}, pp.~613--634.
\newblock Springer, 2013.

\bibitem{DrusvyatskiyWolkowicz:2017}
D.~Drusvyatskiy and H.~Wolkowicz, ``{The Many Faces of Degeneracy in Conic
  Optimization},'' \href{http://dx.doi.org/10.1561/2400000011}{{\em Found.
  Trends Optim.} {\bfseries 3} no.~2, (2017) 77--170}.

\bibitem{PermenterParrilo:2018}
F.~Permenter and P.~A. Parrilo, ``{Partial facial reduction: simplified,
  equivalent SDPs via approximations of the PSD cone},''
  \href{http://dx.doi.org/10.1007/s10107-017-1169-9}{{\em Math. Program.}
  {\bfseries 171} (2018) 1--54}.

\bibitem{ZhuPatakiTranDinh:2019}
Y.~Zhu, G.~Pataki, and Q.~Tran-Dinh, ``{Sieve-SDP: a simple facial reduction
  algorithm to preprocess semidefinite programs},''
  \href{http://dx.doi.org/10.1007/s12532-019-00164-4}{{\em Math. Program.
  Comput.} {\bfseries 11} (2019) 503--586}.

\bibitem{YeToddMizuno:1994}
Y.~Ye, M.~J. Todd, and S.~Mizuno, ``{An $O(\sqrt{n}L)$-Iteration Homogeneous
  and Self-Dual Linear Programming Algorithm},''
  \href{http://dx.doi.org/10.1287/moor.19.1.53}{{\em Math. Oper. Res.}
  {\bfseries 19} (1994) 53--67}.

\bibitem{deKlerkRoosTerlaky:1997}
E.~de~Klerk, C.~Roos, and T.~Terlaky, ``{Initialization in semidefinite
  programming via a self-dual skew-symmetric embedding},''
  \href{http://dx.doi.org/10.1016/S0167-6377(97)00011-4}{{\em Oper. Res. Lett.}
  {\bfseries 20} (1997) 213--221}.

\bibitem{PermenterFribergAndersen:2017}
F.~Permenter, H.~A. Friberg, and E.~D. Andersen, ``{Solving conic optimization
  problems via self-dual embedding and facial reduction: a unified approach},''
  \href{http://dx.doi.org/10.1137/15M1049415}{{\em SIAM J. Optim.} {\bfseries
  27} (2017) 1257--1282}.

\bibitem{Reehorst:rigorous}
M.~Reehorst, ``{Region bounds, error margins and certified positivity in the
  numerical conformal bootstrap; Or the first mathematically rigorous bounds on
  the 3d Ising CFT critical exponents},''
  \href{http://arxiv.org/abs/2610.20701}{{\ttfamily arXiv:2610.20701
  [hep-th]}}.

\bibitem{SDPB-repo}
D.~Simmons-Duffin, W.~Landry, and V.~Dommes, ``{SDPB} 3.1.0.''
  \url{https://github.com/davidsd/sdpb}, release 3.1.0, 2025.

\bibitem{Go:2020ahx}
M.~Go, ``{An Automated Generation of Bootstrap Equations for Numerical Study of
  Critical Phenomena},'' \href{http://arxiv.org/abs/2006.04173}{{\ttfamily
  arXiv:2006.04173 [hep-th]}}.

\bibitem{Chester:2023ehi}
S.~M. Chester, R.~Dempsey, and S.~S. Pufu, ``{Level repulsion in $ \mathcal{N}
  $ = 4 super-Yang-Mills via integrability, holography, and the bootstrap},''
  \href{http://dx.doi.org/10.1007/JHEP07(2024)059}{{\em JHEP} {\bfseries 07}
  (2024) 059}, \href{http://arxiv.org/abs/2312.12576}{{\ttfamily
  arXiv:2312.12576 [hep-th]}}.

\bibitem{DiPietro:2025ozw}
L.~Di~Pietro, S.~R. Kousvos, M.~Meineri, A.~Piazza, M.~Serone, and A.~Vichi,
  ``{A bootstrap study of confinement in AdS},''
  \href{http://dx.doi.org/10.1007/JHEP08(2026)057}{{\em JHEP} {\bfseries 08}
  (2026) 057}, \href{http://arxiv.org/abs/2512.00150}{{\ttfamily
  arXiv:2512.00150 [hep-th]}}.

\bibitem{Reehorst:2019pzi}
M.~Reehorst, E.~Trevisani, and A.~Vichi, ``{Mixed Scalar-Current bootstrap in
  three dimensions},'' \href{http://dx.doi.org/10.1007/JHEP12(2020)156}{{\em
  JHEP} {\bfseries 12} (2020) 156},
  \href{http://arxiv.org/abs/1911.05747}{{\ttfamily arXiv:1911.05747
  [hep-th]}}.

\bibitem{SlaterTools}
M.~Reehorst, ``{sdpb-slater-tools}: {S}later diagnostics and guarantees for
  {SDPB}.'' \url{https://gitlab.com/Reehorst/sdpb-slater-tools}, tag
  \texttt{paper-v1.3}, 2026.

\bibitem{PolyBoot}
M.~Reehorst, ``{PolyBoot}.'' \url{https://gitlab.com/Reehorst/polyboot}, tag
  \texttt{slater-paper-v1.1}, 2026.

\bibitem{Slater:1950}
M.~Slater, ``{Lagrange Multipliers Revisited},'' Discussion Paper: Mathematics
  403, Cowles Commission for Research in Economics, 1950.

\bibitem{Hercek:2026ext}
F.~Her{\v{c}}ek and N.~Levine, ``{Existence of extremal solutions in the 1d
  conformal bootstrap},''. To appear.

\bibitem{Todd:2001}
M.~J. Todd, ``{Semidefinite optimization},''
  \href{http://dx.doi.org/10.1017/S0962492901000071}{{\em Acta Numer.}
  {\bfseries 10} (2001) 515--560}.

\bibitem{Alizadeh:1997}
F.~Alizadeh, J.-P.~A. Haeberly, and M.~L. Overton, ``Complementarity and
  nondegeneracy in semidefinite programming,''
  \href{http://dx.doi.org/10.1007/BF02614432}{{\em Math. Program.} {\bfseries
  77} no.~2, (1997) 111--128}.

\bibitem{NesterovTodd:1997}
Y.~E. Nesterov and M.~J. Todd, ``{Self-Scaled Barriers and Interior-Point
  Methods for Convex Programming},''
  \href{http://dx.doi.org/10.1287/moor.22.1.1}{{\em Math. Oper. Res.}
  {\bfseries 22} (1997) 1--42}.

\bibitem{NesterovTodd:1998}
Y.~E. Nesterov and M.~J. Todd, ``{Primal-Dual Interior-Point Methods for
  Self-Scaled Cones},'' \href{http://dx.doi.org/10.1137/S1052623495290209}{{\em
  SIAM J. Optim.} {\bfseries 8} (1998) 324--364}.

\bibitem{Mehrotra:1992}
S.~Mehrotra, ``{On the Implementation of a Primal-Dual Interior Point
  Method},'' \href{http://dx.doi.org/10.1137/0802028}{{\em SIAM J. Optim.}
  {\bfseries 2} (1992) 575--601}.

\bibitem{Hogervorst:2013sma}
M.~Hogervorst and S.~Rychkov, ``{Radial Coordinates for Conformal Blocks},''
  \href{http://dx.doi.org/10.1103/PhysRevD.87.106004}{{\em Phys. Rev. D}
  {\bfseries 87} (2013) 106004},
  \href{http://arxiv.org/abs/1303.1111}{{\ttfamily arXiv:1303.1111 [hep-th]}}.

\bibitem{Zangwill:1967}
W.~I. Zangwill, ``{Non-linear programming via penalty functions},''
  \href{http://dx.doi.org/10.1287/mnsc.13.5.344}{{\em Manag. Sci.} {\bfseries
  13} no.~5, (1967) 344--358}.

\bibitem{Bertsekas:1975exact}
D.~P. Bertsekas, ``{Necessary and sufficient conditions for a penalty method to
  be exact},'' \href{http://dx.doi.org/10.1007/BF01681332}{{\em Math. Program.}
  {\bfseries 9} (1975) 87--99}.

\bibitem{HanMangasarian:1979}
S.-P. Han and O.~L. Mangasarian, ``{Exact penalty functions in nonlinear
  programming},'' \href{http://dx.doi.org/10.1007/BF01588250}{{\em Math.
  Program.} {\bfseries 17} (1979) 251--269}.

\bibitem{ElShowk:2012hu}
S.~El-Showk and M.~F. Paulos, ``{Bootstrapping Conformal Field Theories with
  the Extremal Functional Method},''
  \href{http://dx.doi.org/10.1103/PhysRevLett.111.241601}{{\em Phys. Rev.
  Lett.} {\bfseries 111} (2013) 241601},
  \href{http://arxiv.org/abs/1211.2810}{{\ttfamily arXiv:1211.2810 [hep-th]}}.

\bibitem{Lofberg:2009}
J.~L{\"o}fberg, ``{Pre- and post-processing sum-of-squares programs in
  practice},'' \href{http://dx.doi.org/10.1109/TAC.2009.2017144}{{\em IEEE
  Trans. Automat. Control} {\bfseries 54} (2009) 1007--1011}.

\bibitem{WakiMuramatsu:2010}
H.~Waki and M.~Muramatsu, ``{A facial reduction algorithm for finding sparse
  SOS representations},''
  \href{http://dx.doi.org/10.1016/j.orl.2010.05.011}{{\em Oper. Res. Lett.}
  {\bfseries 38} (2010) 361--365}.

\bibitem{PermenterParrilo:2014}
F.~Permenter and P.~A. Parrilo,
  \href{http://dx.doi.org/10.1109/CDC.2014.7040427}{``{Basis selection for SOS
  programs via facial reduction and polyhedral approximations},''} in {\em 53rd
  IEEE Conference on Decision and Control}, pp.~6615--6620.
\newblock 2014.

\bibitem{McLinden:1980}
L.~McLinden, ``{An analogue of Moreau's proximation theorem, with application
  to the nonlinear complementarity problem},''
  \href{http://dx.doi.org/10.2140/pjm.1980.88.101}{{\em Pacific J. Math.}
  {\bfseries 88} (1980) 101--161}.

\bibitem{Megiddo:1989}
N.~Megiddo, \href{http://dx.doi.org/10.1007/978-1-4613-9617-8_8}{``{Pathways to
  the optimal set in linear programming},''} in {\em Progress in Mathematical
  Programming: Interior-Point and Related Methods}, N.~Megiddo, ed.,
  pp.~131--158.
\newblock Springer, 1989.

\bibitem{GulerYe:1993}
O.~G{\"u}ler and Y.~Ye, ``{Convergence behavior of interior-point
  algorithms},'' \href{http://dx.doi.org/10.1007/BF01580610}{{\em Math.
  Program.} {\bfseries 60} (1993) 215--228}.

\bibitem{GoldfarbScheinberg:1998}
D.~Goldfarb and K.~Scheinberg, ``{Interior point trajectories in semidefinite
  programming},'' \href{http://dx.doi.org/10.1137/S105262349630009X}{{\em SIAM
  J. Optim.} {\bfseries 8} (1998) 871--886}.

\bibitem{LuoSturmZhang:1998}
Z.-Q. Luo, J.~F. Sturm, and S.~Zhang, ``{Superlinear convergence of a symmetric
  primal-dual path following algorithm for semidefinite programming},''
  \href{http://dx.doi.org/10.1137/S1052623496299187}{{\em SIAM J. Optim.}
  {\bfseries 8} (1998) 59--81}.

\bibitem{HalickaDeKlerkRoos:2002}
M.~Halick{\'a}, E.~de~Klerk, and C.~Roos, ``{On the convergence of the central
  path in semidefinite optimization},''
  \href{http://dx.doi.org/10.1137/S1052623401390793}{{\em SIAM J. Optim.}
  {\bfseries 12} (2002) 1090--1099}.

\bibitem{Sturm:2000}
J.~F. Sturm, ``{Error Bounds for Linear Matrix Inequalities},''
  \href{http://dx.doi.org/10.1137/S1052623498338606}{{\em SIAM J. Optim.}
  {\bfseries 10} (2000) 1228--1248}.

\bibitem{CheungSchurrWolkowicz:2013}
Y.-L. Cheung, S.~Schurr, and H.~Wolkowicz,
  \href{http://dx.doi.org/10.1007/978-1-4614-7621-4_12}{``{Preprocessing and
  regularization for degenerate semidefinite programs},''} in {\em
  Computational and Analytical Mathematics}, vol.~50 of {\em Springer
  Proceedings in Mathematics \& Statistics}, pp.~251--303.
\newblock Springer, 2013.

\bibitem{Erramilli:2022kgp}
R.~S. Erramilli, L.~V. Iliesiu, P.~Kravchuk, A.~Liu, D.~Poland, and
  D.~Simmons-Duffin, ``{The Gross-Neveu-Yukawa archipelago},''
  \href{http://dx.doi.org/10.1007/JHEP02(2023)036}{{\em JHEP} {\bfseries 02}
  (2023) 036}, \href{http://arxiv.org/abs/2210.02492}{{\ttfamily
  arXiv:2210.02492 [hep-th]}}.

\bibitem{Kousvos:2024dlz}
S.~R. Kousvos, A.~Piazza, and A.~Vichi, ``{Exploring replica-Potts CFTs in two
  dimensions},'' \href{http://dx.doi.org/10.1007/JHEP11(2024)030}{{\em JHEP}
  {\bfseries 11} (2024) 030}, \href{http://arxiv.org/abs/2405.19416}{{\ttfamily
  arXiv:2405.19416 [hep-th]}}.

\bibitem{Chang:2024whx}
C.-H. Chang, V.~Dommes, R.~S. Erramilli, A.~Homrich, P.~Kravchuk, A.~Liu, M.~S.
  Mitchell, D.~Poland, and D.~Simmons-Duffin, ``{Bootstrapping the 3d Ising
  stress tensor},'' \href{http://dx.doi.org/10.1007/JHEP03(2025)136}{{\em JHEP}
  {\bfseries 03} (2025) 136}, \href{http://arxiv.org/abs/2411.15300}{{\ttfamily
  arXiv:2411.15300 [hep-th]}}.

\bibitem{Chester:2020iyt}
S.~M. Chester, W.~Landry, J.~Liu, D.~Poland, D.~Simmons-Duffin, N.~Su, and
  A.~Vichi, ``{Bootstrapping Heisenberg magnets and their cubic instability},''
  \href{http://dx.doi.org/10.1103/PhysRevD.104.105013}{{\em Phys. Rev. D}
  {\bfseries 104} no.~10, (2021) 105013},
  \href{http://arxiv.org/abs/2011.14647}{{\ttfamily arXiv:2011.14647
  [hep-th]}}.

\bibitem{Liu:2020tpf}
J.~Liu, D.~Meltzer, D.~Poland, and D.~Simmons-Duffin, ``{The Lorentzian
  inversion formula and the spectrum of the 3d O(2) CFT},''
  \href{http://dx.doi.org/10.1007/JHEP09(2020)115}{{\em JHEP} {\bfseries 09}
  (2020) 115}, \href{http://arxiv.org/abs/2007.07914}{{\ttfamily
  arXiv:2007.07914 [hep-th]}}. [Erratum: JHEP 01, 206 (2021)].

\bibitem{Behan:2016dtz}
C.~Behan, ``{PyCFTBoot: A flexible interface for the conformal bootstrap},''
  \href{http://dx.doi.org/10.4208/cicp.OA-2016-0107}{{\em Commun. Comput.
  Phys.} {\bfseries 22} no.~1, (2017) 1--38},
  \href{http://arxiv.org/abs/1602.02810}{{\ttfamily arXiv:1602.02810
  [hep-th]}}.

\bibitem{simpleboot}
N.~Su, ``{simpleboot}: a {M}athematica framework for bootstrap calculations.''
  \url{https://gitlab.com/bootstrapcollaboration/simpleboot}, 2020.

\bibitem{cboot}
T.~Ohtsuki, ``{cboot}: a {S}age module for the conformal bootstrap.''
  \url{https://github.com/tohtsky/cboot}, 2016.

\bibitem{Poland:2011ey}
D.~Poland, D.~Simmons-Duffin, and A.~Vichi, ``{Carving Out the Space of 4D
  CFTs},'' \href{http://dx.doi.org/10.1007/JHEP05(2012)110}{{\em JHEP}
  {\bfseries 05} (2012) 110}, \href{http://arxiv.org/abs/1109.5176}{{\ttfamily
  arXiv:1109.5176 [hep-th]}}.

\bibitem{Kos:2013tga}
F.~Kos, D.~Poland, and D.~Simmons-Duffin, ``{Bootstrapping the $O(N)$ vector
  models},'' \href{http://dx.doi.org/10.1007/JHEP06(2014)091}{{\em JHEP}
  {\bfseries 06} (2014) 091}, \href{http://arxiv.org/abs/1307.6856}{{\ttfamily
  arXiv:1307.6856 [hep-th]}}.

\bibitem{scalar_blocks}
W.~Landry, P.~Kravchuk, V.~Dommes, and C.~Behan, ``{scalar\_blocks}.''
  \url{https://gitlab.com/bootstrapcollaboration/scalar_blocks}, 2021.
\newblock Release 1.0.1.

\bibitem{Simmons-Duffin:2016wlq}
D.~Simmons-Duffin, ``{The Lightcone Bootstrap and the Spectrum of the 3d Ising
  CFT},'' \href{http://dx.doi.org/10.1007/JHEP03(2017)086}{{\em JHEP}
  {\bfseries 03} (2017) 086}, \href{http://arxiv.org/abs/1612.08471}{{\ttfamily
  arXiv:1612.08471 [hep-th]}}.

\bibitem{CREATE}
{King's College London}, ``{King's Computational Research, Engineering and
  Technology Environment (CREATE)}.'' \url{https://doi.org/10.18742/rnvf-m076},
  2024.
\newblock Retrieved Nov 13, 2024.

\bibitem{GeninHachezNesterovVanDooren:2003}
Y.~Genin, Y.~Hachez, Y.~Nesterov, and P.~Van~Dooren, ``{Optimization Problems
  over Positive Pseudopolynomial Matrices},''
  \href{http://dx.doi.org/10.1137/S0895479803374840}{{\em SIAM J. Matrix Anal.
  Appl.} {\bfseries 25} (2003) 57--79}.

\bibitem{Kos:2015mba}
F.~Kos, D.~Poland, D.~Simmons-Duffin, and A.~Vichi, ``{Bootstrapping the $O(N)$
  Archipelago},'' \href{http://dx.doi.org/10.1007/JHEP11(2015)106}{{\em JHEP}
  {\bfseries 11} (2015) 106}, \href{http://arxiv.org/abs/1504.07997}{{\ttfamily
  arXiv:1504.07997 [hep-th]}}.

\bibitem{Iliesiu:2015qra}
L.~Iliesiu, F.~Kos, D.~Poland, S.~S. Pufu, D.~Simmons-Duffin, and R.~Yacoby,
  ``{Bootstrapping 3D Fermions},''
  \href{http://dx.doi.org/10.1007/JHEP03(2016)120}{{\em JHEP} {\bfseries 03}
  (2016) 120}, \href{http://arxiv.org/abs/1508.00012}{{\ttfamily
  arXiv:1508.00012 [hep-th]}}.

\bibitem{Chester:2015qca}
S.~M. Chester, S.~Giombi, L.~V. Iliesiu, I.~R. Klebanov, S.~S. Pufu, and
  R.~Yacoby, ``{Accidental Symmetries and the Conformal Bootstrap},''
  \href{http://dx.doi.org/10.1007/JHEP01(2016)110}{{\em JHEP} {\bfseries 01}
  (2016) 110}, \href{http://arxiv.org/abs/1507.04424}{{\ttfamily
  arXiv:1507.04424 [hep-th]}}.

\bibitem{Simmons-Duffin:2016gjk}
D.~Simmons-Duffin, \href{http://dx.doi.org/10.1142/9789813149441_0001}{``{The
  Conformal Bootstrap},''} in {\em {Theoretical Advanced Study Institute in
  Elementary Particle Physics}: {New Frontiers in Fields and Strings}},
  pp.~1--74.
\newblock 2017.
\newblock \href{http://arxiv.org/abs/1602.07982}{{\ttfamily arXiv:1602.07982
  [hep-th]}}.

\bibitem{Dymarsky:2017xzb}
A.~Dymarsky, J.~Penedones, E.~Trevisani, and A.~Vichi, ``{Charting the space of
  3D CFTs with a continuous global symmetry},''
  \href{http://dx.doi.org/10.1007/JHEP05(2019)098}{{\em JHEP} {\bfseries 05}
  (2019) 098}, \href{http://arxiv.org/abs/1705.04278}{{\ttfamily
  arXiv:1705.04278 [hep-th]}}.

\bibitem{delaFuente:2019hbl}
A.~de~la Fuente, ``{Bootstrapping mixed correlators in the 2D Ising model},''
  \href{http://arxiv.org/abs/1904.09801}{{\ttfamily arXiv:1904.09801
  [hep-th]}}.

\bibitem{Gimenez-Grau:2020jrx}
A.~Gimenez-Grau and P.~Liendo, ``{Bootstrapping Coulomb and Higgs branch
  operators},'' \href{http://dx.doi.org/10.1007/JHEP01(2021)175}{{\em JHEP}
  {\bfseries 01} (2021) 175}, \href{http://arxiv.org/abs/2006.01847}{{\ttfamily
  arXiv:2006.01847 [hep-th]}}.

\bibitem{Chester:2019wfx}
S.~M. Chester, ``{Weizmann lectures on the numerical conformal bootstrap},''
  \href{http://dx.doi.org/10.1016/j.physrep.2023.10.008}{{\em Phys. Rept.}
  {\bfseries 1045} (2023) 1--44},
  \href{http://arxiv.org/abs/1907.05147}{{\ttfamily arXiv:1907.05147
  [hep-th]}}.

\bibitem{Reehorst:2020phk}
M.~Reehorst, M.~Refinetti, and A.~Vichi, ``{Bootstrapping traceless symmetric
  $O(N)$ scalars},''
  \href{http://dx.doi.org/10.21468/SciPostPhys.14.4.068}{{\em SciPost Phys.}
  {\bfseries 14} no.~4, (2023) 068},
  \href{http://arxiv.org/abs/2012.08533}{{\ttfamily arXiv:2012.08533
  [hep-th]}}.

\bibitem{Baggio:2017mas}
M.~Baggio, N.~Bobev, S.~M. Chester, E.~Lauria, and S.~S. Pufu, ``{Decoding a
  Three-Dimensional Conformal Manifold},''
  \href{http://dx.doi.org/10.1007/JHEP02(2018)062}{{\em JHEP} {\bfseries 02}
  (2018) 062}, \href{http://arxiv.org/abs/1712.02698}{{\ttfamily
  arXiv:1712.02698 [hep-th]}}.

\bibitem{Karateev:2019pvw}
D.~Karateev, P.~Kravchuk, M.~Serone, and A.~Vichi, ``{Fermion Conformal
  Bootstrap in 4d},'' \href{http://dx.doi.org/10.1007/JHEP06(2019)088}{{\em
  JHEP} {\bfseries 06} (2019) 088},
  \href{http://arxiv.org/abs/1902.05969}{{\ttfamily arXiv:1902.05969
  [hep-th]}}.

\bibitem{Kousvos:2022ewl}
S.~R. Kousvos and A.~Stergiou, ``{CFTs with $\mathbf{U(m)\times U(n)}$ global
  symmetry in 3D and the chiral phase transition of QCD},''
  \href{http://dx.doi.org/10.21468/SciPostPhys.15.2.075}{{\em SciPost Phys.}
  {\bfseries 15} no.~2, (2023) 075},
  \href{http://arxiv.org/abs/2209.02837}{{\ttfamily arXiv:2209.02837
  [hep-th]}}.

\bibitem{Chester:2022hzt}
S.~M. Chester and N.~Su, ``{Upper critical dimension of the 3-state Potts
  model},'' \href{http://dx.doi.org/10.1103/vqss-l6f9}{{\em Phys. Rev. D}
  {\bfseries 111} no.~12, (2025) L121701},
  \href{http://arxiv.org/abs/2210.09091}{{\ttfamily arXiv:2210.09091
  [hep-th]}}.

\bibitem{He:2023ewx}
Y.-C. He, J.~Rong, N.~Su, and A.~Vichi, ``{Non-Abelian currents bootstrap},''
  \href{http://dx.doi.org/10.1007/JHEP03(2024)175}{{\em JHEP} {\bfseries 03}
  (2024) 175}, \href{http://arxiv.org/abs/2302.11585}{{\ttfamily
  arXiv:2302.11585 [hep-th]}}.

\bibitem{Kousvos:2025ext}
S.~R. Kousvos and A.~Stergiou, ``{Redundancy channels in the conformal
  bootstrap},'' \href{http://dx.doi.org/10.1007/JHEP01(2026)073}{{\em JHEP}
  {\bfseries 01} (2026) 073}, \href{http://arxiv.org/abs/2507.05338}{{\ttfamily
  arXiv:2507.05338 [hep-th]}}.

\bibitem{Dymarsky:2017yzx}
A.~Dymarsky, F.~Kos, P.~Kravchuk, D.~Poland, and D.~Simmons-Duffin, ``{The 3d
  Stress-Tensor Bootstrap},''
  \href{http://dx.doi.org/10.1007/JHEP02(2018)164}{{\em JHEP} {\bfseries 02}
  (2018) 164}, \href{http://arxiv.org/abs/1708.05718}{{\ttfamily
  arXiv:1708.05718 [hep-th]}}.

\bibitem{Stergiou:2018gjj}
A.~Stergiou, ``{Bootstrapping hypercubic and hypertetrahedral theories in three
  dimensions},'' \href{http://dx.doi.org/10.1007/JHEP05(2018)035}{{\em JHEP}
  {\bfseries 05} (2018) 035}, \href{http://arxiv.org/abs/1801.07127}{{\ttfamily
  arXiv:1801.07127 [hep-th]}}.

\bibitem{Kousvos:2018rhl}
S.~R. Kousvos and A.~Stergiou, ``{Bootstrapping Mixed Correlators in
  Three-Dimensional Cubic Theories},''
  \href{http://dx.doi.org/10.21468/SciPostPhys.6.3.035}{{\em SciPost Phys.}
  {\bfseries 6} no.~3, (2019) 035},
  \href{http://arxiv.org/abs/1810.10015}{{\ttfamily arXiv:1810.10015
  [hep-th]}}.

\bibitem{Stergiou:2019dcv}
A.~Stergiou, ``{Bootstrapping MN and Tetragonal CFTs in Three Dimensions},''
  \href{http://dx.doi.org/10.21468/SciPostPhys.7.1.010}{{\em SciPost Phys.}
  {\bfseries 7} (2019) 010}, \href{http://arxiv.org/abs/1904.00017}{{\ttfamily
  arXiv:1904.00017 [hep-th]}}.

\bibitem{Atanasov:2018kqw}
A.~Atanasov, A.~Hillman, and D.~Poland, ``{Bootstrapping the Minimal 3D
  SCFT},'' \href{http://dx.doi.org/10.1007/JHEP11(2018)140}{{\em JHEP}
  {\bfseries 11} (2018) 140}, \href{http://arxiv.org/abs/1807.05702}{{\ttfamily
  arXiv:1807.05702 [hep-th]}}.

\bibitem{Dowens:2020cua}
M.~T. Dowens and C.~A. Hooley, ``{Multi-fixed point numerical conformal
  bootstrap: a case study with structured global symmetry},''
  \href{http://dx.doi.org/10.1007/JHEP03(2021)147}{{\em JHEP} {\bfseries 03}
  (2021) 147}, \href{http://arxiv.org/abs/2004.14978}{{\ttfamily
  arXiv:2004.14978 [hep-th]}}.

\bibitem{Bae:2016yna}
J.-B. Bae, K.~Lee, and S.~Lee, ``{Bootstrapping Pure Quantum Gravity in
  AdS3},'' \href{http://arxiv.org/abs/1610.05814}{{\ttfamily arXiv:1610.05814
  [hep-th]}}.

\bibitem{Cappelli:2018vir}
A.~Cappelli, L.~Maffi, and S.~Okuda, ``{Critical Ising Model in Varying
  Dimension by Conformal Bootstrap},''
  \href{http://dx.doi.org/10.1007/JHEP01(2019)161}{{\em JHEP} {\bfseries 01}
  (2019) 161}, \href{http://arxiv.org/abs/1811.07751}{{\ttfamily
  arXiv:1811.07751 [hep-th]}}.

\bibitem{Henriksson:2020fqi}
J.~Henriksson, S.~R. Kousvos, and A.~Stergiou, ``{Analytic and Numerical
  Bootstrap of CFTs with $O(m)\times O(n)$ Global Symmetry in 3D},''
  \href{http://dx.doi.org/10.21468/SciPostPhys.9.3.035}{{\em SciPost Phys.}
  {\bfseries 9} no.~3, (2020) 035},
  \href{http://arxiv.org/abs/2004.14388}{{\ttfamily arXiv:2004.14388
  [hep-th]}}.

\bibitem{Afkhami-Jeddi:2020ezh}
N.~Afkhami-Jeddi, H.~Cohn, T.~Hartman, and A.~Tajdini, ``{Free partition
  functions and an averaged holographic duality},''
  \href{http://dx.doi.org/10.1007/JHEP01(2021)130}{{\em JHEP} {\bfseries 01}
  (2021) 130}, \href{http://arxiv.org/abs/2006.04839}{{\ttfamily
  arXiv:2006.04839 [hep-th]}}.

\bibitem{Afkhami-Jeddi:2021iuw}
N.~Afkhami-Jeddi, ``{Conformal bootstrap deformations},''
  \href{http://dx.doi.org/10.1007/JHEP09(2022)225}{{\em JHEP} {\bfseries 09}
  (2022) 225}, \href{http://arxiv.org/abs/2111.01799}{{\ttfamily
  arXiv:2111.01799 [hep-th]}}.

\bibitem{Henriksson:2022gpa}
J.~Henriksson, S.~R. Kousvos, and M.~Reehorst, ``{Spectrum continuity and level
  repulsion: the Ising CFT from infinitesimal to finite
  {\ensuremath{\varepsilon}}},''
  \href{http://dx.doi.org/10.1007/JHEP02(2023)218}{{\em JHEP} {\bfseries 02}
  (2023) 218}, \href{http://arxiv.org/abs/2207.10118}{{\ttfamily
  arXiv:2207.10118 [hep-th]}}.

\bibitem{Gowdigere:2018lxz}
C.~N. Gowdigere, J.~Santara, and Sumedha, ``{Conformal bootstrap signatures of
  the tricritical Ising universality class},''
  \href{http://dx.doi.org/10.1103/PhysRevD.101.116020}{{\em Phys. Rev. D}
  {\bfseries 101} no.~11, (2020) 116020},
  \href{http://arxiv.org/abs/1811.11442}{{\ttfamily arXiv:1811.11442
  [hep-th]}}.

\bibitem{Rong:2020gbi}
J.~Rong and J.~Zhu, ``{On the {\ensuremath{\phi}}$^{3}$ theory above six
  dimensions},'' \href{http://dx.doi.org/10.1007/JHEP04(2020)151}{{\em JHEP}
  {\bfseries 04} (2020) 151}, \href{http://arxiv.org/abs/2001.10864}{{\ttfamily
  arXiv:2001.10864 [hep-th]}}.

\bibitem{Nakayama:2023wrx}
Y.~Nakayama, ``{Functional renormalization group approach to dipolar fixed
  point which is scale invariant but nonconformal},''
  \href{http://dx.doi.org/10.1103/PhysRevD.110.025020}{{\em Phys. Rev. D}
  {\bfseries 110} no.~2, (2024) 025020},
  \href{http://arxiv.org/abs/2309.15307}{{\ttfamily arXiv:2309.15307
  [hep-th]}}.

\bibitem{Nakayama:2024iiw}
Y.~Nakayama, ``{Who told you magnetization is a vector in
  4{\ensuremath{-}}{\ensuremath{\epsilon}} dimensions?},''
  \href{http://dx.doi.org/10.1142/S0217751X25500411}{{\em Int. J. Mod. Phys. A}
  {\bfseries 40} no.~15, (2025) 2550041},
  \href{http://arxiv.org/abs/2404.14669}{{\ttfamily arXiv:2404.14669
  [hep-th]}}.

\bibitem{Manenti:2021elk}
A.~Manenti and A.~Vichi, ``{Exploring $SU(N)$ adjoint correlators in $3d$},''
  \href{http://arxiv.org/abs/2101.07318}{{\ttfamily arXiv:2101.07318
  [hep-th]}}.

\bibitem{Bissi:2020jve}
A.~Bissi, A.~Manenti, and A.~Vichi, ``{Bootstrapping mixed correlators in $
  \mathcal{N} $ = 4 super Yang-Mills},''
  \href{http://dx.doi.org/10.1007/JHEP05(2021)111}{{\em JHEP} {\bfseries 05}
  (2021) 111}, \href{http://arxiv.org/abs/2010.15126}{{\ttfamily
  arXiv:2010.15126 [hep-th]}}.

\bibitem{Albayrak:2021xtd}
S.~Albayrak, R.~S. Erramilli, Z.~Li, D.~Poland, and Y.~Xin, ``{Bootstrapping
  $N_f$=4 conformal QED$_3$},''
  \href{http://dx.doi.org/10.1103/PhysRevD.105.085008}{{\em Phys. Rev. D}
  {\bfseries 105} no.~8, (2022) 085008},
  \href{http://arxiv.org/abs/2112.02106}{{\ttfamily arXiv:2112.02106
  [hep-th]}}.

\bibitem{Chester:2025uxb}
S.~M. Chester, A.~Piazza, M.~Reehorst, and N.~Su, ``{Bootstrapping the simplest
  deconfined quantum critical point},''
  \href{http://dx.doi.org/10.1103/ym81-bksd}{{\em Phys. Rev. D} {\bfseries 113}
  no.~8, (2026) L081701}, \href{http://arxiv.org/abs/2507.06283}{{\ttfamily
  arXiv:2507.06283 [hep-th]}}.

\bibitem{Ghosh:2026dse}
K.~Ghosh and Z.~Zheng, ``{Bootstrapping 3D Conformal Field Theories with
  Product Analytic Functionals},''
  \href{http://arxiv.org/abs/2608.06497}{{\ttfamily arXiv:2608.06497
  [hep-th]}}.

\bibitem{Erramilli:2020rlr}
R.~S. Erramilli, L.~V. Iliesiu, P.~Kravchuk, W.~Landry, D.~Poland, and
  D.~Simmons-Duffin, ``{blocks{\_}3d: software for general 3d conformal
  blocks},'' \href{http://dx.doi.org/10.1007/JHEP11(2021)006}{{\em JHEP}
  {\bfseries 11} (2021) 006}, \href{http://arxiv.org/abs/2011.01959}{{\ttfamily
  arXiv:2011.01959 [hep-th]}}.

\bibitem{Bonifacio:2021aqf}
J.~Bonifacio, ``{Bootstrapping closed hyperbolic surfaces},''
  \href{http://dx.doi.org/10.1007/JHEP03(2022)093}{{\em JHEP} {\bfseries 03}
  (2022) 093}, \href{http://arxiv.org/abs/2111.13215}{{\ttfamily
  arXiv:2111.13215 [hep-th]}}.

\bibitem{Bonifacio:2021msa}
J.~Bonifacio, ``{Bootstrap bounds on closed hyperbolic manifolds},''
  \href{http://dx.doi.org/10.1007/JHEP02(2022)025}{{\em JHEP} {\bfseries 02}
  (2022) 025}, \href{http://arxiv.org/abs/2107.09674}{{\ttfamily
  arXiv:2107.09674 [hep-th]}}.

\bibitem{Karateev:2025sjw}
D.~Karateev, P.~Kravchuk, A.~Manenti, M.~Reehorst, and A.~Vichi, ``{Bounds on
  Abelian Currents in 4d CFTs},''
  \href{http://arxiv.org/abs/2512.20710}{{\ttfamily arXiv:2512.20710
  [hep-th]}}.

\bibitem{Erramilli:2026ykj}
R.~S. Erramilli and M.~S. Mitchell, ``{Towards 3D CFT Cartography with the
  Stress Tensor Bootstrap},'' \href{http://arxiv.org/abs/2602.13383}{{\ttfamily
  arXiv:2602.13383 [hep-th]}}.

\bibitem{Lanzetta:2025xfw}
R.~A. Lanzetta, S.~Liu, and M.~A. Metlitski, ``{The beginning of the endpoint
  bootstrap for conformal line defects},''
  \href{http://arxiv.org/abs/2508.14964}{{\ttfamily arXiv:2508.14964
  [cond-mat.str-el]}}.

\bibitem{Bonifacio:2020xoc}
J.~Bonifacio and K.~Hinterbichler, ``{Bootstrap Bounds on Closed Einstein
  Manifolds},'' \href{http://dx.doi.org/10.1007/JHEP10(2020)069}{{\em JHEP}
  {\bfseries 10} (2020) 069}, \href{http://arxiv.org/abs/2007.10337}{{\ttfamily
  arXiv:2007.10337 [hep-th]}}.

\bibitem{Manenti:2019kbl}
A.~Manenti, A.~Stergiou, and A.~Vichi, ``{Implications of ANEC for SCFTs in
  four dimensions},'' \href{http://dx.doi.org/10.1007/JHEP01(2020)093}{{\em
  JHEP} {\bfseries 01} (2020) 093},
  \href{http://arxiv.org/abs/1905.09293}{{\ttfamily arXiv:1905.09293
  [hep-th]}}.

\bibitem{Correia:2022dyp}
M.~Correia, J.~Penedones, and A.~Vuignier, ``{Injecting the UV into the
  bootstrap: Ising Field Theory},''
  \href{http://dx.doi.org/10.1007/JHEP08(2023)108}{{\em JHEP} {\bfseries 08}
  (2023) 108}, \href{http://arxiv.org/abs/2212.03917}{{\ttfamily
  arXiv:2212.03917 [hep-th]}}.

\bibitem{Li:2025tub}
W.~Li and X.~Su, ``{Bootstrapping Yang-Mills matrix integrals},''
  \href{http://arxiv.org/abs/2510.06704}{{\ttfamily arXiv:2510.06704
  [hep-th]}}.

\bibitem{Kull:2022wof}
I.~Kull, N.~Schuch, B.~Dive, and M.~Navascu{\'e}s, ``{Lower Bounds on
  Ground-State Energies of Local Hamiltonians through the Renormalization
  Group},'' \href{http://dx.doi.org/10.1103/PhysRevX.14.021008}{{\em Phys. Rev.
  X} {\bfseries 14} no.~2, (2024) 021008},
  \href{http://arxiv.org/abs/2212.03014}{{\ttfamily arXiv:2212.03014
  [quant-ph]}}.

\end{thebibliography}\endgroup
